\documentclass[pdflatex,sn-basic]{sn-jnl}

\usepackage{graphicx}%
\usepackage{multirow}%
\usepackage{amsmath,amssymb,amsfonts}%
\usepackage{amsthm}%
\usepackage{mathrsfs}%
\usepackage[title]{appendix}%
\usepackage{xcolor}%
\usepackage{textcomp}%
\usepackage{manyfoot}%
\usepackage{booktabs}%
\usepackage{algorithm}%
\usepackage{algorithmicx}%
\usepackage{algpseudocode}%
\usepackage{listings}%

\theoremstyle{thmstyleone}%
\theoremstyle{thmstyletwo}%

\theoremstyle{thmstylethree}%

\definecolor{kh}{cmyk}{1.0, 0., 0.25, 0.45}

\defcitealias{protection1991icrp}{ICRP 60 (1991)}
\defcitealias{valentin2007icrp}{ICRP 103 (2007)}
\defcitealias{dietze2013icrp}{ICRP 123 (2013)}

\begin{document}

\title[Chapter4]{Stellar impact on exoplanetary atmospheric evolution and habitability}

\author*[1]{\fnm{Antígona} \sur{Segura}}\email{antigona@nucleares.unam.mx}

\author[2,3]{\fnm{Manuel} \sur{Güdel}}\email{manuel.guedel@univie.ac.at}

\author[4]{\fnm{Konstantin} \sur{Herbst}}\email{konstantin.herbst@geo.uio.no}

\author[5,6]{\fnm{Dibyendu} \sur{Nandy}}\email{dnandi@iiserkol.ac.in}

\author[7]{\fnm{Arghyadeep} \sur{Paul}}\email{arghyadeepp@gmail.com}

\author[8]{\fnm{Raissa} \sur{Estrela}}\email{raissa.estrela@jpl.nasa.gov}

\author[9]{\fnm{Allison} \sur{Youngblood}}\email{allison.a.youngblood@nasa.gov}

\affil*[1]{\orgdiv{Instituto de Ciencias Nucleares}, \orgname{Universidad Nacional Autónoma de México}, \orgaddress{\street{Circuito de la Investigac\'on Cient\'ifica S/N}, \city{Coyoac\'an}, \postcode{04510}, \state{Ciudad de M\'exico}, \country{Mexico}}}

\affil[2]{\orgdiv{Department of Astrophysics}, \orgname{University of Vienna}, \orgaddress{\street{T\"urkenschanzstr. 17}, \postcode{1180} \city{Vienna}, \country{Austria}}}

\affil[3]{\orgdiv{ASTRON}, \orgname{Netherlands Institute for Radio Astronomy}, \orgaddress{\street{Oude Hoogeveensedijk 4}, \postcode{7991}, \city{PD Dwingeloo}, \country{The Netherlands}}}

\affil[4]{\orgdiv{Centre for Planetary Habitability (PHAB), Department of Geosciences}, \orgname{University of Oslo}, \orgaddress{\street{Sem Sælands vei 2A}, \postcode{0371},  \city{Oslo}, \country{Norway}}}

\affil[5]{\orgdiv{Center of Excellence in Space Sciences India}, \orgname{Indian Institute of Science Education and Research Kolkata}, 
\orgaddress{\street{Mohanpur 741246}, \postcode{741246}, \city{West Bengal}, \country{India}}}

\affil[6]{\orgdiv{Department of Physical Sciences}, \orgname{Indian Institute of Science Education and Research Kolkata}, \orgaddress{\street{Mohanpur 741246}, \postcode{741246}, \city{West Bengal}, \country{India}}}

\affil[7]{\orgdiv{Université Paris Cité, Université Paris-Saclay}, \orgname{CEA, CNRS, AIM},\orgaddress{\street{Gif-sur-Yvette cedex}, \postcode{F-91191}, \city{}, \country{France}}}

\affil[8]{\orgdiv{Jet Propulsion Laboratory}, \orgname{California Institute of Technology}, \orgaddress{\street{4800 Oak Grove Drive}, \postcode{91109}, \city{Pasadena, CA}, \country{USA}}}

\affil[9]{\orgdiv{Goddard Space Flight Center}, \orgname{NASA}, \orgaddress{\street{Mail Code 667}, \postcode{20771}, \city{Greenbelt, MD}, \country{USA}}}


\abstract{This chapter will review the deep connection of planetary habitability and stellar irradiation. We present the long-term stellar evolution as one of the drivers of atmospheric escape and climate changes on exoplanets, as well as the chemistry driven by stellar UV and stellar energetic particles. Habitability is presented in the context of short and long-term stellar variability and evolution to layout what we understand and what we need to know about stellar irradiation to constrain our planetary atmospheric models and choose the best targets for future missions that may characterize those exoplanets. }




\maketitle

\tableofcontents

\section{Introduction}\label{Intro}

Habitability is defined as the “capability to sustain life or for the origin of life” \citep{montoya2023habitability}. The three key elements for life are energy sources, bio-elements know as CHONPS (carbon, hydrogen, oxygen, nitrogen, phosphorus and sulfur), and, for carbon-based life, liquid water. 

The concept of the habitable zone (HZ) was derived from the requirement of liquid water, being the circumstellar zone where a rocky planet with atmosphere may retain liquid water on its surface \citep{kopparapu2013}. Here, a rocky planet is any solid planet composed mostly of silicates and iron. For exoplanets, planets around other stars, we can only make assumptions about the possible budget of elements required for life based on the star’s metallicity \citep[e.g.][]{wangetal-2019}. Planetary mass and radius combined with models of exoplanet interiors can help to constrain the composition of an exoplanet, including its possible water budget \citep{baumeister2025fundamentals}
but there are caveats. The composition of planets with sizes $>1$ Earth radius (R$_\oplus$) and $<4$ R$_\oplus$ is not restricted by the mass-radius relation 
\citep{valencia2007detailed, zeng2008computational}, which poses a challenge for identifying potentially habitable rocky planets when the only data is the radius and/or the mass of the planet. 

Planets in the HZ with masses lower than 1$M_{\oplus}$ are unlikely to keep their atmospheres over very long times. Mars' atmospheric pressure has dropped massively over geological timescales \citep{kite2019, warren2019} even though it orbits within the limits of the solar HZ \citep{kopparapu2013}. On the other hand, we do not have yet enough data to constrain the volcanic activity on rocky exoplanets with masses $>$ M$_{\oplus}$ that are relevant for recycling CO$_2$ atmospheres \citep{rushby2018long, oosterlo-etal2021}

A planet in the habitable zone requires an atmosphere to maintain liquid water on the surface. Such atmosphere should be dense enough to provide the pressure required for liquid water but also should have greenhouse gases to warm the surface. Usually, the atmospheres assumed in models to calculate the habitable zone are composed of CO$_2$, H$_2$O and N$_2$ \citep{kastingHZ1993, kopparapu2013}, all of them degassed from the interior of planets made of silicates. In turn, the retention of a planetary atmosphere implies an active carbonates-silicates cycle to replenish the CO$_2$ sequestered by the surface, and that the interaction with the stellar wind and high energy radiation does not result in a catastrophic atmospheric loss \citep[e.g.][]{Kasting-Catling2003}.

\begin{figure}[t]
\begin{center}
\includegraphics[width=\textwidth]{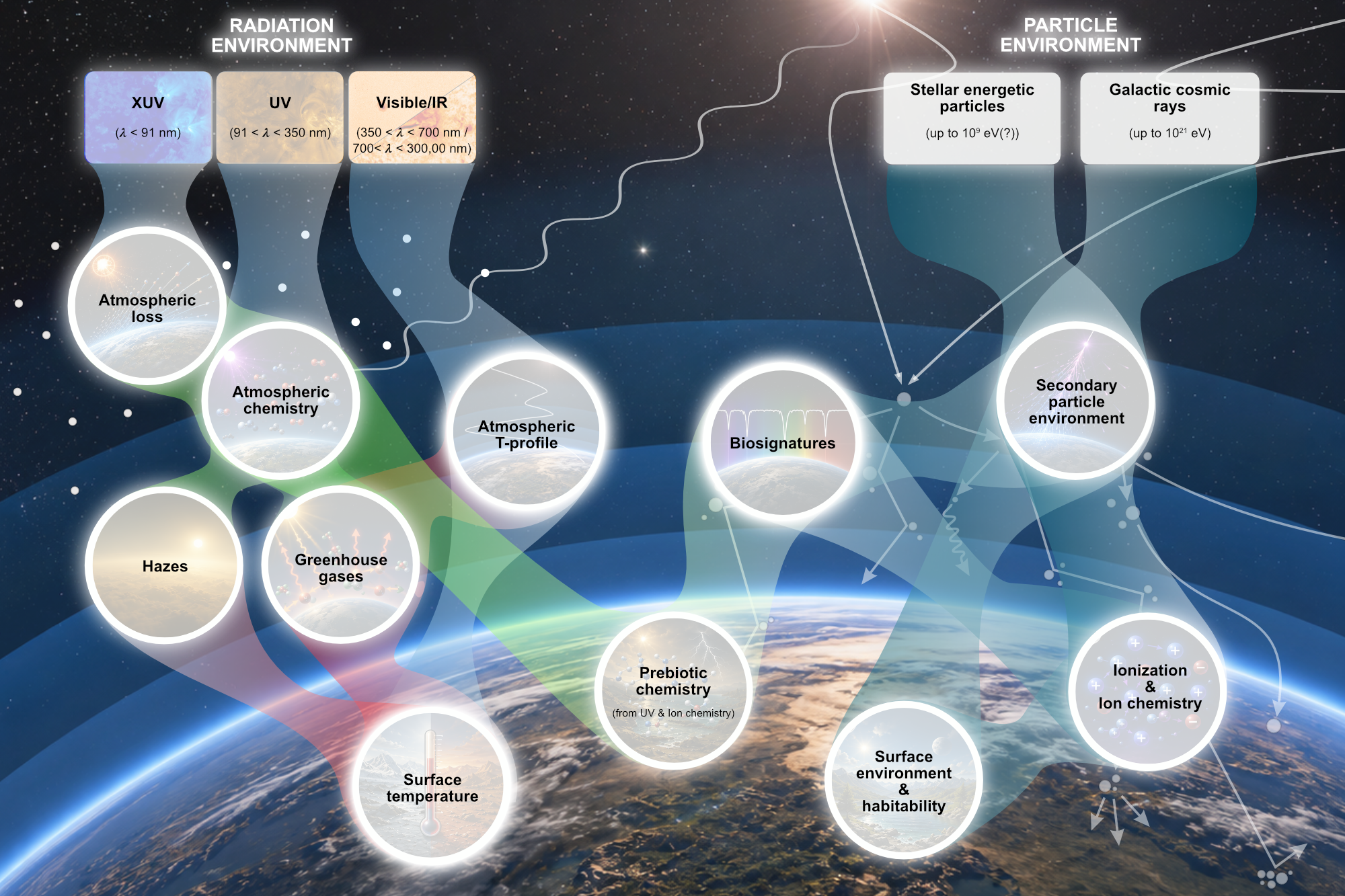} 
\caption{Stellar influence on planetary habitability. Figure credit: solar images from
NASA/ESA and KPNO/NSO/AURA/NSF, background pictures generated through OpenAI's image-generation system (i.e., \href{https://firefly.adobe.com/}{Adobe Firefly} and \href{https://openai.com/index/introducing-chatgpt-images-2-0/}{ChatGPT Images 2.0}). Illustration: K. Herbst}  \label{fig:stellarinfluence}
\end{center}
\end{figure}

The star provides the energy to warm the planet and to drive photosynthetic metabolisms but their role for planetary habitability is more complex. X rays and extreme ultraviolet (identified together as XUV, $\lambda <$ 91 nm) heat the top of the atmosphere and can lead to atmospheric loss. UV radiation (91 nm $< \lambda <$350 nm) and high energy particles emitted by the star, have the potential of promoting the chemistry to create compounds relevant for the origins of life and also can sterilize the planetary surface. Atmospheric photochemistry driven by UV affects the abundance of greenhouse gases, compounds that could be signatures of life and creates hazes that impact the surface temperature. Visible (350 nm $< \lambda <$ 700 nm) and infrared (IR, 700 nm $< \lambda <$ 300,000 nm) influence the atmospheric and surface temperature \ref{fig:stellarinfluence}. Note that the limits for each wavelength range may change by a few nanometers across the literature.

The emission of the star changes on different timescales. Stellar luminosity has a long-term evolution as a result of the interior changes in the star, while the stellar photosphere and chromosphere undergo phenomena that last from minutes to hours emitting high energy particles and radiation. In this chapter we review how the star influences planetary habitability and the abundances of compounds that are potential signatures of life. 

\section{The Habitable Zone and the Stellar Evolution and Activity}\label{StellarEvol}

The canonical habitable zone defined by \citet{kastingHZ1993} and later recalculated by \citet{kopparapu2013} depends on the stellar luminosity, the most optimistic limits are located, roughly, where the stellar flux received by the planet fall between 0.3 $S_{\odot 0}$ and 1.8 $S_{\odot 0}$, being  $S_{\odot 0}=$ the solar flux at the top of the Earth’s atmosphere. These limits are a good initial indicator for selecting possible habitable planets, but other factors should be accounted for when studying specific cases. For example, when the HZ falls within 0.1 au, tidal forces may start a runway greenhouse that results in a catastrophic loss of water \citep[e.g.][]{barnes2013tidal}, or, XUV radiation could be high enough to initiate the atmospheric loss until de planet is not longer habitable \citep[][and references therein]{Colombo2025}. In this section we present how the short- and long-term stellar phenomena may impact planets located at the canonical HZ.

\subsection{Long term evolution of stellar luminosity }\label{luminosity-evol}

\begin{figure}[ht]
\begin{center}
\includegraphics[width=0.7\textwidth]{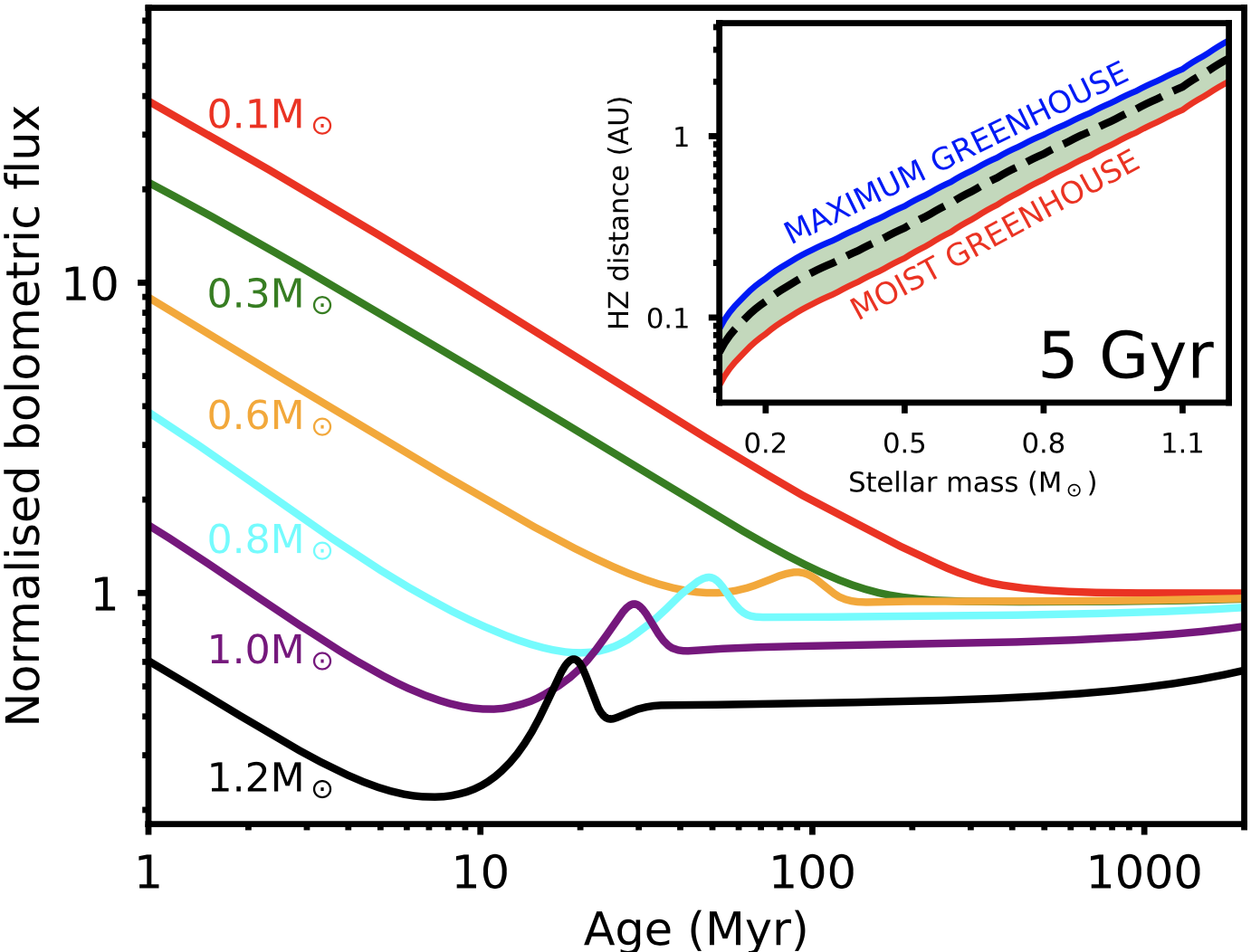} 
\caption{Evolution of bolometric luminosity for stars of different masses. The curves are normalized to values in the HZ of a 5~Gyr old star. The inset shows the location of the HZ (range, in au) as a function of stellar mass at 5~Gyr. (From \citealt{johnstone2021a}.)}  \label{fig:Lbol}
\end{center}
\end{figure}

An important aspect in evolutionary studies is the time dependent radius of the HZ because of the evolution of the stellar bolometric luminosity. With a view on stable habitable conditions on a planet and also on the log-term evolution of potential life, we adopt an appropriate HZ for a given star when its age is 5~Gyrs. How variable $L_{\rm bol}$ is before that age is shown in Fig.~\ref{fig:Lbol}. Stars of solar mass and higher stay significantly ``underluminous'' in the first few Gyrs, while lower-mass stars are initially overluminous and show a long decline owing the the contraction of the star in  the pre-main sequence. This contraction phase is particularly strong and long for M dwarfs (masses less than $\sim 0.5M_{\odot}$). The radius of the HZ scales with the square root of the normalized bolometric flux in the figure. For example, the HZ of a 0.1$M_{\odot}$ star shrinks by a factor of about 6 from 1~Myr to 1~Gyr, and still by a factor of 1.5 from 100~Myr onward \citep{ramirez2014habitable}.

\subsection{Rotational evolution and stellar XUV activity}\label{sect:rotXUVevol}

\subsubsection{Stellar rotation and magnetic activity}
It has been known for long that stellar magnetic activity of cool main-sequence stars is an expression of an internal magnetic dynamo
that is driven by differential rotation and convection \citep{kraft1967, skumanich1972}. Empirical evidence reveals a close relation between the rotation period proper and magnetic activity: More rapidly rotating stars are more magnetically active. Because stars spin down with age -- again a finding from observations of large stellar samples with known ages -- we are confronted with a set of relations between activity, age, and rotation, the so-called \textit{age-rotation-activity paradigm} that needs a theoretical explanation.

Early work suggested that a feedback loop is at work, involving stellar mass loss in a magnetized wind (for our Sun: the \textit{solar wind}; see, for example, \citealt{kraft1967} or \citealt{skumanich1972}). The principal mechanisms at work are the following: A magnetized cool star loses, by a process involving plasma heating close to the star, hot plasma along open magnetic field lines that can be considered to be rigidly rotating close to the stellar surface. At some distance related to the Alfv\'en radius, the mass flow gets decoupled from the star and thus removes angular momentum from it \citep{weber1967}; it consequently spins down. The spin-down weakens the internal dynamo and thus all expressions of magnetic activity on the stellar surface, including the total magnetic flux. Wind mass loss is also a consequence of magnetic activity, and hence the angular momentum transport away from the star declines with time. The coverage of magnetic spots, plage, and the coronal X-ray luminosity decline accordingly. 

However, it is important to realize that in the first 1--3 Gyrs magnetic activity is not a function of age alone but strongly depends on the initial stellar rotation rate after the protoplanetary disk phase. These initial rotation rates vary by up to two orders of magnitude in young, co-eval stellar samples. The evolutionary tracks of rotation rate therefore also vary widely; at the same time they depend on stellar mass (or spectral type). The negative feedback loop mentioned above, however, eventually leads to a convergence of rotation; at ages past a mass-dependent convergence age, the rotation rate becomes essentially a function of stellar mass $M_*$ and age, while the star ``loses memory'' about its initial rotation rate.

Elements of an angular momentum-stellar wind-spin down theory have been developed over the past six decades; a theory was summarized in Chapter 2 of this book, confined to single stars not affected by any magnetic or rotational interactions with close-in planets. Here, we are interested in the results of such studies to understand the long-term influence of stellar activity on planetary atmospheres. We repeat briefly the salient ingredients to the theory:

Wind mass loss exerts a torque $\tau$ on the star that can be expressed by an equation (\citealt{johnstone2015a}, see also \citealt{johnstone2021a})
\begin{equation}\label{tau}
\tau  = f(B, \dot{M}, R_*, \Omega_*) \propto B^{0.87}\dot{M}^{0.56}R_*^{2.87}\Omega_*
\end{equation}
where $B$ is the average stellar surface magnetic filed strength, $\dot{M}$ is the wind-induced mass-loss rate, $R_*$ is the stellar radius, and $\Omega_*$ is the momentary rotation rate. The numerical approximation is based on a torque law from magnetohydrodynamical simulations \citep{matt2012} and can be replaced with other realistic formulations. Then, the spin-down rate is given by
\begin{equation}\label{torque}
\frac{{\rm d}\Omega_*}  {{\rm d}t}  = 
\frac{1}{I_*}  \left(\tau - \frac{{\rm d}I_*}{{\rm d}t}\right)\Omega_*, 
\end{equation}
where $I_*$ is the star's moment of inertia. We also require two further equations that are empirically supported:
\begin{align}
    B &= B(\Omega_*, \tau) \propto (\Omega_*\tau)^{1.32} \\
    \dot{M} &= f(R_*, \Omega_*, M_*) = \dot{M}_{\odot}\left(\frac{R_*}{R_{\odot}}\right)^2\left(\frac{\Omega_*}{\Omega_{\odot}}\right)^a\left(\frac{M_*}{M_{\odot}}\right)^b
\end{align}
where the right-hand side of the first equation was given by \citet{johnstone2021a} and the exponent for rotation by \citet{vidotto2014}, and the parameters $a$ and $b$ have to be found by fitting the theory to stellar samples at different ages; \citet{johnstone2015a} found $a = 1.33, b = -3.36$.

\subsubsection{X-ray and extreme ultraviolet stellar emission}
In addition, we need to ``translate'' rotation rate into an activity indicator, for which we choose the X-ray luminosity $L_{\rm X}$ in the following. X-ray observations of stellar samples have identified a relation between rotation rate and $L_{\rm X}$, which is however better formulated in terms of $R_{\rm X} \equiv L_{\rm X}/L_{\rm bol}$ being a function of the Rossby number $R_o = P_{\rm rot}/\tau$, where $L_{\rm bol}$ is the stellar bolometric luminosity and $P_{\rm rot} = 2\pi/\Omega_*$ is the stellar rotation period. A recent, convenient function between the two parameters was given by \citet{wright2011} that says that up to a saturation value of $R_o \approx 0.06$, $R_{\rm X}$ is nearly constant at a ``saturation value'' of $R_{\rm X} \approx 10^{-3}$, while at higher $R_o$ values $R_{\rm X}$ decays as 
$R_{\rm X} \propto R_o^{-1.89}$ \citep{johnstone2021a}; exponents of --2.18 and --2.7 were given by \citet{wright2011}.

\begin{figure}[ht]
\begin{center}
\includegraphics[width=1.0\textwidth]{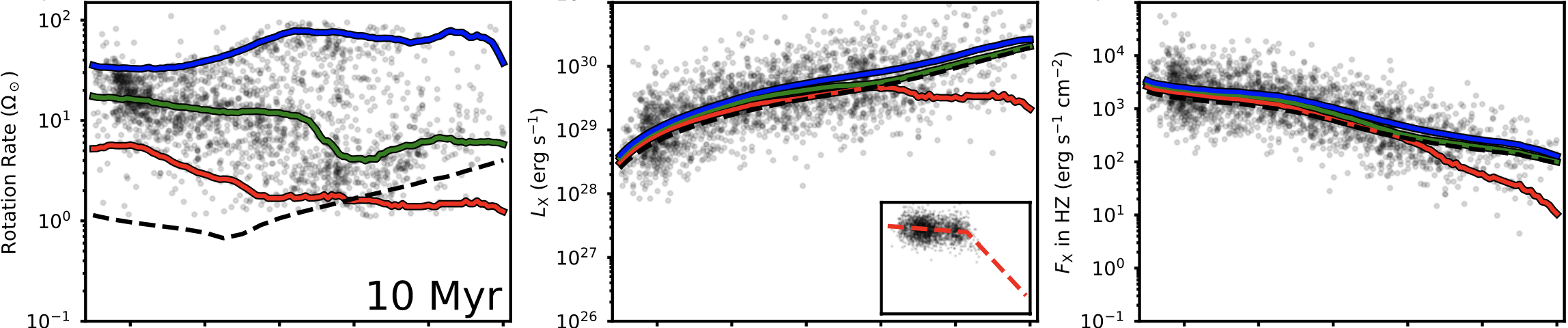} 
\includegraphics[width=1.0\textwidth]{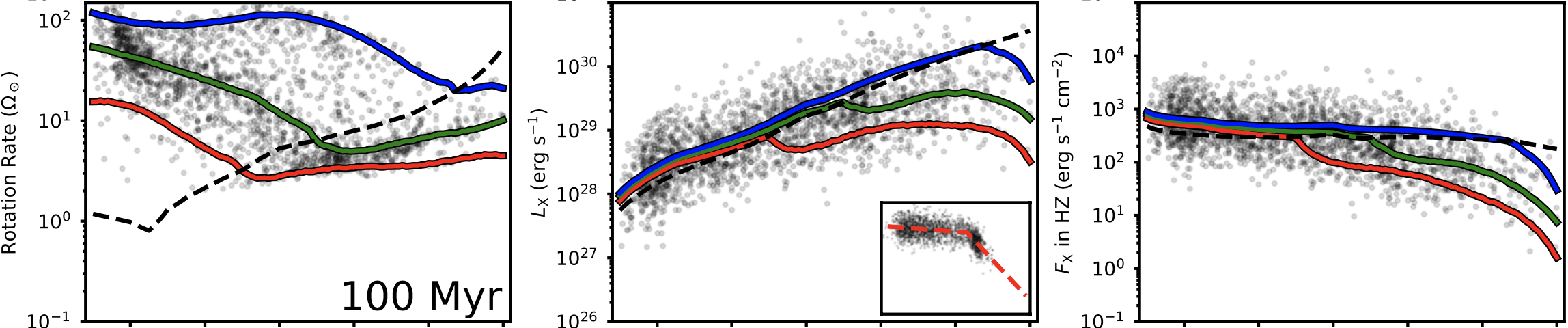} 
\includegraphics[width=1.0\textwidth]{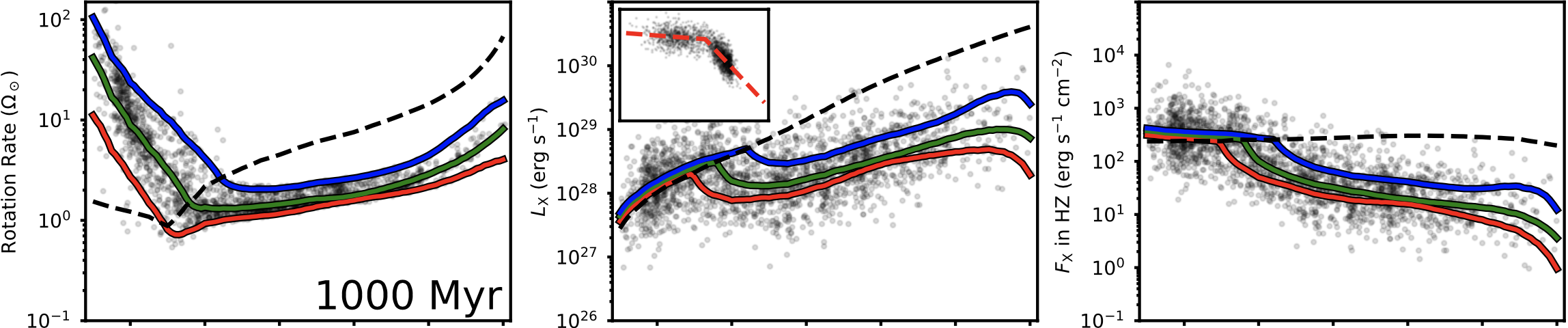} 
\includegraphics[width=1.0\textwidth]{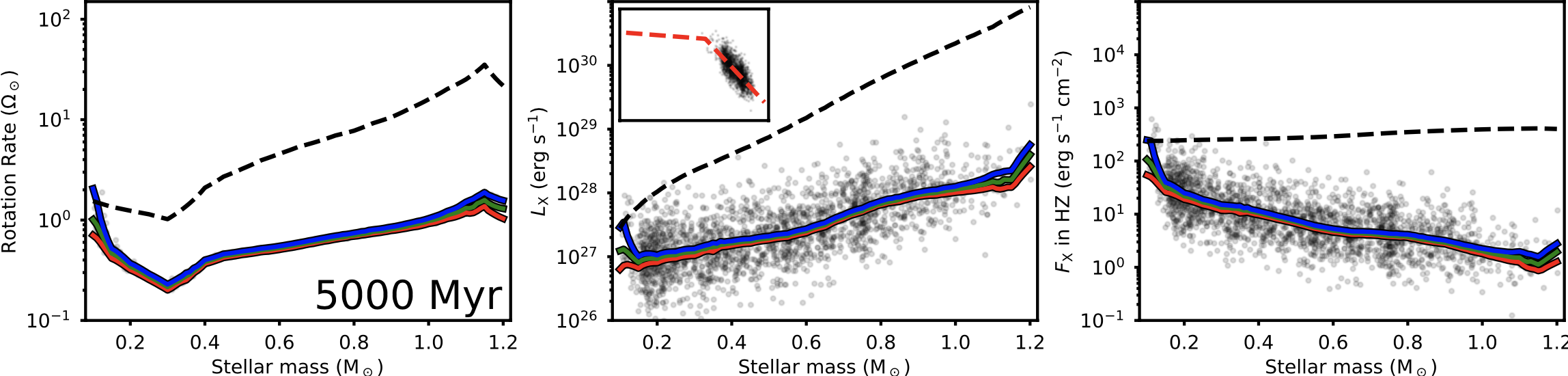} 
\caption{Rotational and X-ray evolution as a function of stellar mass and initial rotation rate. Left column: Rotation rate. Middle column: $L_{\rm X}$. Right column: $F_{\rm X}$ in the HZ at 5~Gyrs. The red, green, and blue tracks are for initially (post-disk) slow, medium, and fast rotators, where these are determined as the 5th, 50th, and 95th percentile of the rotation distribution from samples of $\sim$150~Myrs of age. The dashed curves indicate the rotation rate above which X-rays saturate, resp. the saturation levels in $L_{\rm X}$ or $F_{\rm X}$. The gray dots represent stars from a stellar sample with known rotation periods. The insets in the middle column show $R_{\rm X}$ against $R_o$ for the model distribution, the dashed red lines showing the empirical relation given by \citet{wright2011}. (From \citealt{johnstone2021a}.)}  \label{fig:OmegaLXFX}
\end{center}
\end{figure}
With this in hand, we can integrate the equation system and describe both the rotational evolution and the $R_{\rm X}$ evolution on long time scales, as a function of stellar mass and initial rotation rate. In Fig.~\ref{fig:OmegaLXFX} we combine the evolutionary tracks for rotation rate, $L_{\rm X}$, and X-ray flux $F_{\rm X}$ in the habitable zone of the star once it reaches an age of 5~Gyr. The red, green, and blue lines show the values for a slow, intermediate, and fast rotator (at the 5th, 50th, and 95th percentile of the distribution at 150~Myrs of age. As rotational spin-down proceeds, the more massive stars fall out of saturation earlier than lower-mass stars. Stars with masses $< 0.4M_{\odot}$ stay saturated for at least 1~Gyr. We also see that the rotational spin-down takes much longer for these stars than for solar-mass stars. At 5~Gyrs, however, essentially all stars are unsaturated in $L_{\rm X}$ and there is no longer a distinction between (initially) slow, medium, and fast rotators. Concerning $F_{\rm X}$ in the HZ of a star once it reaches 5~Gyrs of age,  this flux is generally higher in the HZ of lower-mass stars. This is a consequence of more massive stars falling out of saturation earlier, and, at least for ages $< 1$~Gyr, the HZ of lower-mass stars is still shrinking to the final location it will attain at 5~Gyrs (lower-mass stars take longer to reach the main sequence).

\begin{figure}[ht]
\begin{center}
\hbox{  
\includegraphics[width=0.49\textwidth]{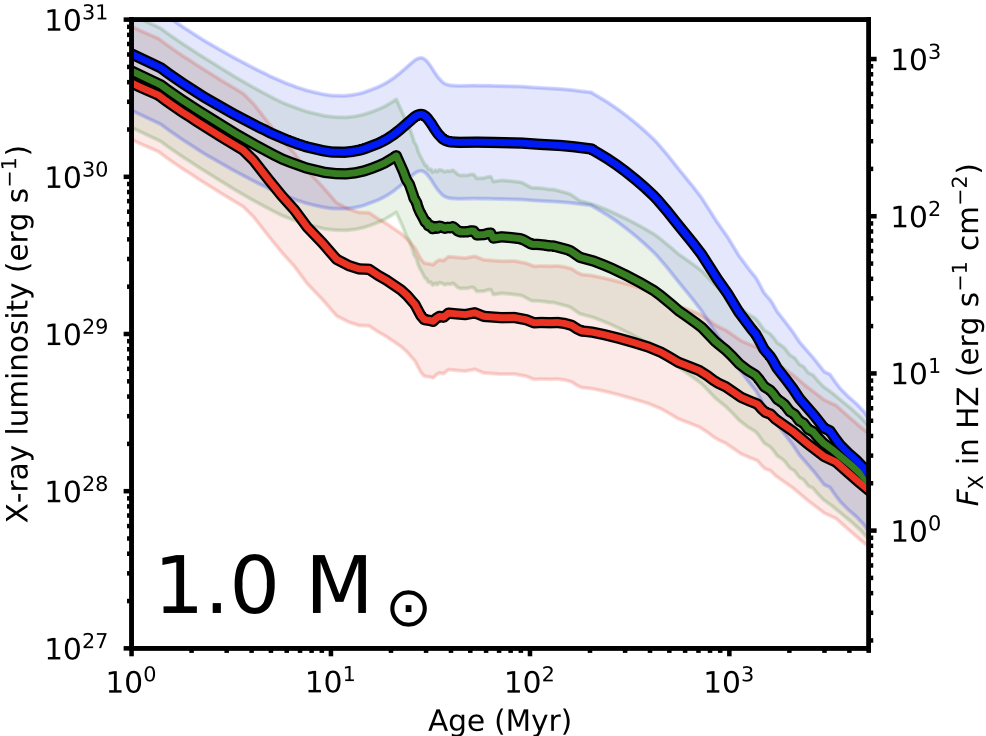} 
\includegraphics[width=0.49\textwidth]{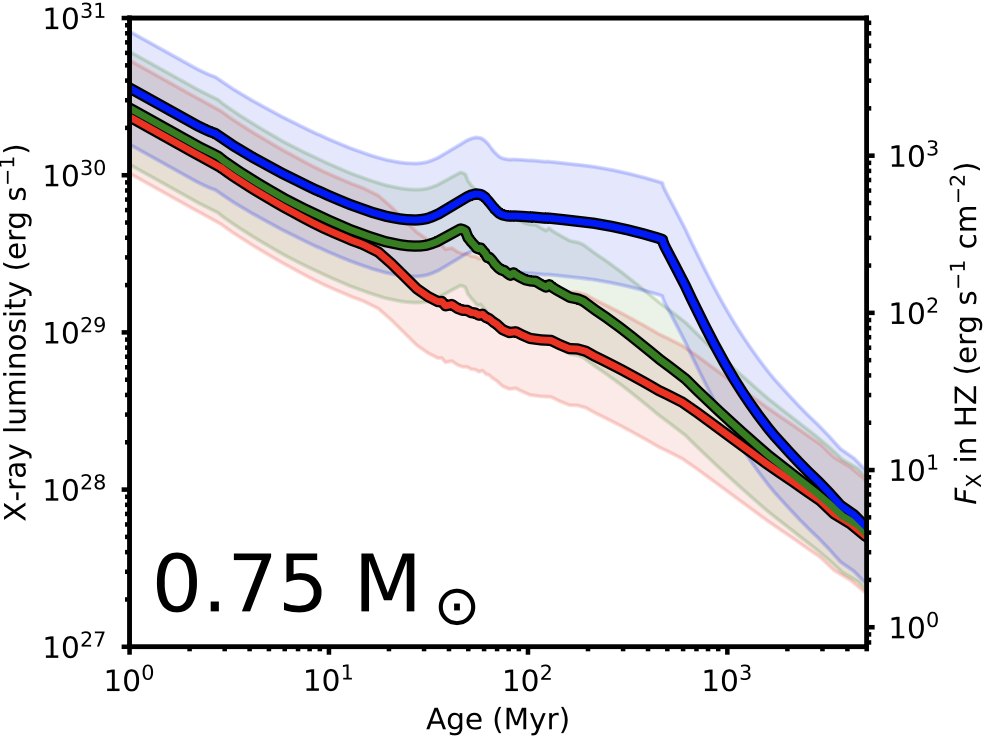} 
}
\hbox{
\includegraphics[width=0.49\textwidth]{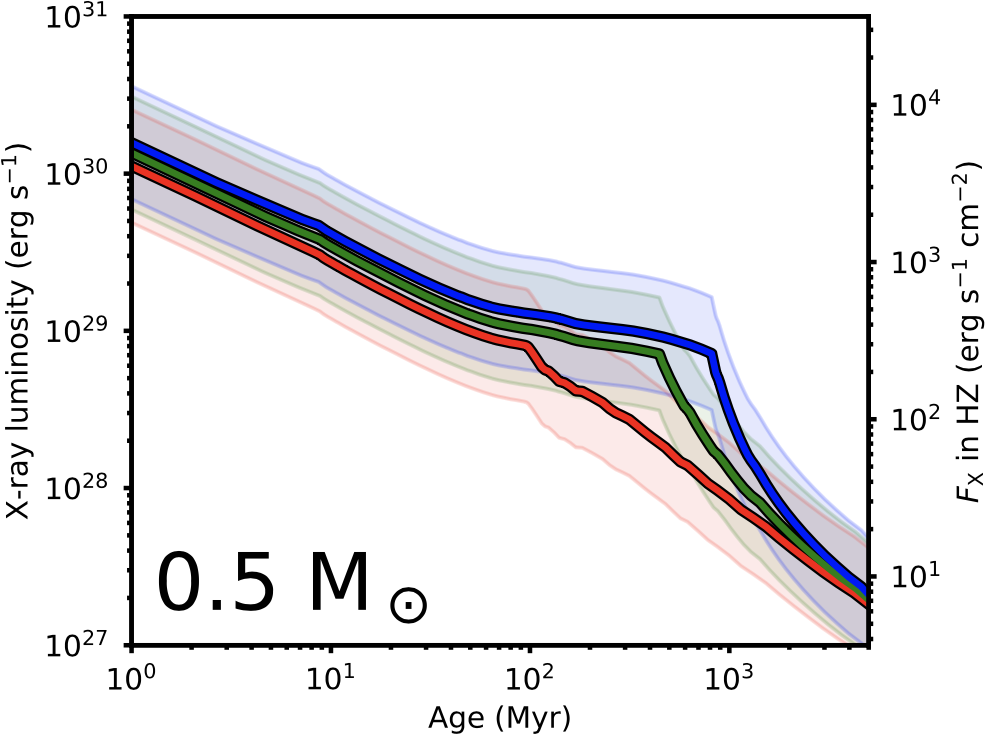} 
\includegraphics[width=0.49\textwidth]{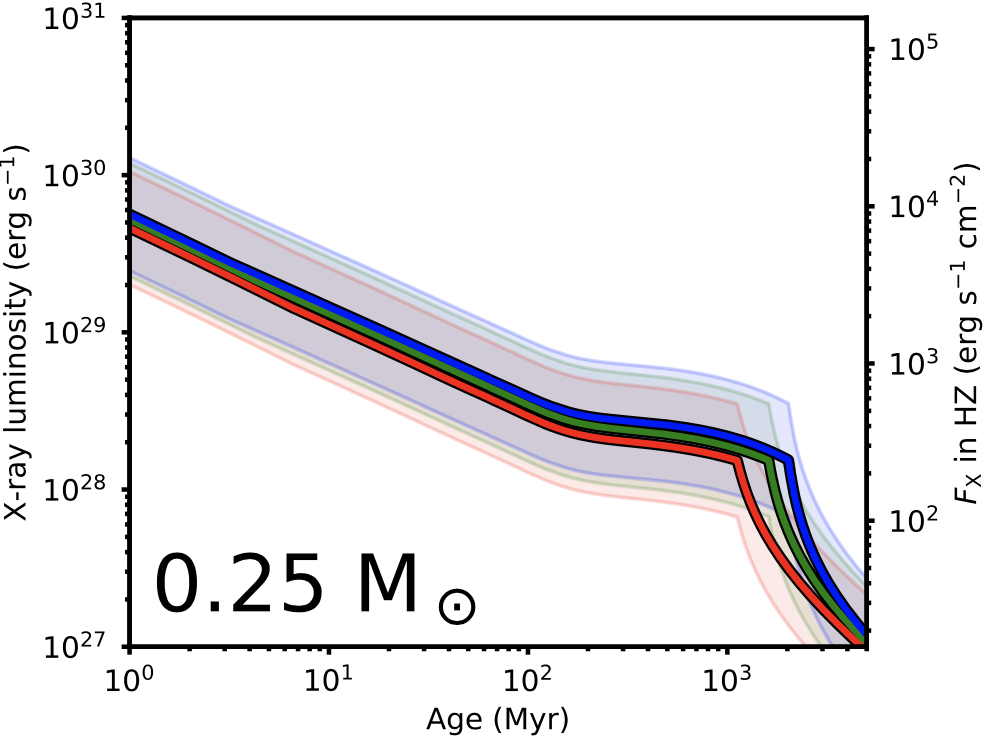} 
}
\caption{X-ray evolutionary tracks for slow (red), intermediate (green), and fast (blue) rotators, for stellar masses of 1.0, 0.75, 0.5, and 0.25~$M_{\odot}$. (From \citealt{johnstone2021a}.)}  \label{fig:LXtracks}
\end{center}
\end{figure}
We can use these relations to extract evolutionary tracks for example for $L_{\rm X}$ or $F_{\rm X, HZ}$ for different stellar masses. Fig.~\ref{fig:LXtracks} shows that slow, medium, and fast rotators follow distinctly different evolutionary tracks over a significant period of their evolution. Here, we distinguish different evolutionary periods. First, during the pre-main sequence contraction phase, the bolometric luminosity declines, and because most stars are still saturated, so does $L_{\rm X}$. There is therefore initially no distinction between slow/medium/fast rotators although in the course of the contraction phase, the slower rotators drop out of saturation. We note that this phase lasts only a few tens of Myr for a solar-mass star but takes many 100~Myrs for a 0.25$M_{\odot}$ star. Once on the main sequence, fast rotators may stay in saturation for an extended period of time while $L_{\rm X}$ of slower rotators declines as the rotation rate declines due to angular momentum loss in the wind. We highlight that for the lowest-mass stars, the saturation phase lasts at least 1~Gyr. Finally, convergence of the tracks is reached at about (1--2)~Gyr except for stars of $\sim 0.25M_{\odot}$ or lower for which convergence is reached only after a few Gyrs. \textit{It is important to note that neither are there universal X-ray decay laws nor can the tracks be described by simple power laws}. For a solar-mass star, the fast and slow tracks differ by up to a factor of $\sim$15 in $L_{\rm X}$ over hundreds of Myr, and in particular during a period when terrestrial planets in the solar system built up their atmospheres and water oceans (e.g. Earth and Mars), and on Earth crucial steps toward the formation of life were probably taken.

\begin{figure}[ht]
\begin{center}
\hbox{   
\includegraphics[width=0.49\textwidth]{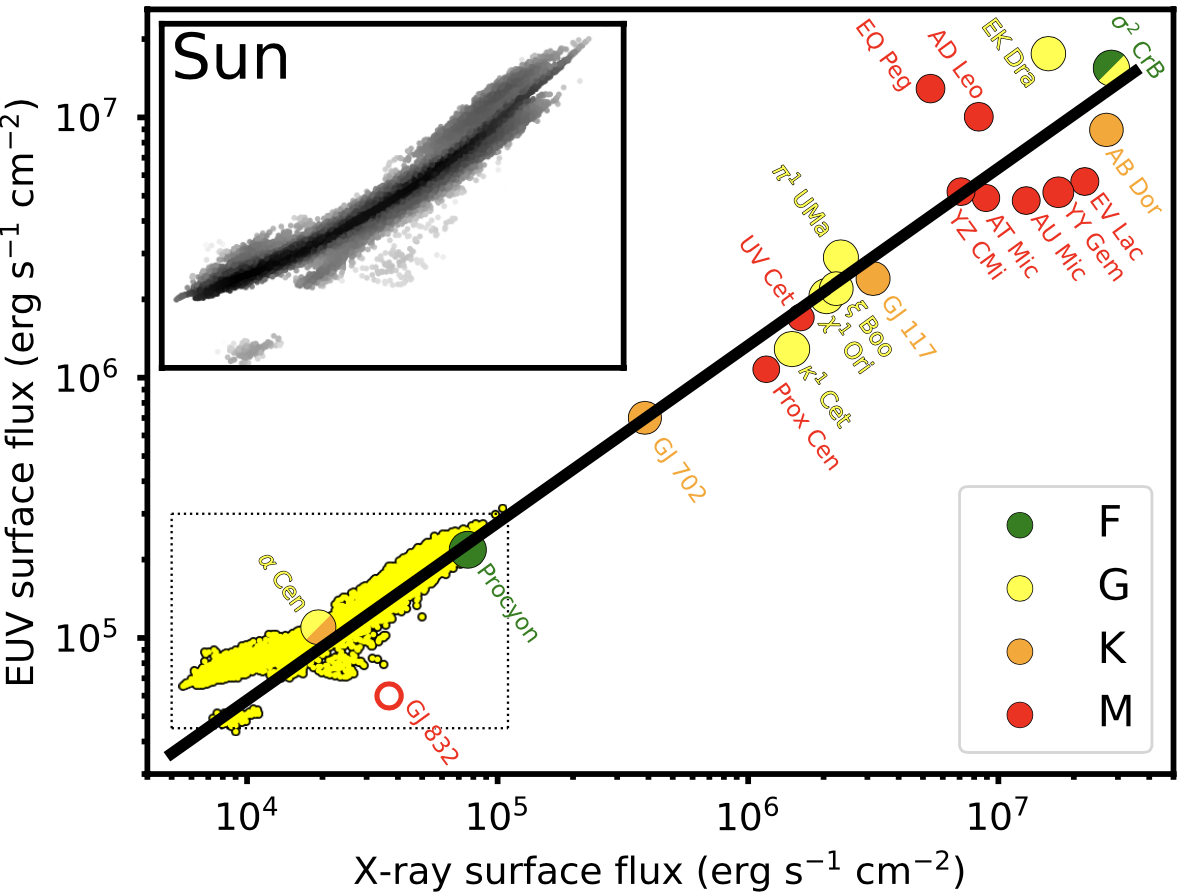} 
\includegraphics[width=0.49\textwidth]{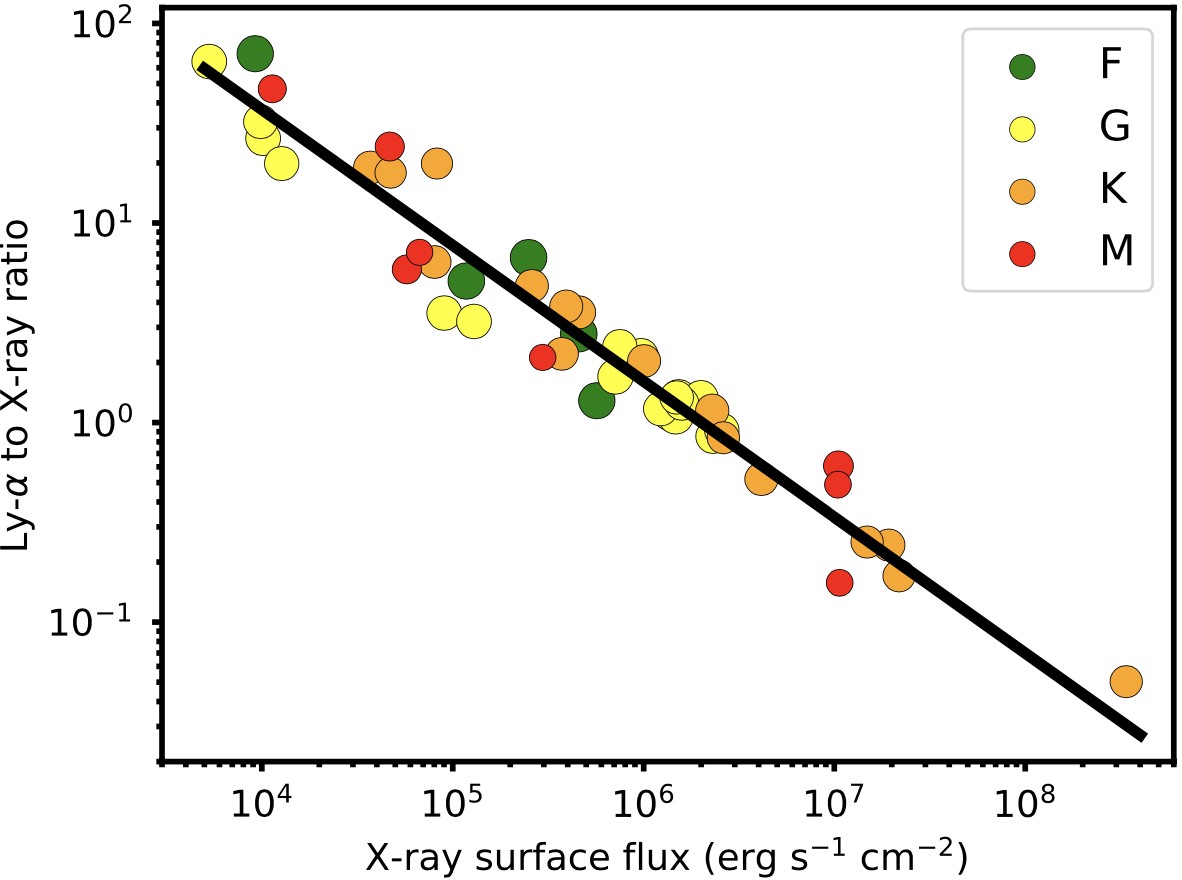} 
}
\caption{\textit{Left:} Relation between EUV (wavelength range: 10--36~nm) and X-ray (wavelength range: 0.517--12.4~nm) surface flux for different spectral classes (F, G, K, M, see inset). The yellow area and the inset show values from a large number of solar observations across the solar cycle. -- \textit{Right:} The ratio of Ly$\alpha$ flux and X-ray surface flux as a function of EUV surface flux, for different spectral classes (see inset). (From \citealt{johnstone2021a}.)}  \label{fig:XvsEUVvsLya}
\end{center}
\end{figure}

For planetary atmospheres, the extreme ultraviolet (EUV) part of the spectrum is more important to drive ion chemistry and heating the upper atmosphere than X-rays. We show in Fig.~\ref{fig:XvsEUVvsLya}a a correlation between the stellar surface fluxes of X-rays (0.517--12.4~nm) and in the EUV, here defined for the shorter wavelength part of 10--36~nm.  It is interesting to note that for the more active stars in the upper-right part of the figure the XUV ratio is significantly higher ($\sim 2$)  than for Sun-like stars ($\sim 0.25$). This is a consequence of a relation between activity and average coronal X-ray temperature \citep{johnstone2015b}: coronae of stars at higher activity levels (in the sense of $L_{\rm X}/L_{\rm bol}$) are hotter than low-activity stars.  In Fig.~\ref{fig:XvsEUVvsLya}b we show the ratio between the flux in the Ly$\alpha$ line, usually by far the brightest far-UV line, and the X-ray flux, as a function of the X-ray surface flux. Again, we see that for higher-activity stars, the relative importance of X-rays is much higher than in low-activity stars, which indicates a more important role in coronal radiation for active stars. The following relations are useful for applications\footnote{The second equation (for $\log F_{\rm EUV,2}$) corresponds to Eq.~(22) in \citet{johnstone2021a}; the related Eq.~(21) in that paper appears to contain a typographical error in the constant (-0.341) that should be 10$\times$ smaller.  \\ The third equation (for $\log F_{\rm Ly\alpha}$) corresponds to Eq.~(24) and the best-fit regression in Fig.~15 of \citet{johnstone2021a}; the related Eq.~(23) in that paper slightly deviates from the regression line in their Fig.~15.}:
\begin{align}
      \log F_{\rm EUV,1}    &=  \phantom{-}2.04\phantom{4}  + 0.681\log F_{\rm X}, \\
      \log F_{\rm EUV,2}    &= -0.034 + 0.920\log F_{\rm EUV,1}, \\
      \log F_{\rm Ly\alpha} &=  \phantom{-}4.29\phantom{4}  + 0.319\log F_{\rm X},
\end{align}
where $F_{\rm EUV, 1}$ and $F_{\rm EUV, 2}$ are EUV surface fluxes in the bands 
of 10--36~nm and 36--92~nm, respectively, while X-rays are given for the band of 0.517--12.4~nm (corresponding to energies of 0.1--2.4~keV), and all fluxes are given in erg~cm$^{-2}$~s$^{-1}$, averaged over the stellar surface and short-term variability \citep{johnstone2021a}. $F_{\rm Ly\alpha}$ is the flux in the far-ultraviolet hydrogen Ly$\alpha$ line.

The evolutionary trends discussed above are modified if we consider the flux above the atmosphere of a planet in the Habitable Zone (HZ) of a star. With this in mind, we show in Fig.~\ref{fig:OmegaLXFX} (right column) the evolution of X-ray fluxes incident on a planet in the respective HZ$_{\rm 5Gyr}$. While for solar-mass stars, these fluxes decline in relation to $L_{\rm X}$ at ages greater than about 40~Myrs, lower-mass stars show much elevated fluxes even at 5~Gyrs; this is a consequence of the slow decline of $L_{\rm bol}$ for the lowest-mass stars even after 1~Gyr due to contraction, but also on the slow decline of $L_{\rm X}$ due to spin-down. The saturated $L_{\rm X}$ values decline, but for the HZ$_{5Gyr}$ these values are ``overluminous''.

\begin{figure}[ht]
\begin{center}
\includegraphics[width=1.0\textwidth]{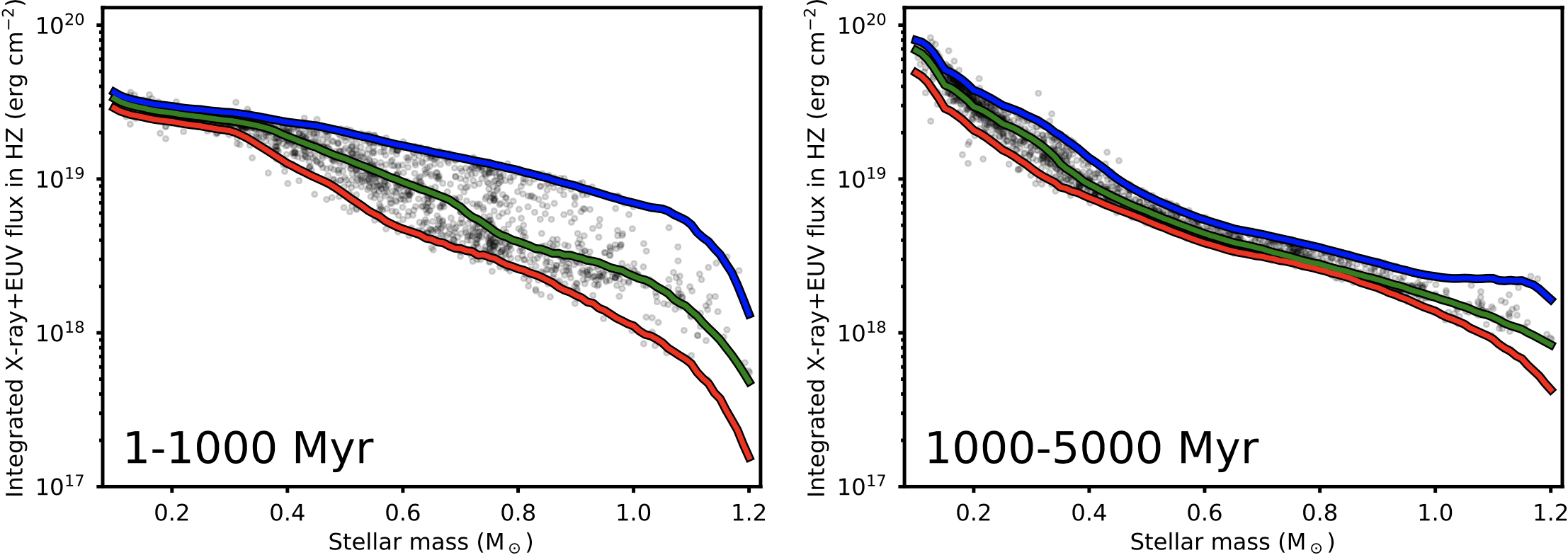} 
\caption{X-ray fluence in the HZ$_{\rm 5Gyr}$ as a function of stellar mass, for the age range of 1--1000~Myr (left) and 1000-5000~Myr (right). The red, green, and blue lines refer to fast, medium, and slow rotators. (From \citealt{johnstone2021a}.)}  \label{fig:XUVFluence}
\end{center}
\end{figure}
Since the total effect on a planetary atmosphere such as  mass loss is the result of the integrated stellar X-ray and EUV flux in time, we show in Fig.~\ref{fig:XUVFluence}ab the integrated energy or ``fluence'' for the age ranges of 1--1000~Myrs and 1--5~Gyrs, respectively. We see that the
fluence into an HZ atmosphere is much higher for an M dwarf planet compared to G dwarf planet, regardless of the rotational evolutionary track taken, thiS is the result of the the HZ shrinking strongly but only slowly for the lower-mass stars, and the spin-down induced decline of $L_{\rm X}$ on the main-sequence taking longer.

\subsubsection{Stellar flares}
Finally, we analyze the role of X-ray or EUV flares in the course of stellar evolution. It is often said that M dwarfs flare energetically and this should be the primary cause for planets in the HZ around M dwarfs potentially not 
being conducive to life. Flares and their statistics are described in Chapter 2 of this book. We briefly recall that the distribution of flare radiative energies on a given star is, as found empirically including the Sun, a power law of the form:

\begin{equation}\label{flareenergies}
    \frac{{\rm d}N}{{\rm d}E} \propto E^{-\alpha}
\end{equation}

\noindent where the left side indicates the number of flares ${\rm d}N$ per energy interval ${\rm d}E$, and $\alpha$ is an empirical power-law  index often found to be 2.0--2.8 in the X-ray and EUV range (\citealt{audard2000}, \citealt{kashyap2002}, \citealt{guedel2003}, \citealt{arzner2004}, or \citealt{stelzer2007}; indices for optical flare energy \textit{do not} need to be the same as the two bands do not need to correlate linearly). We emphasize that to compare ``flare rates'' between stars, the flare count needs to use the same lower energy threshold as otherwise, arbitrarily large numbers of flares could be obtained. The absolute normalization of such distributions indicates that more massive stars produce much stronger flares than lower-mass stars (\citealt{yang2019}, their Fig. 3 for optical flares; the flare frequency above a given energy is the highest for G stars). This is equivalent to the observation that stars with a higher X-ray luminosity flare more often above a given flare energy threshold (\citealt{audard2000} for the EUV and X-ray ranges). 
The observation that M dwarfs seem to flare more often is very likely the result of an observational contrast bias: Given the much smaller surface area of an M dwarf compared to a G dwarf, a flare in an M dwarf active region produces a larger fractional perturbation of the corona than in a G dwarf (e.g., \citealt{reale2004}, their Fig.~8; the analogous situation holds notoriously for the optical range where in addition the temperature contrast between flare footprints and the photosphere is much larger, see, e.g., \citealt{balona2015}). 

\begin{figure}[ht]
\begin{center}
\hbox{  
\includegraphics[width=0.35\textwidth]{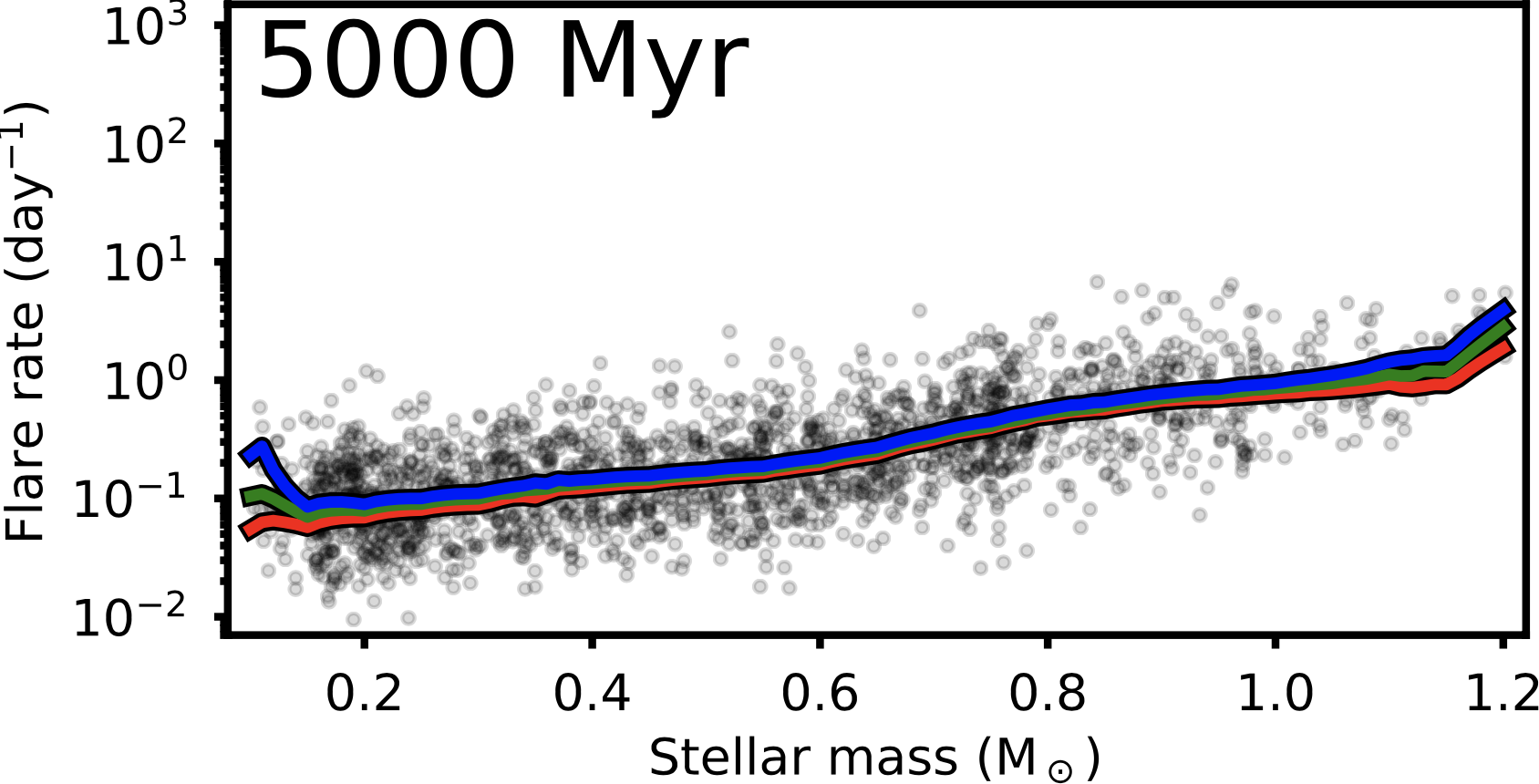} 
\includegraphics[width=0.31\textwidth]{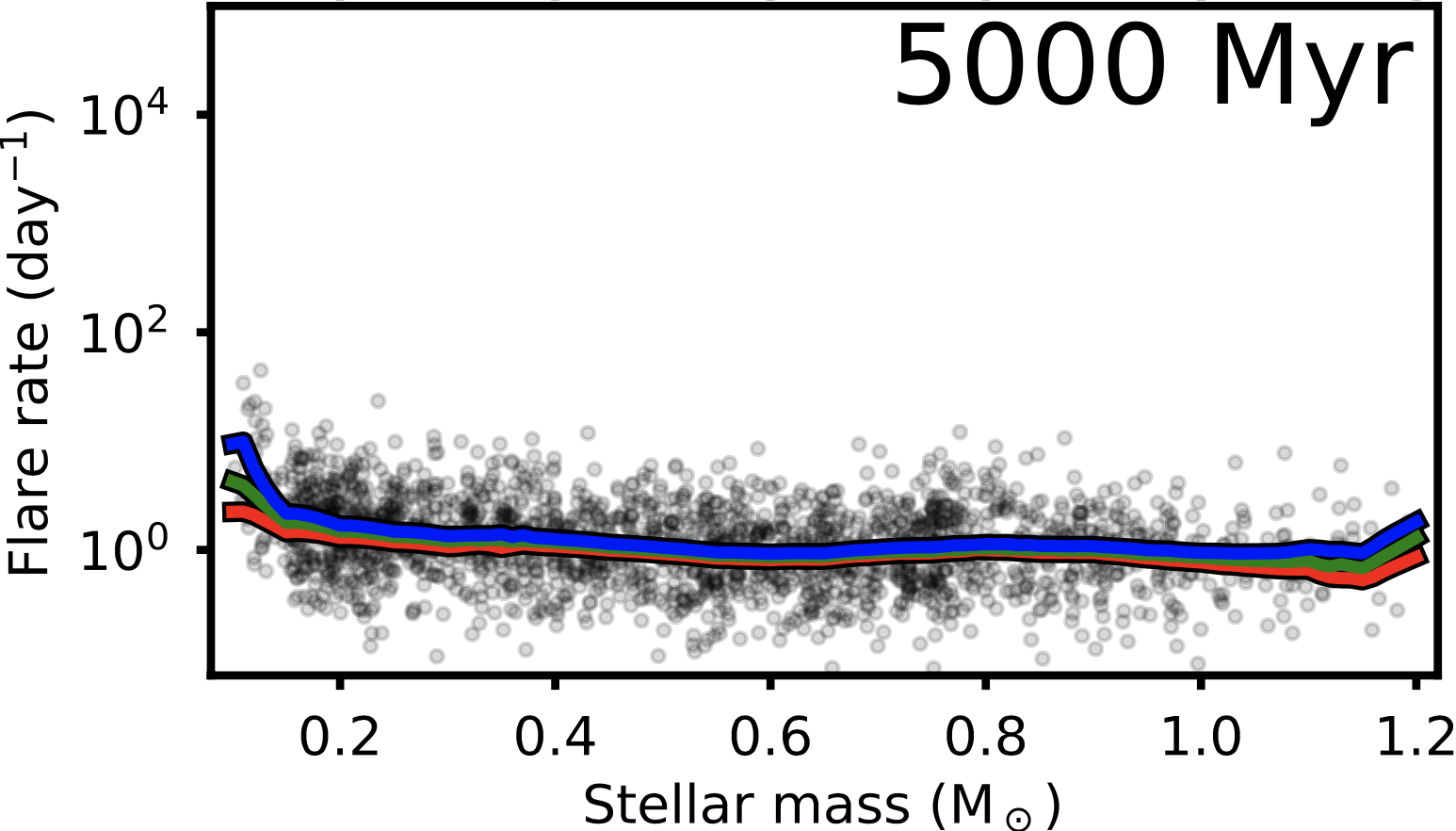} 
\includegraphics[width=0.31\textwidth]{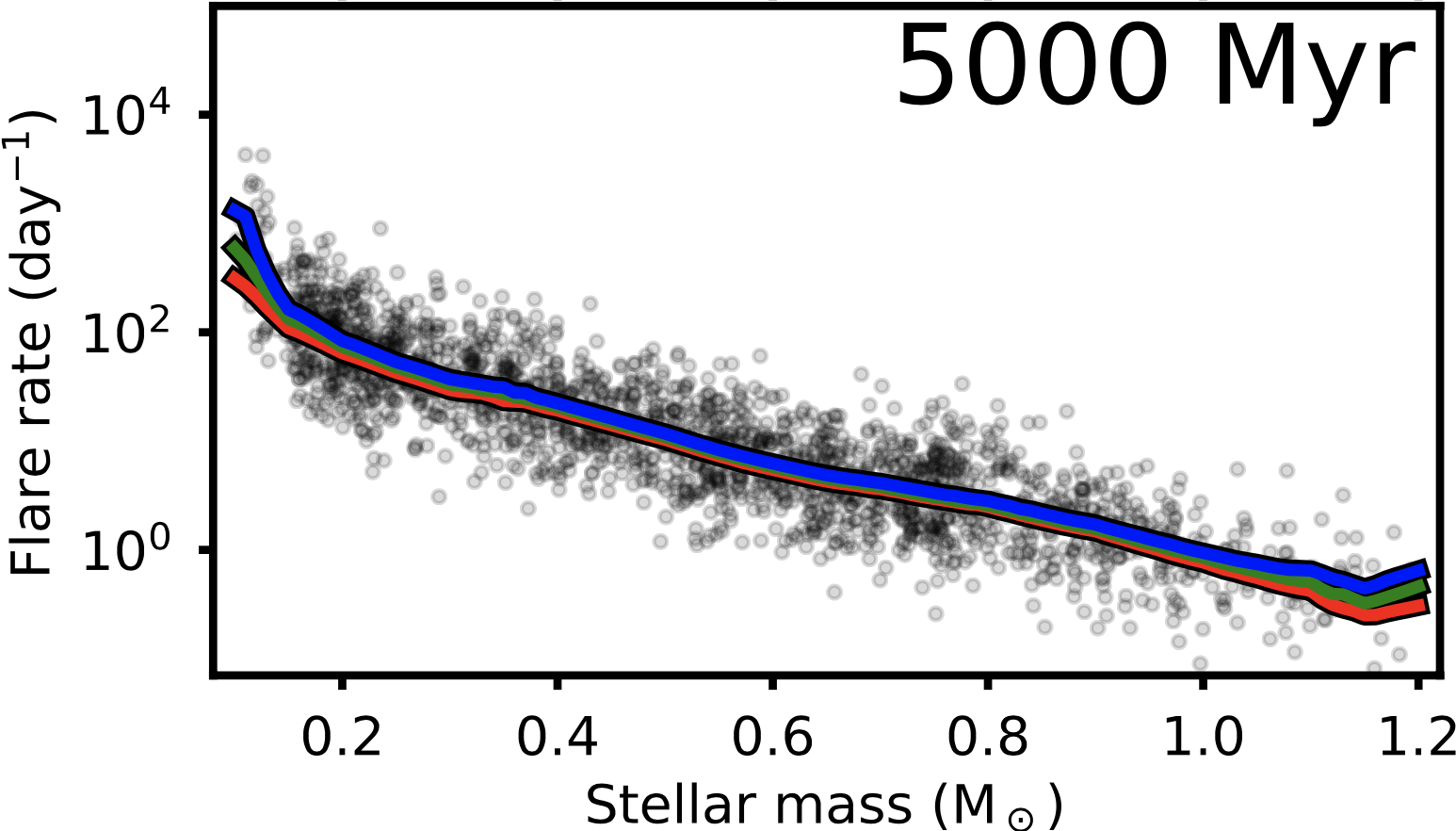} 
}
\caption{\textit{Left:} Rate of flares exceeding $10^{32}$~erg in radiated XUV energy as a function of stellar mass, for an age of 5~Gyr; red, green, and blue curves refer to fast, medium, and slow rotators that have converged at this age; the gray dots show the complete model distribution derived from observations of rotation rates. -- \textit{Middle:} Rate of flares exceeding a \textit{fluence} of $1.8\times 10^4$~erg~cm$^{-2}$ at a planet in the midddle of the HZ$_{\rm 5Gyr}$, at 5~Gyrs, for $\alpha = 1.6$. -- \textit{Right:} Same, but for $\alpha = 2.4$. (From \citealt{johnstone2021a}.)} 
\label{fig:FlareRate}
\end{center}
\end{figure}

Without any doubt, a fraction of the so-called quiescent X-ray and EUV emission from stars is due to the sum of unresolved, superposed small flares not individually detected; with $\alpha$ values as quoted above, the entire quiescent emission could be due to superposed flares \citep{guedel2003, telleschi2005}. \citet{audard2000} found, for a sample of stars observed in the EUV, that the flare rate above a fixed threshold increases linearly with $L_{\rm X}$. For this reason, the summary on X-ray and EUV evolution given in this subsection did not take flares separately into account; flares enter observations as stochastic short-term variability and thus contribute to the scatter in X-ray observations that are included in the luminosity and flux statistics. We nevertheless study how important flares may be if we restrict the statistics to flares above a \textit{fluence threshold} in the HZ of the star. Because the HZ radius is smaller for lower-mass stars, the corresponding luminosity threshold is also lower than in higher-mass stars, allowing more flares to enter the statistics even though in absolute terms, more massive stars flare more frequently. The result depends strongly on $\alpha$. A very low value of $\alpha = 1.6$, often observed in the optical, leads to a nearly mass independent relevant flare rate in the HZ (Fig.~\ref{fig:FlareRate} middle), while $\alpha = 2.4$ results in a much more prominent role of flares for lower-mass stars (Fig.~\ref{fig:FlareRate} right), because a larger $\alpha$ gives more weight to smaller flares.

\subsection{Evolution of Far and Near Ultraviolet Stellar Emission}\label{stellarUV} 

The time evolution of longer wavelength UV emission ($\lambda$ $>$ 91 nm) is largely similar to the evolution of X-ray and EUV emission because all three are tied to the magnetic dynamo driven by rotation and convection. An important caveat is that the $>$170 nm emission becomes increasingly dominated by photospheric emission with increasing stellar effective temperature as the Wien's tail of the stellar blackbody emission shifts into the ultraviolet. 

With a small sample of solar analogs spanning ages 0.1-7 Gyr, \citet{Ribas2005} found that UV emissions of young G-type stars are 10-50$\times$ stronger than the present-day Sun (5 Gyr). Similar studies for K dwarfs \citet{RicheyYowell2019,RicheyYowell2022} and M dwarfs \citet{Pineda2021,Loyd2021} reveal similar trends to the X-ray. 

\citet{RicheyYowell2022} find from a sample of K dwarfs at three distinct ages (40 Myr, 650 Myr, 5 Gyr) that transition region and chromospheric UV emission lines remain saturated beyond 650 Myr, which may be attributed to the stalling of rotational spin down that K dwarfs experience. 

By comparison, using a sample of young and old early M dwarfs in the same three age bins (40 Myr, 650 Myr, and 5 Gyr), \citet{Loyd2021} find that transition region and chromospheric emission normalized by the stellar surface area remains saturated for $\sim$250 Myr before decaying. They also find that certain features (Mg II and chromospheric continuum) decay more slowly than other far-UV emission lines. In a separate experiment, \citet{Pineda2021} use spectroscopic observations of a sample of young and old M dwarfs to show that transition region emission normalized by bolometric luminosity remains saturated until $\sim$1 Gyr, although mid M dwarfs remain saturated slightly longer than early M dwarfs. Post-saturation, most far-UV emission lines decay at similar rates with decreasing rotation, but Ly$\alpha$ appears to decline more slowly.

Stellar UV has a two-fold role for life, on one hand, it breaks organic molecules, particularly nucleic acids (DNA and RNA), on the other hand UV promotes chemistry relevant for the origins of life . \citep[Sect.\ref{UV-life}, ][and references therein]{Segura2025-RadiativeEffects}. Following the work by \citet{Cockell1999}, \citet{Buccino2006, Buccino2007} proposed limits for the HZ considering the minimum number of photons to drive prebiotic chemistry and the maximum number of photons before the planetary surface is sterilized by UV. Their work was later revised by \citet{spinelli2023ultraviolet} who used the abiogenesis zone from \citet{rimmer2018origin} that establish the UV (200–280 nm) flux necessary to drive prebiotic chemistry in 6.8 $\times 10^{9}$ photons~cm$^{-2}$~s$^{-1}$~ \AA$^{-1}$. Their results favor planets around K-type stars as host of habitable environments.

\section{Stellar impact on planetary atmospheres}\label{AtmEvol}

The defining condition for a planet to be in the liquid-water habitable zone is its potential to keep an atmosphere of sufficiently high pressure and within a surface temperature range  that allow water to be in its liquid state. Therefore, atmospheric mass loss could prevent a planet from being habitable. The star also drives the chemistry of planets around potentially habitable planets promoting the formation of compounds relevant for the origins of life and, once life has emerged, impacts the abundance of chemical species produced by life, called biosignatures. In this section we review the planetary atmospheric loss and the atmospheric chemistry relevant for prebiotic chemistry and biosignatures. 

For life to form, persist, and evolve, habitable conditions must remain similar over long time periods of millions to billions of years. Atmospheric mass loss must be small enough that atmospheric conditions at most vary moderately around otherwise stable averages over tens to hundreds of Myr. In this section we present the current available observations of exoplanetary atmospheres within a general frame called the "cosmic shoreline" and models that include chemistry, radiative transport and several mechanisms for atmospheric loss that provide time-dependent scenarios for atmospheric retention.
 
The ability of a planet to retain its atmosphere depends on i) the planetary mass, ii) the atmospheric composition, iii) the stellar XUV irradiation driving thermal loss of the upper atmosphere, and iv) the solar wind particle flux and the presence of additional energetic (accelerated) particles  interacting with the upper atmospheres via ``non-thermal'' mechanisms, leading to further loss. At the same time, atmospheres can be replenished via magma ocean outgassing (at the beginning of planetary evolution) and volcanism.

\subsection{The Cosmic Shoreline and Exoplanet Atmospheric Observations}

The concept of the ``Cosmic Shoreline,'' introduced by \citet{Zahnle2017}, provides a semi-empirical framework for understanding which planetary bodies are capable of retaining substantial atmospheres. The shoreline is defined as a boundary in the space of stellar irradiation versus planetary escape velocity (or, equivalently, surface gravity), separating planets that could potentially have retained atmospheres from those that have not. In the solar system, this boundary successfully distinguishes atmosphere-bearing bodies such as Earth, Venus, Titan, and Triton from atmosphere-free bodies such as the Moon, Mercury, and most small moons. The underlying physics reflects the competition between atmospheric escape — driven primarily by stellar XUV irradiation — and gravitational retention.

The Cosmic Shoreline has predictive implications for rocky exoplanets, particularly those orbiting M~dwarf stars. M~dwarfs are known to be active, especially in their early evolution, emitting high levels of XUV radiation that can drive vigorous hydrodynamic escape and ion sputtering on close-in planets \citep{France2016, Bourrier2017}. Planets in the habitable zones of M~dwarfs receive relatively low bolometric flux but may have been exposed to cumulative XUV doses orders of magnitude higher than Earth, placing them in a potentially unfavorable region near or beyond the Cosmic Shoreline.

\begin{figure}[ht]
    \centering
    \includegraphics[width=\textwidth]{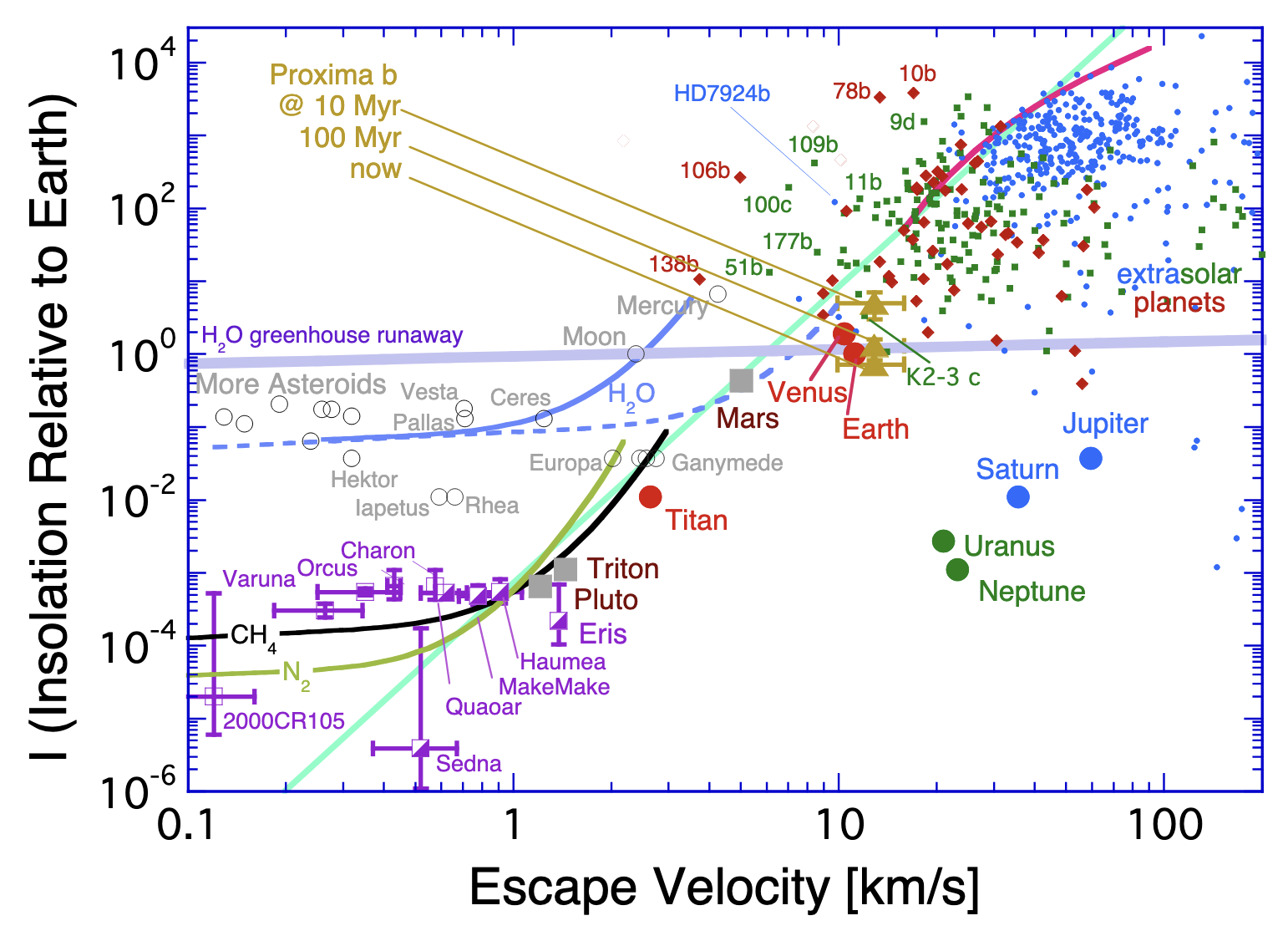}
    \caption{Atmospheric loss and retention for exoplanets and planetary bodies in the solar system. The planetary capacity of retaining an atmosphere by gravity is represented with the escape velocity, thermal escape is accounted for with the insolation. Planetary bodies at the left of $I \propto v^{4}_{esc}$ the power law (turquoise line) may not retain an atmosphere, being $v_{esc}$, the escape velocity. See the original figure in \citet{Zahnle2017} for more details.}
    \label{fig:Cosmic_shore}
\end{figure}

Observational evidence from transmission and emission spectroscopy with JWST is beginning to test these theoretical predictions directly. The TRAPPIST-1 system—comprising seven Earth-sized planets, three of which reside within the habitable zone—has emerged as a fundamental benchmark for terrestrial atmospheric retention. Early JWST observations of TRAPPIST-1\,b and TRAPPIST-1\,c, using secondary eclipse photometry in the mid-infrared, indicate high dayside temperatures and a lack of significant absorption features. These data suggest the absence of thick, CO$_{2}$-dominated atmospheres, favoring either a bare-rock scenario or extreme atmospheric depletion \citep{Greene2023, Zieba2023} These results are broadly consistent  with the Cosmic Shoreline prediction that highly irradiated, low-gravity rocky planets are unlikely to retain substantial secondary atmospheres.

Two additional targets, GJ\,1132\,b and LHS\,1140\,b, illustrate complementary aspects of this picture and highlight the complexity of interpreting atmospheric observations for M-dwarf rocky planets. GJ\,1132\,b is a rocky super-Earth (1.1 $\pm$ 0.04 $R_\oplus$, 1.84 $\pm$ 1.19\,$M_\oplus$), slightly denser than Earth ($5.97^{+0.96}_{-0.79}$ g/cm$^{3}$), orbiting an M4-type dwarf on a 1.6-day orbit, placing it well within the highly irradiated regime above the Cosmic Shoreline. Early claims of atmospheric detection based on \textit{HST}/WFC3 near-infrared transits \citep{Swain2021} suggested the presence of aerosol scattering, HCN, and CH$_4$ in a low mean molecular weight atmosphere being supplied outgassing. These observations were subsequently challenged by independent reanalyses of the same data, which found no evidence for molecular absorption and preferred a flat, featureless spectrum  \citep{Mugnai2021, LibbyRoberts2022}. Follow-up \textit{JWST}/NIRSpec G395H transmission spectroscopy revealed discrepant results between two independent transit visits: one consistent with a possible H$_2$O-bearing atmosphere, and the other featureless — leaving the question unresolved and raising the need of repeatability requirements for rocky planet atmospheric claims \citep{May2023}. Follow-up with \textit{JWST}/MIRI thermal emission observations, which measured a dayside brightness temperature within $\sim$1$\sigma$ of the maximum bare-rock value, ruling out atmospheres with substantial CO$_2$ or H$_2$O content and supporting the conclusion that GJ\,1132\,b is most likely a bare rock \citep{Xue2024}. Its position well above the Cosmic Shoreline in the irradiation-versus-escape-velocity plane is consistent with this outcome \citep{Xue2024}.

LHS\,1140\,b stands out as a more compelling target in the context of atmospheric retention. With a radius of 1.730$\pm{0.025}$ R$_\oplus$, in the middle of the radius valley \citep{Fulton2017}, and orbiting within the habitable zone of its M-dwarf host at $\sim$43\% of Earth's incident flux, the planet receives substantially lower irradiation than GJ\,1132\,b or the TRAPPIST-1 inner worlds, placing it in a more favorable region with respect to the Cosmic Shoreline \citep{Cadieux2024, Zahnle2017}. Its bulk density (5.9 $\pm$0.3 g cm$^{3}$) revised downward from earlier estimates, is inconsistent with a purely rocky Earth-like interior and instead suggests that LHS\,1140\,b may be a water world, with roughly 9--19\% of its mass in water \citep{Cadieux2024}, or a rocky world with a retained thin H$_{2}$ envelope ($\sim$0.1\% H/He by mass). \textit{JWST}/NIRISS transmission spectroscopy has excluded a hydrogen-dominated mini-Neptune atmosphere at high significance ($>10\sigma$), and the data show tentative evidence — at the $\sim$2$\sigma$ level — for a nitrogen-rich, high-mean-molecular-weight atmosphere \citep{Cadieux2024}. If confirmed, this would represent the first detection of a secondary atmosphere on a potentially habitable exoplanet. However, stellar contamination from unocculted faculae may be mimicking these spectral features. Disentangling genuine atmospheric signals from such stellar activity remains a significant challenge; consequently, further JWST observations are essential to validate these preliminary hints \citep{Cadieux2024}.

Another rocky world, 55\,Cancri\,e offers a distinct and particularly instructive case that extends the Cosmic Shoreline discussion into a regime not well captured by the original framework: the ultra-short-period lava world. With a radius of $\sim$1.95\,$R_\oplus$ and a mass of $\sim$8.8\,$M_\oplus$, orbiting a Sun-like G-type star on a 0.74-day orbit, 55\,Cnc\,e has an equilibrium temperature of $\sim$2000\,K, placing it among the most highly irradiated rocky planets known. Unlike the M-dwarf targets discussed above, the extreme irradiation of 55\,Cnc\,e is not the result of a habitability-zone orbit around a cool star but of an extremely tight orbit around a bright host. Its position on the Cosmic Shoreline diagram is well above the boundary defined by solar system bodies, placing it firmly in the regime where primary atmospheric retention is implausible and where the persistence of any secondary atmosphere would require continuous replenishment.

JWST thermal emission spectroscopy with NIRCam and MIRI/LRS, spanning 4--12\,$\mu$m, has provided the most compelling evidence to date for a secondary volatile atmosphere on a rocky exoplanet \citep{Hu2024}. The measured spectrum rules out the scenario in which the planet is a bare lava world shrouded only by a tenuous atmosphere of vaporized silicate rock, and instead indicates the presence of a bona fide volatile atmosphere, likely rich in CO$_2$ or CO \citep{Hu2024}. The key physical interpretation is that this atmosphere is not a remnant of accretion but is actively outgassed from — and sustained by — a persistent magma ocean driven by the intense stellar heating and tidal dissipation \citep{Hu2024}. This positions 55\,Cnc\,e as a living example of the outgassing side of the escape-versus-outgassing balance. Rather than losing its atmosphere irreversibly, the planet may maintain a dynamic 
equilibrium in which volatile loss to space is continuously offset by volcanic replenishment from below.

However, substantial uncertainties remain. Independent \textit{JWST}/NIRCam transit observations revealed strong variability in eclipse depths on sub-weekly timescales \citep{Patel2024}, suggesting that the atmosphere — if present — is not in a steady state but undergoes rapid changes, possibly driven by dynamic interactions between the atmosphere and the underlying magma ocean \citep{Zilinskas2025}. Forward modeling exercises using extensive grids of self-consistent atmospheric compositions find that, while a nitrogen-dominated, hydrogen-poor atmosphere enriched in PO and CO$_2$ is statistically favored, a wide range of alternative compositions — including H$_2$O-, CO-, or Si-bearing atmospheres — cannot currently be excluded \citep{Zilinskas2025}. The heat redistribution factor and surface pressure remain degenerate with composition at the current signal-to-noise level, underscoring the need for additional observations.

In the context of the Cosmic Shoreline, 55\,Cnc\,e stands up as an interesting case because it appears to host an atmosphere despite sitting well above the boundary defined by solar system analogues. This suggests that the traditional Cosmic Shoreline, which was calibrated primarily against bodies where escape dominates a fixed volatile inventory, does not capture the full picture for planets where outgassing from a magma ocean can continuously regenerate a lost atmosphere \citep{Hu2024, Zilinskas2025}. As noted by \citet{Hu2024}, both Venus and early Earth may have passed through a similar lava-world stage, making 55\,Cnc\,e a potential observational window into the atmospheric conditions of terrestrial planets during their earliest evolution. The implication is that the Cosmic Shoreline may need to be complemented by a second boundary — an ``outgassing shoreline'' delineating conditions under which volcanic replenishment can sustain an atmosphere against escape to fully describe the diversity of rocky planet atmospheric states.

This conceptual need for a more dynamic framework is addressed quantitatively by \citet{Ji2025}, whose recent work has prompted a significant reassessment of the Cosmic Shoreline itself. \citet{Ji2025} computed time-integrated mass loss to determine the critical insolation required for retention as a function of planetary mass, stellar mass, and initial volatile inventory using the atmospheric loss rates as a function of XUV flux; the atmospheric loss models used in this compilation were calculated by other authors for diverse atmospheric compositions (CO$_2$: \citealt{Tian2009thermal,Tianetal2009}; H$_2$O: \citealt{johnstone2020}, see Sect.~\ref{AtmRet} below; N$_2$-dominated: \citealt{nakayama2022-survival} using a derivative of the \citealt{johnstone2018} model, and \citealt{chatterjee-pierrehumbert2026}). Under this expanded multi-dimensional framework, a carbon-rich atmosphere on 55~Cnc~e becomes theoretically plausible despite its extreme irradiation. Specifically, the planet falls below the revised retention threshold for CO$_2$- and N$_2$-dominated atmospheres, provided there is a sufficient initial volatile reservoir, even though it would remain ``above the line'' for lighter CH$_4$-dominated cases \citep{Ji2025}.

A similar retention analysis was performed by \citet{vanlooveren2025} for CO$_2$+N$_2$ atmospheres based on the atmospheric model of \citet{johnstone2018} in the context of stellar evolution (see Sect.~\ref{AtmRet}  below). Both works carry direct implications for future observational programs, particularly for JWST and the Hubble Space Telescope's Rocky Worlds DDT Program. Although they come to different conclusions, the revised shoreline calculated by \citet{Ji2025} indicates that dozens of additional known planets may be capable of retaining atmospheres, provided they began with sufficient volatile inventories. Furthermore, \citet{Ji2025} suggest that the Cosmic Shoreline is better characterized as a ``transition zone" rather than a binary boundary. If planets form with diverse initial volatile contents, we should expect a gradual statistical transition between airless worlds and those that retain substantial secondary atmospheres. On the other hand \citet{vanlooveren2025} model predicts that none of the planets observed by JWST during Cycle 1 and 2 that are currently located in the HZ of their respective stars, would be able to retain atmospheres dominated by N$_2$ or CO$_2$. Planets with masses larger than 1 M${_\oplus}$ could be the best candidates for atmospheric retention an characterization \citep{pass2025receding}.

More broadly, these works underscore that the traditional two-dimensional Cosmic Shoreline (irradiation vs.\ escape velocity) is a simplification; a comprehensive treatment must incorporate atmospheric composition, initial volatile budgets, stellar type, and system age. Some of this is addressed by numerical models of atmospheric loss described in Sect.~\ref{AtmRet} below. 

\subsection{Atmospheric retention}\label{AtmRet}

In the following discussion, we consider models of atmospheric loss driven by stellar radiation; if loss occurs on time scales of order Myrs, we consider it as catastrophic and not supportive of evolving life. Also, we will entirely focus on secondary atmospheres, that is, atmospheres degassed from planetary mantles after the primordial hydrogen-rich atmosphere has been lost. 
For the sake of making meaningful models, we briefly discuss the parameter range likely of interest for habitable planets. Planets with masses much below 1$M_{\oplus}$ are unlikely to keep their atmospheres over very long times. Mars’ atmospheric pressure has dropped massively over geological timescales even though it orbits at the outer boundary of the solar HZ \citep{kite2019, warren2019}. We therefore limit our discussion to planets with masses of order 1$M_{\oplus}$.

The present-day Earth's atmosphere would not have survived if subject to XUV irradiation levels expected from the Sun in the Hadean or Archean epoch. Subject to an irradiation in X-rays of $\sim 35$~erg~cm$^{-2}$s$^{-1}$, that corresponds to the slow-rotator track at an age of 100~Myr and the medium rotator track at 1~Gyr, photodissociation of molecules, partial ionization and a transonic wind will dominate in the upper atmosphere (\citealt{johnstone2019}; see also Chapter 3 of this book). The entire atmosphere would have been lost within 0.1~Myr, which is essentially instantaneous. Evidently, the atmospheric composition at those early times must have been different, including efficiently cooling molecules such as CO$_2$ or CH$_4$ \citep{kulikov2007, johnstone2018, lammer2018, catling2020}. 

Similar results hold for an H$_2$O-rich atmosphere under stellar XUV irradiation conditions of young planets. If a pure H$_2$O atmosphere develops early in a Earth-like planet's life, for example around a pure ocean planet, H$_2$O will photodissociate and its products ionize under the XUV irradiation. \citet{johnstone2020} simulated such an atmosphere for EUV irradiation in the range of $\sim$100--5600~erg~cm$^{-2}$~s$^{-1}$ (X-ray fluxes being $\sim 5-8$ times lower; see also Chapter 3 of this book). The upper atmosphere will heat up to 19,000~K and will develop a transonic hydrodynamic oxygen+hydrogen wind at altitudes of several $10^4$~km. During the first few 100~Myrs of a Sun-like star's evolution, an Earth-mass planet at 1~au loses the amount of water corresponding to between 1 Earth ocean (for the Sun's slow rotator track corresponding to a rotation period of $P_{\rm rot} \approx 6$~d on the zero-age main sequence, see \citealt{tu2015} and Sect.~\ref{sect:rotXUVevol}) and 40 Earth oceans, if such a reservoir is available. An Earth-like planet is therefore at risk to lose any H$_2$O atmosphere in its early evolution, and the situation would be worse for planets around M dwarfs for which the high-activity period is prolonged (see Sect.~\ref{sect:rotXUVevol} above).

Evidently, then, stellar XUV evolution requires appropriate compositions for atmospheres to survive the early, intense irradiation levels. Most importantly, gases cooling the upper atmospheres radiatively are needed.
To avoid the complex chemistry involving sulfur and methane for the moment, we assume CO$_2$ + N$_2$ atmospheres, both gases being abundant around rocky bodies in the solar system. These molecules behave very differently regarding atmospheric cooling. While CO$_2$ strongly cools atmospheres by radiation, N$_2$ is a very poor radiator and therefore helps keeping atmospheres hot. 

Retention models must therefore be thermo-chemical including chemical reaction networks, radiative transport, mechanisms of mass-loss to space (Jeans, hydrodynamic, or non-thermal, see Chapter 2 of this book), and internal outgassing. \citet{johnstone2021b} studied CO$_2$+N$_2$ atmospheres and their evolution on early Earth (Archean Earth, 3.8--2.5~Gyr before present) taking into account various possible evolutionary tracks of the solar XUV radiation (Sect.~\ref{sect:rotXUVevol}). They used the thermo-chemical 1-D upper-atmosphere code \textit{Kompot} \citep{johnstone2018} that considers solar XUV and IR heating, non-thermal photoelectron heating, Joule heating, heating due to exothermic reactions, radiative cooling via IR lines, conduction, and energy exchange between electrons, ions, and neutrals. Chemical reaction networks, eddy diffusion and molecular diffusion are included, as is radiative transport. To study conservative lower limits to the mass loss, only Jeans escape was considered in the study (but no hydrodynamic escape).

\begin{figure}[ht]
\begin{center}
\includegraphics[width=0.8\textwidth]{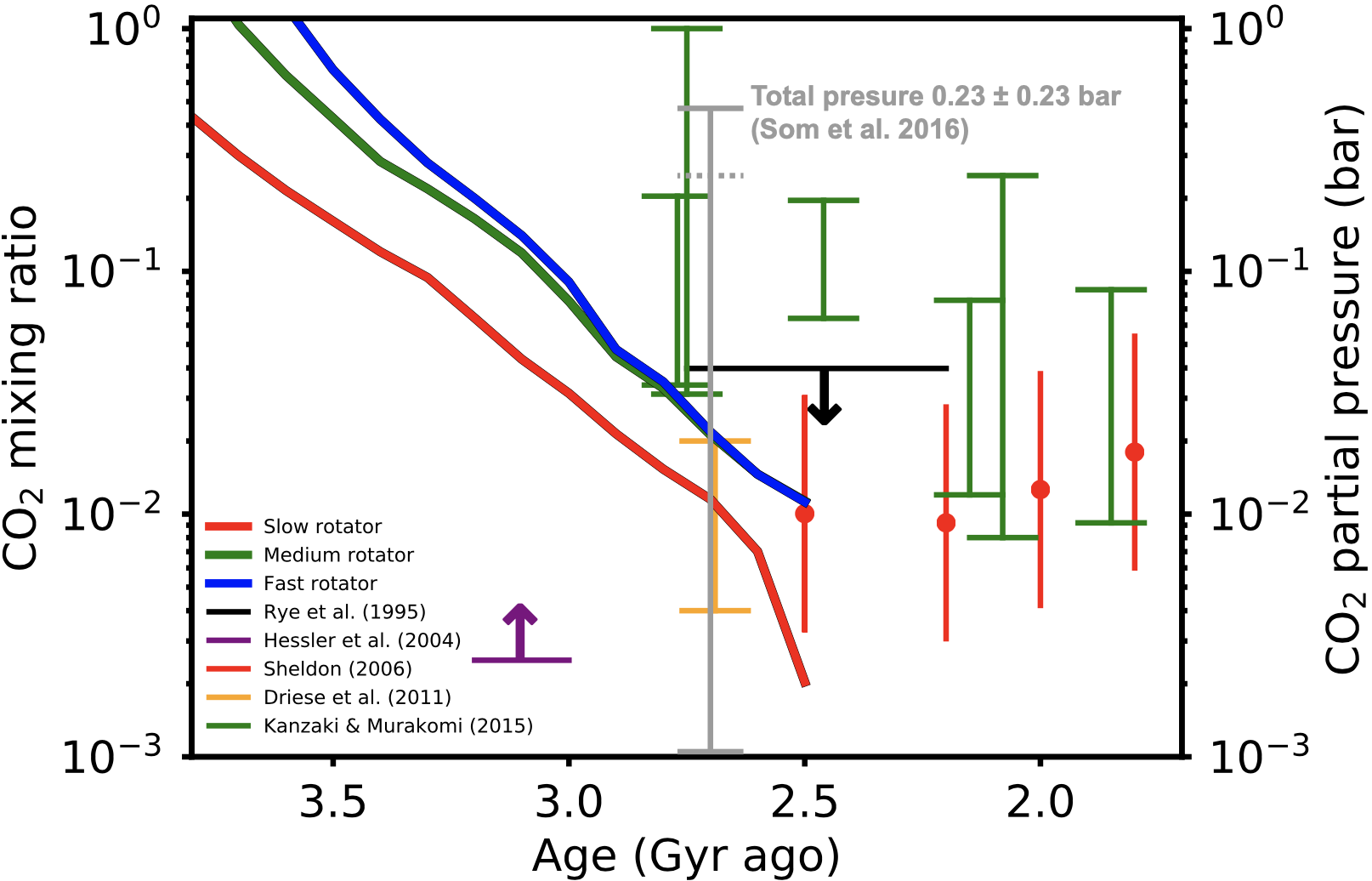} 
\caption{Minimum CO$_2$ mixing ratio required for the Archean Earth's atmosphere for three solar XUV evolutionary tracks (red/green/blue curves for initially slow, medium, and fast rotators, respectively; see Sect.~\ref{sect:rotXUVevol}). The uncertainty error bars or upper limits  are from geochemical measurements. The grey error bar shows the total pressure of $0.23\pm 0.32$~bar at 2.7~Gyr before present, inferred from gas bubbles in basaltic lava flows by \citet{som2016earth}. (From \citealt{johnstone2021b} where the references for the geochemical measurements are listed.)}  \label{fig:ArecheanCO2}
\end{center}
\end{figure}

Going back in time, the Sun's XUV luminosity was higher, although the unknown initial solar rotation period makes it impossible to determine the XUV evolutionary path from first principles (Sect.~\ref{sect:rotXUVevol}). In any case, the upper atmosphere was heated more efficiently, and the exobase altitude where the atmosphere becomes non-collisional and Jeans escape becomes effective, was higher. The latter helps making escape more efficient. Raising levels of XUV flux make CO$_2$ cooling increasingly important; if the CO$_2$ mixing ratio is too low, the upper atmosphere heats sufficiently to get lost entirely on short time scales. The model results in Fig.~\ref{fig:ArecheanCO2} \citep{johnstone2021b} show that between 2.5~Gyr and 3.8~Gyr before present, the minimum CO$_2$ mixing ratio to keep Earth's atmosphere varied from $<1\%$ (at 2.5~Gyr) to at least 40\% (3.8~Gyr). For the latter period, distinguishing between fast/medium/slow rotation tracks for the Sun is important. Only a slow rotator track allows the CO$_2$ mixing ratio to stay below 100\% for atmospheric retention at 3.8~Gyrs. \textit{This excludes a medium or fast rotator track for the Sun.}
The required CO$_2$ mixing ratios for atmospheric retention in the Archean are also found to be sufficient to solve the Faint Young Sun paradox (for the latter, see \citealt{sagan1972, feulner2012}).

This model can be generalized to all spectral types and stellar masses on the cool main sequence. To make retention estimates more reliable, ongoing outgassing through volcanism must be considered that may replenish secondary atmospheres, beyond the initial buildup of a secondary atmosphere through the solidification of a magma ocean. For example, the present Earth's atmosphere gains $\sim 10^4$~kg~s$^{-1}$ of CO$_2$ and  $\sim 5.4\times 10^4$~kg~s$^{-1}$ of H$_2$O \citep{catling2017}. Stagnant-lid planets like early Earth can outgass about $\sim 1.4\times 10^4$~kg~s$^{-1}$ of CO$_2$ and some H$_2$O and CO \citep{guimond2021}. Even over multiple Gyrs, outgassing of 100~kg~s$^{-1}$ should be common \citep{baumeister2023}. We note that for CO$_2$, an outgassing rate of $\sim 1.7\times 10^4$~kg~s$^{-1}$ corresponds to 10\% of the Earth's atmosphere in 1 Myr. 

To retain the atmosphere, the outgassing rate must exceed mass-loss to space. \citet{vanlooveren2025} studied atmospheric retention of a CO$_2$+N$_2$ atmosphere around Earth-mass planets in the HZ considering outgassing and thermal loss throughout much of the main-sequence evolution of late-type stars with masses $(0.1-1)M_{\odot}$. They considered an atmosphere to be catastrophically lost if the mass-loss rate exceeds $\sim 1.6\times 10^4$~kg~s$^{-1}$ or 1 Earth atmosphere in 10~Myrs. This rate is similar to the higher outgassing rates mentioned above, thus providing a conservative estimate favoring atmospheric retention. The study is also conservative in favor or retention assuming a slowly rotating star, considering only Jeans-mass loss (many model atmospheres would lead to much higher hydrodynamic loss), only day-side loss (potentially lower rates on the night side would add more loss), ignoring additional non-thermal loss processes, and ignoring H$_2$O (which would be easily lost after photodissociation). The evolution of XUV irradiation of the star is fully accounted for.

\begin{figure}[ht]
\begin{center}
\includegraphics[width=0.8\textwidth]{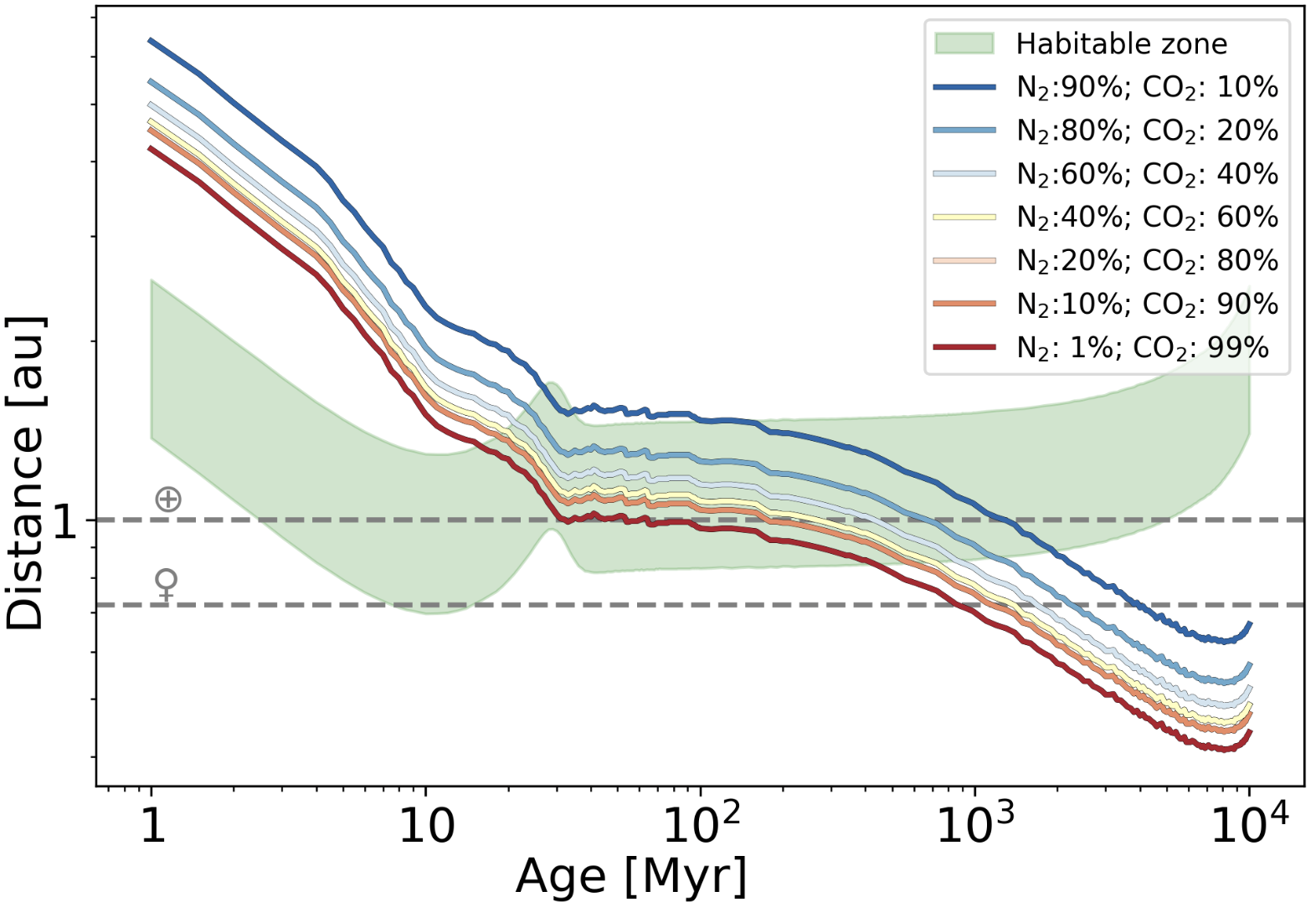} 
\caption{Evolution of atmospheric retention distance for a CO$_2$ + N$_2$ atmospheres on a $1M_{\oplus}$ planet around an slow-rotator solar-mass star, from an age of 1~Myr after the protoplanetary disk phase to 10~Gyr. Above the colored lines (referring to different CO$_2$/N$_2$ mixing ratios indicated in the inset), the momentary atmosphere will be retained, below the line, it will be lost (retention time criterion: 10~Myr). The orbital distances of Venus and Earth are marked with gray dashed lines. (From \citealt{vanlooveren2025}.)}  \label{fig:Retetion1Msun}
\end{center}
\end{figure}

Figure~\ref{fig:Retetion1Msun} shows the evolution of the \textit{atmospheric retention distance} (ARD) for a $1M_{\oplus}$ planet around an initially slowly rotating $1M_{\odot}$  star (Sect.~\ref{sect:rotXUVevol}) for various CO$_2$:N$_2$ mixing ratios. Planets located above the colored lines (referring to different mixing ratios, see inset) can retain their atmospheres, those below will lose it. The green band shows the location of the HZ, which varies somewhat over time because of stellar evolution. The ARDs initially contract steeply because of the decline of the saturated $L_{\rm X} \propto L_{\rm bol}$ during the stellar contraction phase. Later (after $\sim$40~Myrs), they drop further due to the decline of $L_{\rm X}$ owing to stellar spin-down. In the age range 10--100~Myrs retention is problematic for all mixing ratios. For the Earth's distance, retention is possible at $\sim$100~Myrs only for nearly pure CO$_2$ atmospheres. Only at 1~Gyr will retention be possible for any mixing ratio. These results echo those of the \citet{johnstone2021b} study. For Venus, no mixing ratio guarantees retention before an age of 1~Gyr. Outgassing over extended periods of billions of years under stagnant-lid conditions is a viable remedy \citep{orourke2015}.

\begin{figure}[ht]
\begin{center}
\includegraphics[width=1.0\textwidth]{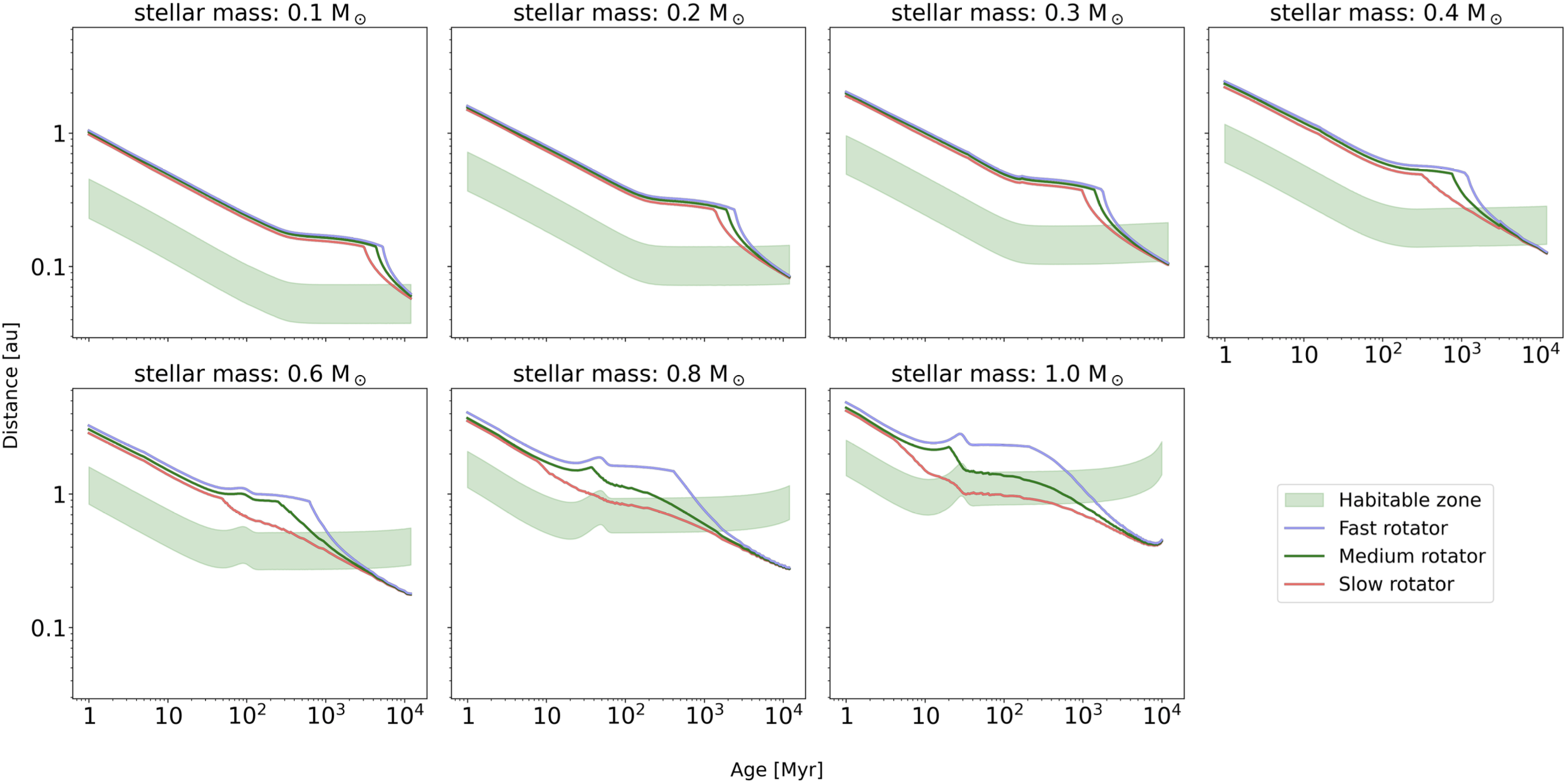} 
\vspace*{-0.3 cm}
\caption{ARDs (colored lines) as a function of age for different stellar masses,  for CO$_2$ mixing ratio of 99\%
(retention time criterion: 10~Myr). Red, green, and blue lines refer to an initially  slow, medium, and fast rotator. The green band shows the location of the HZ. (From \citealt{vanlooveren2025}.)}  \label{fig:RetentionLowMass}
\end{center}
\end{figure}

Fig.~\ref{fig:RetentionLowMass} shows similar plots for different stellar masses, but instead of different mixing ratios, we plot ARDs for an initially slow (red), medium (green), and fast (blue) rotator, while we conservatively adopt a CO$_2$:N$_2$ mixing ratio
of 99\%:1\% for all simulations. As in the study of \citet{johnstone2021b}, we see that for a solar-mass star atmospheric retention is problematic for medium and fast rotators up to about 1~Gyr at 1~au even for a pure CO$_2$ atmosphere. The situation becomes progressively worse for decreasing stellar mass. For 0.4$M_{\odot}$ atmospheric retention in the HZ is possible only past 1~Gyr regardless of the atmospheric mixing ratio. For a 0.1~$M_{\odot}$ star, retention can be achieved at ages beyond 7--9~Gyrs only. This is the situation for the Trappist-1 planets. The Trappist-1  age is estimated to be 7.46$^{+2.01}_{-2.10}$~Gyr \citep{fleming2020} or $7.6\pm 2.2$~Gyr \citep{burgasser2017}, and simulations performed for these planets show that presently (and at any time in the past) the ARDs are located significantly outside the HZ \citep{vanlooveren2024}.

\begin{figure}[ht]
\begin{center}
\includegraphics[width=1.0\textwidth]{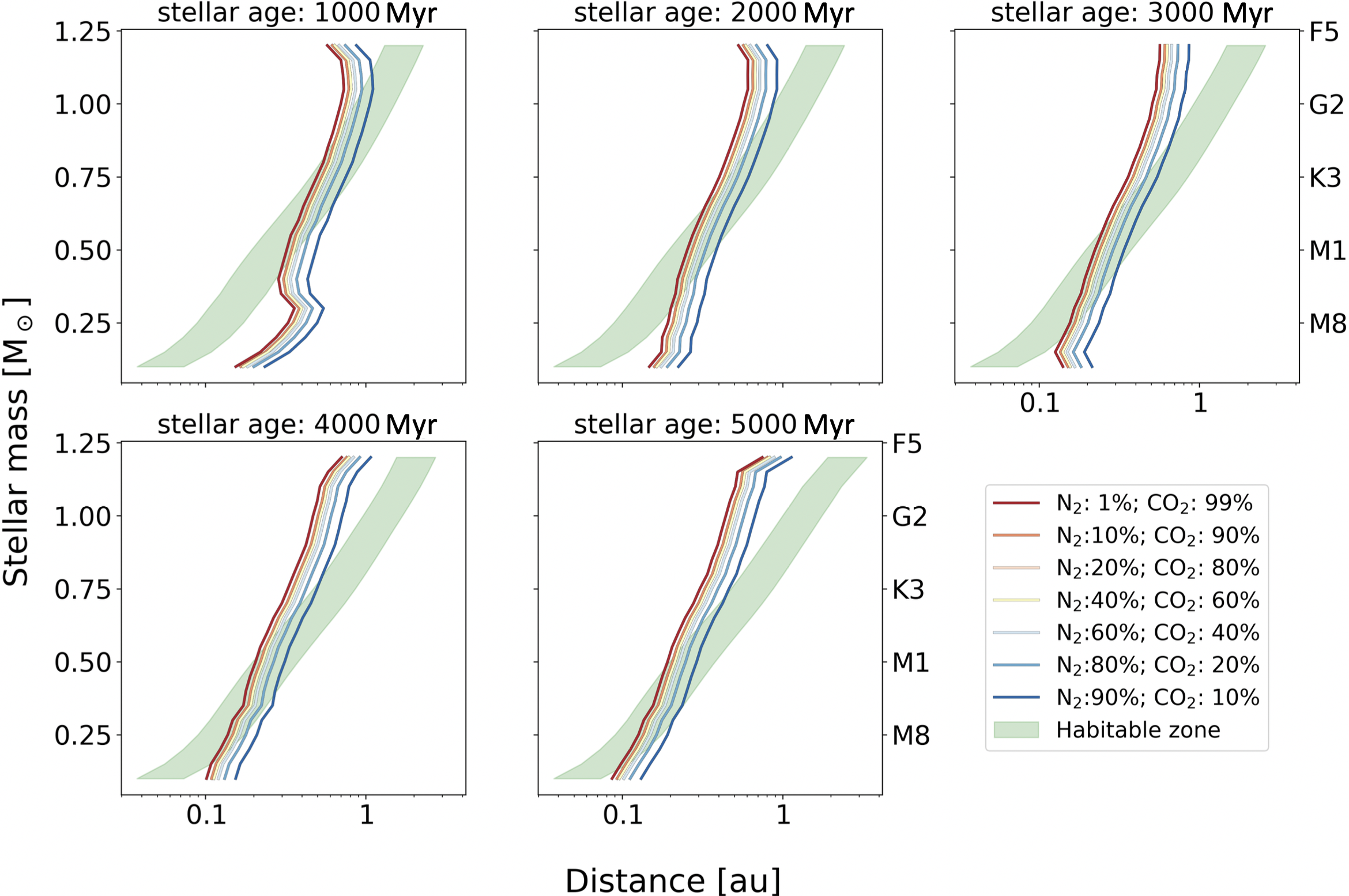} 
\vspace*{-0.3 cm}
\caption{Retention distances versus stellar mass for different ages. The colors refer to different CO$_2$/N$_2$ mixing ratios (see box at lower right). An initially slowly rotating star has been assumed. Atmospheres can be retained to the right of the ARDs (retention time criterion: 10~Myr). (From \citealt{vanlooveren2025}.}  \label{fig:RetentionAge}
\end{center}
\end{figure}

Finally, Fig.~\ref{fig:RetentionAge} shows ARDs as a function of stellar mass at different ages for an initially slowly rotating star with a mass of $1M_{\odot}$. Here, the colored lines again refer to different CO$_2$:N$_2$ mixing ratios. Atmospheric retention is possible to the right of these lines. At 1~Gyr all ARDs for stars below $0.4M_{\odot}$ are outside the HZ (green band). N$_2$-rich atmospheres have, as expected, larger ARDs and it requires a higher stellar mass for a given age, or a higher age for a given mass to allow for retention. The knee at 1~Gyr for $<0.35M_{\odot}$ stars toward large distances is a consequence of the slow spin-down and activity decline of late M dwarfs, and for the lowest-mass stars partly a consequence of the long, ongoing contraction phase (Sect.~\ref{sect:rotXUVevol}). As time proceeds, all ARDs move inward as stellar activity declines for all stellar masses, cross the HZ, and eventually are located inside the HZ. The high-CO$_2$ mixing ratio atmospheres are always the most favorable ones. However, the ARD migration inside the HZ takes longest for M dwarfs for which even past 5~Gyrs retention critically depends on atmospheric composition. 

An interesting prediction from atmospheric loss models is the existence of planets with massive O$_2$ atmospheres around active M dwarfs. After a planet has initiated a runaway greenhouse, the cold trap at the tropopause moves towards the upper atmosphere until the atmosphere becomes isothermal. During this process water molecules are photolyzed and hydrogen is lost to space, leaving oxygen behind. With this mechanism, \citet{luger2015extreme} predicted atmospheres of the order of 1000 bars of O$_2$, but oxygen can be dragged along hydrogen in that case atmospheres would be less dense \citep{Bolmont2017-waterloss,johnstone2020}. The other highly relevant ingredient is atmosphere-surface interactions, \citet{Wordsworth2018-redox} found that before the catastrophic loss of water, the surface temperature produced by a the water vapor atmosphere would be high enough to melt the crust. As a result, oxygen produced by water photolyisis would be absorbed by the magma ocean avoiding the accumulation in the atmosphere above a few tenths of bars \citep{Wordsworth2018-redox}. Currently, planets in the TRAPPIST-1 system are some of the best candidates to hold O$_2$ dominated atmospheres \citep{Lincowski2019-oceanlost,turbet2020review}.

To summarize, we emphasize the necessity to consider stellar and planetary evolution to assess the presence of atmospheres and habitability based on models: From our discussion, we see that habitable conditions on a planet cannot be derived from the presently observed stellar properties and fundamental parameters of a planet such as mass, radius, and orbital distance from the host star. Unless explicit observations can prove habitable climate conditions and the presence of liquid water, \textit{model calculations need to include the entire evolutionary history of a planetary atmosphere under the radiative [bolometric, UV, XUV] and wind evolution of the host star.} The shrinking of the HZ of M dwarfs over long time periods (hundreds of Myr to over a Gyr) during which planetary secondary atmospheres may be built up, and the consequent excessive radiative forcing of their atmospheres and the risk of water loss especially in the early stages of evolution remain a fundamental problem. But also the initial rotation periods of stars matter; rapid initial rotation may also lead to excessive forcing in the first Gyr of evolution, potentially making habitability impossible for later times. At the time we observe planets in the HZ, the atmosphere and surface water may be long gone owing to unfavorable stellar conditions during the previous evolution, even if the present stellar conditions were favorable for a habitable planet. On the other hand, planets may reach habitable conditions later if outgassing of a secondary atmosphere is continuing sufficiently long or the formation of a secondary atmosphere starts only once the external forcing has declined sufficiently; direct observational verification remains indispensable.

\subsection{Atmospheric photochemistry}\label{AtmPhoto}
The main atmospheric components of rocky planets in the habitable zone are usually assumed to be CO$_2$, N$_2$ and H$_2$O, but models predict that also CO and H$_2$ could be abundant in terrestrial planetary atmospheres \citep[e.g.][]{oosterlo-etal2021,ortenzietal-2020}. The composition depends on the mantle redox state, surface sinks, atmospheric loss, the distance to the star and the planetary size, among other processes \citep[see][for a review]{wordsworth-kreidberg2022}. These compounds are degassed by volcanoes and they are sources of life-bearing elements, C, N, H and O. 

Atmospheric chemistry is driven by the UV photons and high energy particles emitted by the star, that break molecules which products participate in thermochemical and third-body reactions. In this section we focus on the effect of photons in the atmospheric chemistry, for the effect of particles see \ref{particles-habitability}.  The atmospheric chemistry relevant for a potentially habitable exoplanets are:
\begin{enumerate}
    \item CO\textsubscript{2} photolysis:
    \begin{subequations}
    \begin{equation}
        \mathrm{CO}_2 + \mathrm{h}\nu \longrightarrow \mathrm{CO} + \mathrm{O}\quad (\lambda \lesssim 200\,\mathrm{nm})
    \end{equation}
    \begin{equation}
        \mathrm{CO}_2 + \mathrm{h}\nu  \longrightarrow \mathrm{CO} + \mathrm{O}(^{1}\mathrm{D})\quad (\lambda < 167\,\mathrm{nm})
    \end{equation}
    \end{subequations}
Where photons are represented by their energy ($h\nu$), and O(\textsuperscript{1}D) is an oxygen radical that is highly reactive and thus short lived, and as such can return to a ground state via collision O(\textsuperscript{1}D) + M $\rightarrow$ O + M or react with other chemical species. M is any atmospheric atom or molecule.

\item O\textsubscript{2} formation:
    \begin{equation}
        \mathrm{O} + \mathrm{O} + \mathrm{M} \longrightarrow \mathrm{O}_2 + \mathrm{M}
    \end{equation}
Thus, the net result of the CO\textsubscript{2} photolysis is 2CO\textsubscript{2} $\rightarrow$ 2CO + O\textsubscript{2}.
    
 \item The Chapman cycle:
    \begin{subequations}
    \begin{equation}
        \mathrm{O}_2 + \mathrm{h}\nu \rightarrow \mathrm{O} + \mathrm{O} \quad(175\,\mathrm{nm} < \lambda < 242\, \mathrm{nm})
    \end{equation}
     \begin{equation}
        \mathrm{O}_2 + \mathrm{h}\nu \longrightarrow \mathrm{O} + \mathrm{O}(^{1}\mathrm{D})\quad (\lambda < 175\, \mathrm{nm})
    \end{equation}
    \end{subequations}   
    \begin{equation}
        \mathrm{O} + \mathrm{O}_2 + \mathrm{M} \longrightarrow \mathrm{O}_3 + \mathrm{M}
    \end{equation}

    \item The HO\textsubscript{x} (H, OH, HO$_2$) chemistry that starts with the generation of H and OH via the reactions:
    \begin{equation}
        \mathrm{H}_{2}\mathrm{O} + h\nu \longrightarrow \mathrm{OH} + \mathrm{H}
    \end{equation}
    \begin{equation}
        \mathrm{H}_{2}\mathrm{O} + \mathrm{O}(^{1}\mathrm{D}) \longrightarrow \mathrm{OH} + \mathrm{OH}
    \end{equation}
    
\item The NO\textsubscript{x} (N, NO, NO$_2$) chemistry initiated by high energy particles or atmospheric lightning breaking the triple bond of N$_2$:
    \begin{equation}
        \mathrm{N}(^{2}\mathrm{D})+ \mathrm{O}_{2} \longrightarrow \mathrm{NO} + \mathrm{O}
    \end{equation}
    \begin{equation}
        \mathrm{NO} + \mathrm{O}_3 \longrightarrow \mathrm{NO}_2 + \mathrm{O}_2
    \end{equation}
For present Earth another source of NO$_x$ is reaction of N$_2$O produced by life with O(\textsuperscript{1}D):
     \begin{equation}\label{reaction:H2Ophoto}
        \mathrm{N}_{2}\mathrm{O} +\mathrm{O}(^{1}\mathrm{D}) \longrightarrow \mathrm{NO} + \mathrm{NO}
    \end{equation}
    
\item The catalytic cycles that destroy O\textsubscript{3}:
\begin{subequations}
    \begin{equation}
        \mathrm{XO} + \mathrm{O}_3 \longrightarrow \mathrm{XO} + \mathrm{O}_2
    \end{equation}
    \begin{equation}
        \mathrm{XO} + \mathrm{O} \longrightarrow \mathrm{X} + \mathrm{O}_2
    \end{equation}
\end{subequations}
    where X can be any free radical, usually N or H, other atoms\textemdash for example, Cl and Br\textemdash  are less important for the general case of CO$_2$-N$_2$ dominated terrestrial atmospheres. The net result of these catalytic reactions is O\textsubscript{3} + O $\rightarrow$ 2O\textsubscript{2}.

\item CO$_2$ recombination via CO + O is is spin forbidden \citep{yung_photochemistry_1999}. The responsible of recombining CO$_2$ are catalytic reactions with HO$_x$, this mechanism is relevant for Venus-like atmosphere were small abundances of water ($\sim$ 1 ppm) are able to recombine  CO back to CO$_2$:
    \begin{equation}
        2 ( \mathrm{H} + \mathrm{O}_2 + \mathrm{M} \longrightarrow \mathrm{HO}_2 + \mathrm{M} )
   \end{equation}
    \begin{equation}
        \mathrm{HO}_2 + \mathrm{HO}_2 \longrightarrow \mathrm{H}_2\mathrm{O}_2 + \mathrm{O}_2
    \end{equation}
    \begin{equation}
       \mathrm{H}_2\mathrm{O}_2 + h\nu \longrightarrow \mathrm{OH} + \mathrm{OH}
    \end{equation}
    \begin{equation}
       2(\mathrm{OH} + \mathrm{CO} \longrightarrow \mathrm{CO}_2 + \mathrm{H})
    \end{equation}
The net result of these reactions is 2CO + O\textsubscript{2}~$\rightarrow$~2CO\textsubscript{2}. When O\textsubscript{3} becomes more abundant then, the former pathway shifts to a set of reactions with the net result of CO + O\textsubscript{3}~$\rightarrow$~CO\textsubscript{2} + O\textsubscript{2} \citep{grenfell2013,gao_stability_2015}:
    \begin{equation}
     \mathrm{H} + \mathrm{O}_2 + \mathrm{M} \longrightarrow \mathrm{HO}_2 + \mathrm{M}
     \end{equation} 
     \begin{equation}
        \mathrm{HO}_2 + \mathrm{O}_3 \longrightarrow \mathrm{OH} + 2\mathrm{O}_2
    \end{equation} 
     \begin{equation}
       \mathrm{OH} + \mathrm{CO} \longrightarrow \mathrm{CO}_2 + \mathrm{H}
    \end{equation}
\end{enumerate}

Methane is not expected to be abundant in terrestrial atmosphere \citep[][ and references therein]{thompson2022-CH4}, except for ocean-vaporizing impacts \citep[e.g.,][]{zahnle2020creation}, but it is nonetheless, a potential biosignature (Sect. \ref{Biosig}, and it participates on reactions relevant for prebiotic chemistry \citep[e.g.][]{Rimmer-Rugheimer-2019} and the formation of hazes \citep[e.g.][]{arney_pale_2017}.

Along with the photochemistry other processes are relevant for the abundance of trace chemical species. Sulfur-bearing compounds and HO$_x$, for example, are highly soluble in water, this means that they will not accumulate on planets with active hydrological cycles. For Earth-sized planets, H$_2$ is lost by diffusion because its low molecular weight. 

\section{Planetary habitability and biosignatures}\label{Hab}

The requirement of liquid water for a habitable planet imposes restrictions on the interior evolution of the planet, its formation history, and the stellar-planet interactions  \citep{Meadows-Barnes2018-Habitability}. 
Once the conditions necessary for water to remain liquid on a planet over a long period of time (ranging from millions to billions of years) have been established, the planet is considered habitable but is not yet inhabited. Being Earth the only planet with life, our tools to infer if there is life on others planets are chemistry principles, our knowledge about the characteristics of our planet when life arose, experiments and numerical models. 
Our main purpose is to identify potentially habitable planets and ultimately, to find signatures of life on those planets. In this section we review how the star impacts the habitability potential and the abundance of chemical compounds that may indicate the presence of life.

\subsection{Stellar UV: damage vs prebiotic chemistry}\label{UV-life}

The chemistry that leads to the formation of the building blocks necessary for the origins of life is called prebiotic chemistry \citep{cleaves2013prebiotic}. The idea that life can arise from simple molecules present in the early Earth proposed, independently by Oparin \citep{oparin-originlife}, Urey \citep{urey1952early} and Bernal \citep{bernal1949physical}, lead to the first experiment of prebiotic chemistry performed by Stanley Miller \citep{miller1953production}. In this experiment, H$_2$O, H$_2$, CH$_4$, and NH$_3$ under electric discharges generated a mixture of compounds relevant for life, such as amino acids \citep{miller1953production}. For this famous experiment the simulated energy source was lightning but other possible sources are stellar UV photons and high energy particles. From the possible energy sources for prebiotic chemistry on exoplanets, UV is the only one that can be directly constrained from observations.

The UV radiation leads chemical pathways by breaking certain molecules, increasing the relative abundance of those that are UV photo-resistant. The new molecules that resulted from the photolysis of others, participate in new reactions \citep{rapf2016sunlight}. Presently, our atmosphere is opaque for wavelengths $<$ 280 nm due to the absorption by O$_2$ and O$_3$. During the Archean, before life appearance, surface UV radiation may have been $\sim3$ times larger than that of present Earth in the 200-315 nm wavelength range \citep{cockell1998biological}. Although Miller-type experiments have been performed for almost a century, UV as a source of energy for prebiotic chemistry has not been extensively explored \citep{rapf2016sunlight, ranjan2016influence}. In addition, the experiments focused on reproducing the conditions of early Earth may no be applicable to planets around other stars \citep{ranjan2017surfaceUV} because although all planets at the HZ receive nearly the same visible and near IR integrated stellar flux, the UV spectrum would depend on the contribution of the photosphere and the chromospheric activity of the star (Sect. \ref{stellarUV}).
 
\citet{rimmer2018origin} proposed an abiogenesis zone, the distance from a star where a planet would receive enough energy in the UV to build a relevant inventory of molecules for the origins of life. That energy was calculated considering a chemical pathway of seven reactions that ended in the production of RNA pyrimidine nucleotides. Their analysis concluded that M dwarfs would not provide enough UV for prebiotic chemistry except on a few very active cases, even considering the energy delivered by flares. \citet{ranjan2023uv} revised the abiogenesis zone using a lower, more realistic, single-scattering albedo to calculate the surface UV flux, finding that young or active M dwarfs as well as some quiescent M dwarfs would fall in the abiogenesis zone. 

Once life has emerged the next concern is the possible UV damage that may result in the sterilization of the planet. Damage has been calculated using the DNA inactivation action spectra \citep[e.g.][]{Cockell1999,segura2003ozone}, which provides a useful comparison for DNA damage under different UV spectra. K-type stars were deemed as the ones that provide the best UV radiation environment considering planets with anoxic atmospheres \citep{Cockell1999}. An esterilization zone was later proposed by \citet{gunther2020stellar} extrapolating the calculations by \citet{segura2010effect} when all the ozone layer of a planet with an atmosphere of 0.21 O$_2$ would be depleted by the particles emitted during a high energy flare. The sterilization zone fails to consider that the planet simulated by \citet{segura2010effect} has O$_2$ produced by photosynthesis, in other words, there will be no O$_3$ to be destroyed if we assumed a planet already sterilized. On the other hand, the high-energy particle flux associated to the 10$^{32}$ ergs UV flare, responsible of depleting the ozone layer was calculated using solar flare relations between the UV, X rays and particle flux that are poorly constrained and may not apply to M dwarfs \citep{herbst2019-solar2stellar}.

While prebiotic chemistry scenarios depend on chemistry and our ability to reproduce plausible geologic scenarios, either in the laboratory or with numerical models, UV damage scenarios require the assumption that life in other planets evolved similar protection mechanisms against UV. Thus we advocate to focus the role of UV radiation on prebiotic chemistry over scenarios that heavily overly on terrestrial biology.

\subsection{The role of high energy particles on habitability}\label{particles-habitability}

\citet{2017A&A...603A..96R} showed that stellar energetic particles (StEPs) of young, active stars play a key role in shaping the chemistry and ionization of planet-forming disks, and likely also affect the initial conditions and evolution of exoplanetary atmospheres and surfaces. Therefore, understanding high-energy particle effects is essential for interpreting observations of exoplanets and assessing their habitability, especially for those around young, active stars.
\begin{figure}[ht]
    \centering
    \includegraphics[width=\textwidth]{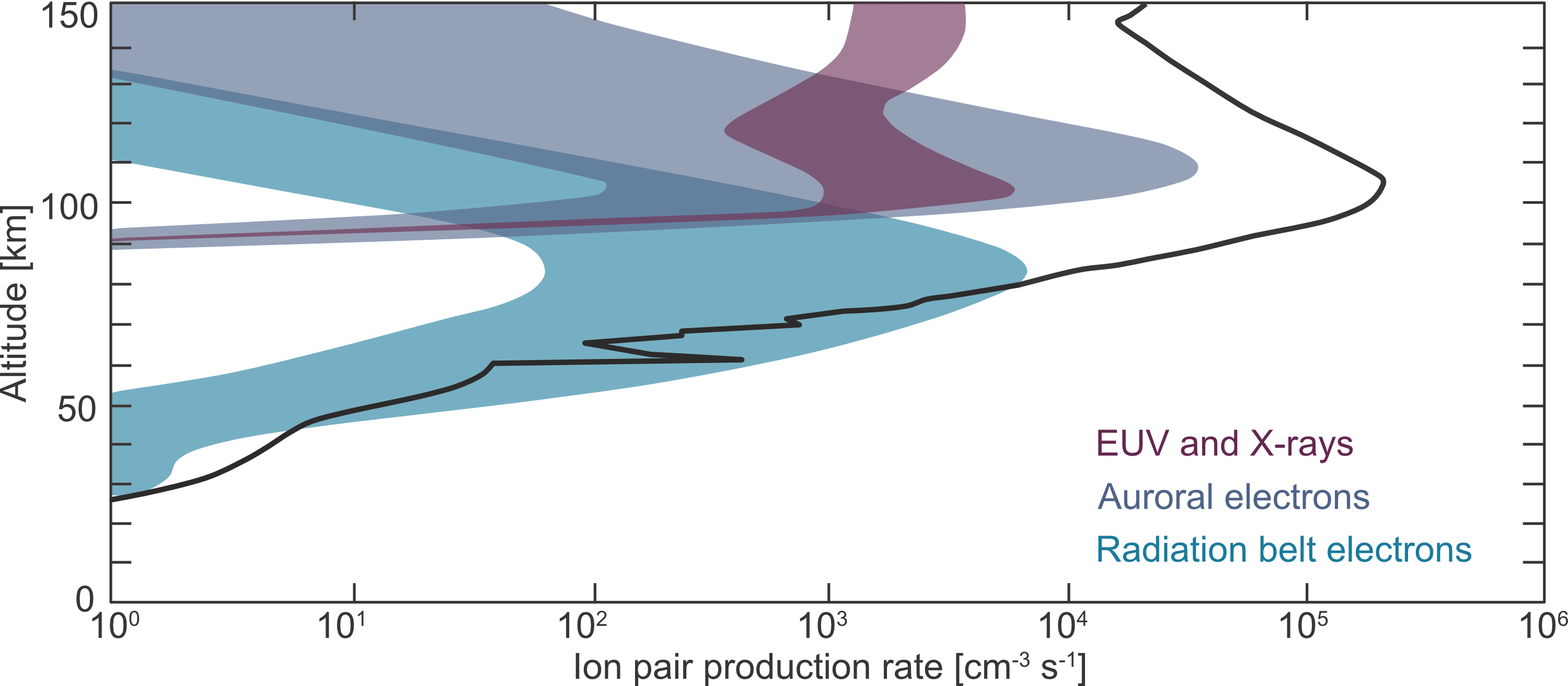}
    \caption{Altitude dependent terrestrial atmospheric ionization due to solar EUV and X-rays (down to 90 km), auroral electrons (down to 90 km), radiation belt electrons (down to 30 km). In addition, the radiation belt contribution including Bremsstrahlung according to \citet{Frahm-1997} (black line). Figure after \citet{BakerEA12}.}
    \label{fig:5.4.2}
\end{figure}

Within the Earth's atmosphere, interactions of solar wind and solar high-energy radiation (i.e., EUV and X-ray) lead to atmospheric heating, expansion, non-thermal escape processes, photodissociation, and particle precipitation in the upper atmospheres of solar system planets \citep[e.g.,][]{Airapetian2020, GronoffEA20}. EUV radiation also triggers photoionization and affects the temperature of the upper atmosphere \citep[e.g.,][]{2004plae.book.....B}. As shown in Fig.~\ref{fig:5.4.2}, these precipitation populations form the atmospheric high-altitude ionization profile \citep[e.g.,][]{BakerEA12}. 

However, to describe the full range of solar/stellar radiation also energetic particles have to be addressed. Solar system observations have revealed that solar flares are often accompanied by coronal mass ejections (CMEs) characterized by the eruption of relatively dense, magnetized material to the outer corona and the stellar wind \citep{2012LRSP....9....3W, 2017LRSP...14....2B}, and that both act as sources of charged particles (i.e., protons and electrons). These solar energetic particles (SEPs) are accelerated to energies in the keV to GeV regime via magnetic reconnection in flares or fermi acceleration at CME shock fronts. In general, SEPs emerging from flare sites are less energetic than their CME-driven counterparts \citep[e.g.,][]{2016LRSP...13....3D, Reames-1999, 2021LNP...978.....R}. Together with galactic cosmic rays (GCRs) originating far outside our solar system by being accelerated at the shock fronts of supernovae \citep{2006Natur.439...45D}, supernova remnants \citep{2005ApJ...619..314B}, and pulsars \citep{j.asr.2007.05.051}, these particles form the planetary high-energy particle environment. An in-depth introduction to the stellar particle environment is given in Alvarado-Gómez et al. (Chapter 2 of this book).

It is the high-energy cosmic rays (i.e., GCRs and SEPs) that dominate the ionization of the lower atmospheres, the chemistry and climate changes at these altitudes, as well as the radiation exposure on the surface. Causes and consequences will be discussed in the following.

\subsubsection{The atmospheric energy loss of charged particles} 

\begin{figure}
    \centering
    \includegraphics[width=\textwidth]{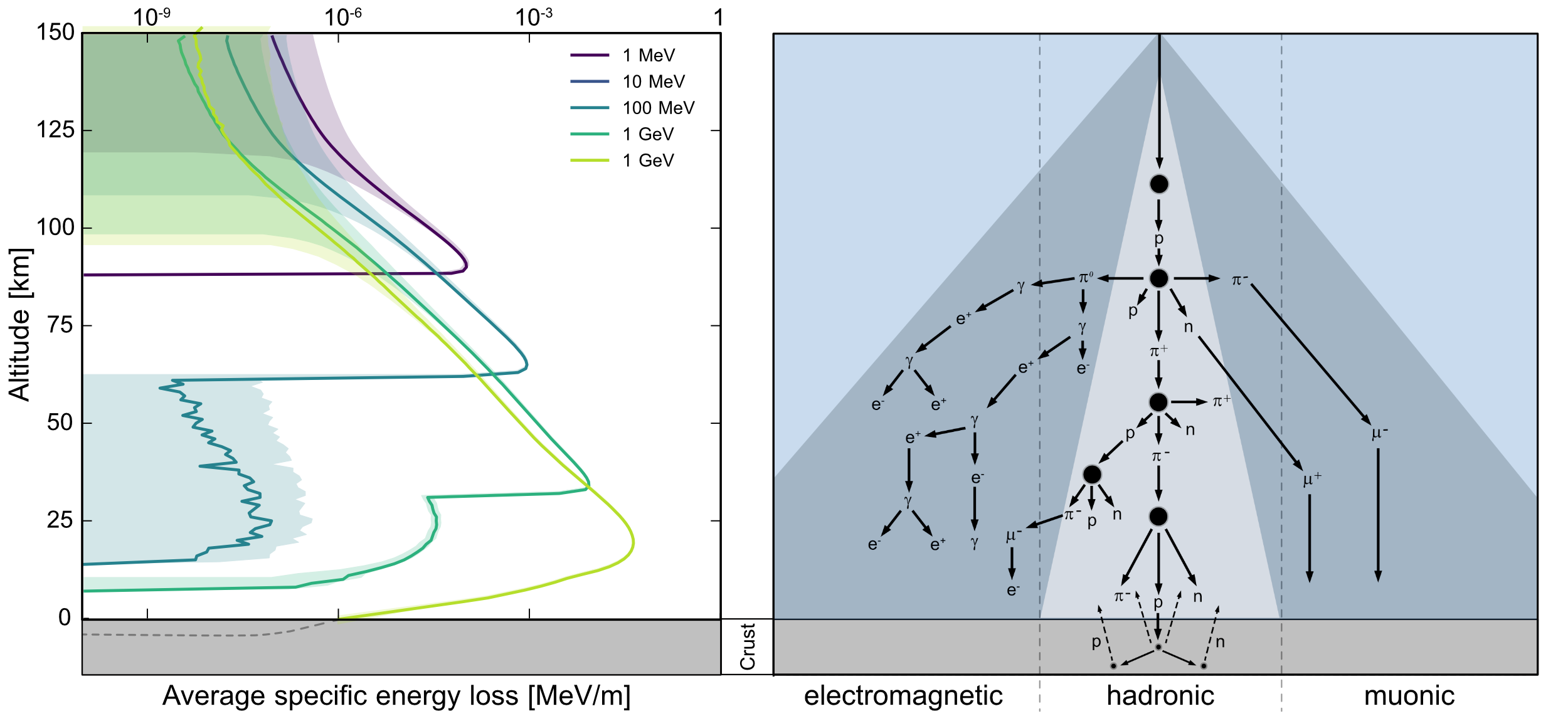}
    \caption{Left panel: Modeled average specific energy loss of primary protons with energies of 1 MeV (purple), 10 MeV (petrol), 100 MeV (green), and 1 GeV (light green). Figure adapted from \citet{Banjac-etal-2019a}. Right panel: Evolution of the complex hadronic–muonic–electromagnetic cascades triggered by high-energy charged particles that produce secondary (and tertiary) particles that propagate towards the ground.}
    \label{fig:5.4.3}
\end{figure}

As charged particles enter the atmosphere they are prone to inelastic scattering processes while interacting with the atmospheric atoms and molecules. Thereby, the particles experience a characteristic energy loss (d$E$/d$x$) that is depending on the charge and velocity of the particle. This energy loss can be described by the Bethe-Bloch equation going back to the original works by \citet{1930AnP...397..325B} and \citet{1933AnP...408..285B}. Including the quantum-mechanical updates by \citet{Leo-1994} the characteristic energy loss can be described as

\begin{equation}
-\biggl\langle\frac{dE}{dx}\biggr\rangle = 2\pi N_A r_e^2 c^2\rho\frac{Z}{A}\frac{z^2}{\beta^2}\left[\ln\left(\frac{2 m_e \gamma^2 v^2 W_{max}}{I^2}\right)-2\beta^2-\delta-2\cdot\frac{C}{Z}\right],
\label{bethe2}
\end{equation}
with
\begin{tabular}{l l}
$2\pi N_A r_e^2 c^2$ = 0.1535 MeV cm$^2$ g$^{-1}$ & $r_e$=2.817$\cdot 10^{-13}$ cm, e$^-$ radius\\
$m_e$: the e$^-$ rest mass & $\rho$: density of the target material\\
$N_A$ = 6.022$\cdot 10^{23}$ mol$^{-1}$, Avogadro number & $\beta = \frac{v}{c}$\\
I: mean excitation potential & z: projectile charge\\
Z: atomic number of the absorber & A: atomic mass\\
$\gamma = \frac{1}{\sqrt{1-\beta^2}}$ & $\delta$ = density correction\\
\end{tabular}
\\
\\
and $W_{max}$ as the maximum energy transfer in a single collision given as
\begin{equation}
W_{max}=\frac{2 m_e c^2 \eta^2}{1+2 s \sqrt{1+\eta^2}+s^2},
\end{equation}

\noindent where $s=\frac{m_e}{M}$, $M$ the projectile mass and $\eta=\beta \gamma$. By solving Eq.~(\ref{bethe2}) numerically, the energy loss of an ion along its atmospheric path can be tracked, as exemplarily shown in the left panel of Fig.~\ref{fig:5.4.3} for primary protons of different energies between 1 MeV (purple) and 1 GeV (light green). While low-energy cosmic rays lose most of their energy due to elastic collisions with atmospheric neutrals before being stopped and absorbed at high atmospheric altitudes charged particles with energies above 1 GeV can reach the surface. Further, because of the $v^2$-dependence of the energy loss, the deposited energy is increasing with increasing particle energy and most of the primary energy is deposited close to stopping altitudes.  Note that the energy loss and the atmospheric ionization are directly linked. According to \citet{Kallenrode-2004}, in a first order approximation, the atmospheric ionization rate is given as the product of the energy loss and the flux of the incident primary particles. Thus, the atmospheric ionization due to GCRs and SEPs strongly depends on the collisional ionization rate, the energy of the ions, the interaction cross sections, and the atmospheric number density. The energy loss further strongly depends on the excitation potential $I$, and therewith on the atmospheric composition.

The ionization is further linked to the evolution of secondary particle cascades that are also triggered by high-energy cosmic rays undergoing inelastic scattering with the atmospheric nuclei. As sketch in the right panel of Fig.~\ref{fig:5.4.3}, in general, between the electromagnetic, the hadronic, and the muonic branch of a cascade can be distinguished. Thereby, the hadronic branch mainly consists of pions ($\pi^0$, $\pi^\pm$), kaons, protons, and neutrons. Although being the branch with the smallest number of produced secondary particles, this branch is the main core of the cascade and feeds both the electromagnetic as well as the muonic branch, because most of the neutral pions decay into a pair of photons before interacting with atmospheric nuclei. The latter further produce electrons and positrons through pair production, generating photons due to bremsstrahlung. Since both processes directly feed each other, an electromagnetic sub-cascade develops. Charged mesons, on the other hand, decay into muons, feeding the muonic branch. Thereby, the composition of the secondary particle cascade changes with altitude. While being proton-dominated at the top of the terrestrial atmosphere, a neutron-dominated composition occurs at sea level \citep[see, e.g.,][]{Masarik-Beer-1999}. Note that towards lower altitudes, the cascade becomes more complex until a maximum in the secondary particle flux is reached, after which the average secondary particle energy is too low to further drive the evolution of the cascade.

The induced particle cascades can lead to drastic changes in the atmospheric evolution, the atmospheric climate, as well as the atmospheric chemistry, and with that on atmospheric biosignatures \citep[e.g.,][]{grenfell2019exoplanetary, Scheucher_2020}, as well as the altitude-dependent atmospheric radiation dose \citep[e.g.,][]{Atri-2020, Herbst-etal-2019c} and habitability.

\subsubsection{Cosmic ray induced atmospheric ionization} 

Numerically speaking, the planetary atmospheric ionization $Q$ can be described as:
\begin{equation}
    Q(E_c,x) = \sum_i \int_{E_c}^{\infty} j_T(E)\cdot
     \underbrace{2\pi \int \cos(\theta)\sin(\theta)\,d\theta \cdot \frac{1}{E_{ion}} \frac{\mathrm{d} E_i}{\mathrm{d} x}}_{Y_i(E,x)}~dE,
\label{eq:Q}
\end{equation}
\begin{wrapfigure}{r}{0.53\textwidth}
\vspace{-0.5cm}
\includegraphics[width=0.52\textwidth]{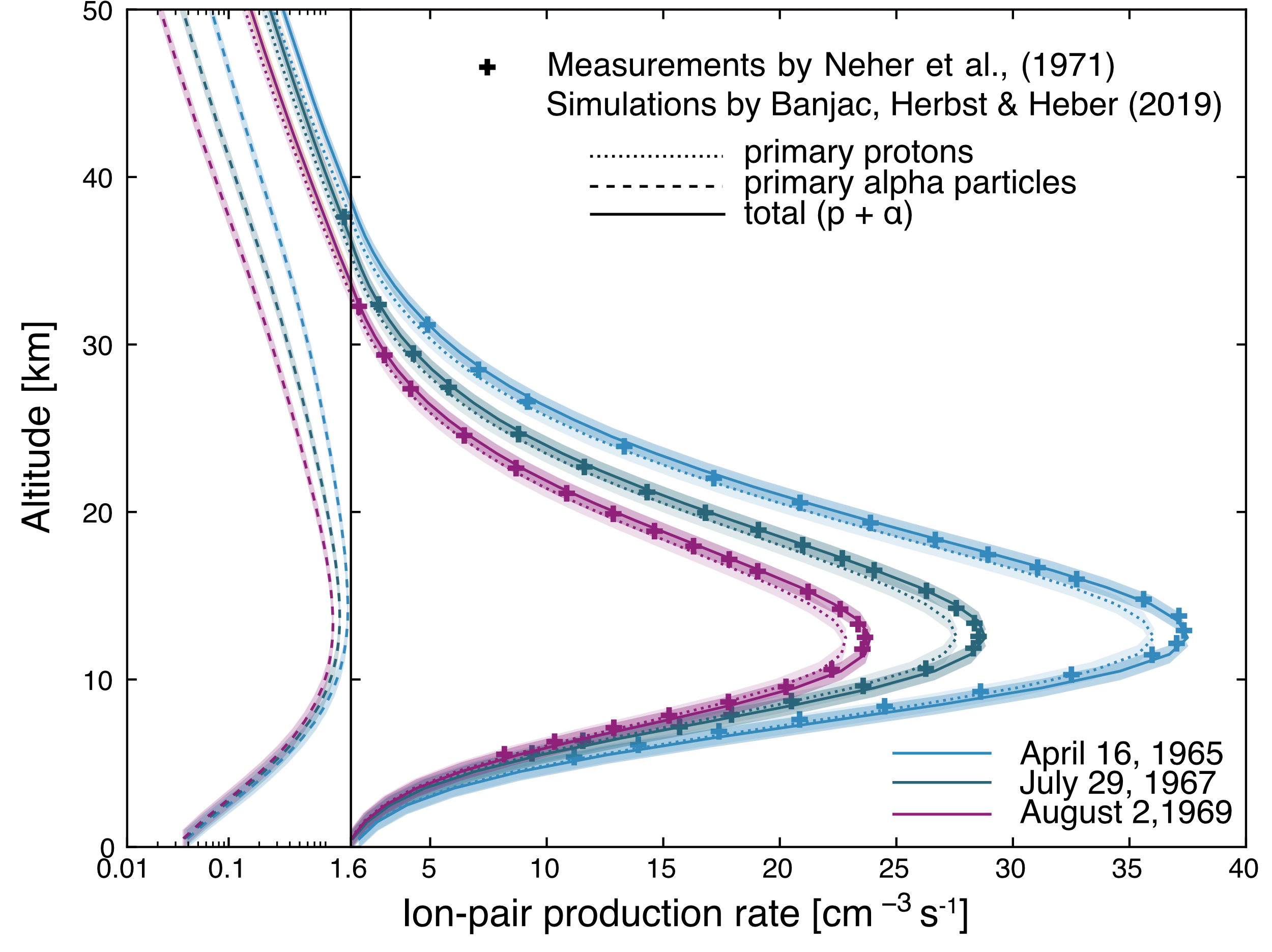}
\vspace{-0.2cm}
\caption{Comparison of observed terrestrial ion pair production rates \citep[at Thule,
Greenland, colored symbols, e.g.,][]{neher1971cosmic} with simulations based on primary protons (dotted lines) and alpha particles(dashed lines). Figure adapted from \citet{Banjac-etal-2019a}.}\label{fig:5.4.5}
\vspace{-0.8cm}
\end{wrapfigure}
where $i$ refers to the primary particle type, $x$ the atmospheric depth/altitude, and $E_c$ is the cutoff energy of the primary particle. The latter is the energy needed to reach a specific locations within a planetary magnetic field correlated to the cutoff rigidity via $E_c=\sqrt{(R_c \cdot q)^2c^2+m^2c^4}-mc^2$, with $q$ the charge of the particle and $m$ its mass. $Y_i(E,x)$ refers to the atmospheric ionization yield, with $E_{ion}$ as the average atmospheric ionisation energy necessary to produce an ion-electron pair. 

Figure~\ref{fig:5.4.5} shows observations \citep[symbols,][]{neher1971cosmic} in comparison to the modeling efforts by \citet{Banjac-etal-2019a}. Both agree within $\pm$ 5$\%$. The ion-pair production rates thereby strongly depend on the atmospheric density as well as the mean atmospheric ionization potential. While Earth with an N$_2$-O$_2$ dominated atmosphere, for example, has an average atmospheric ionisation energy of 31.7$\pm$1.7 eV, Venus and Mars with CO$_2$ dominated atmospheres have an $E_i$ of 28.7$\pm$4.3 eV and 28.4 $\pm$ 4.3 eV, respectively \citep[][]{Wedlund-etal-2011}. 

\begin{figure}[!t]
    \centering
    \includegraphics[width=\textwidth]{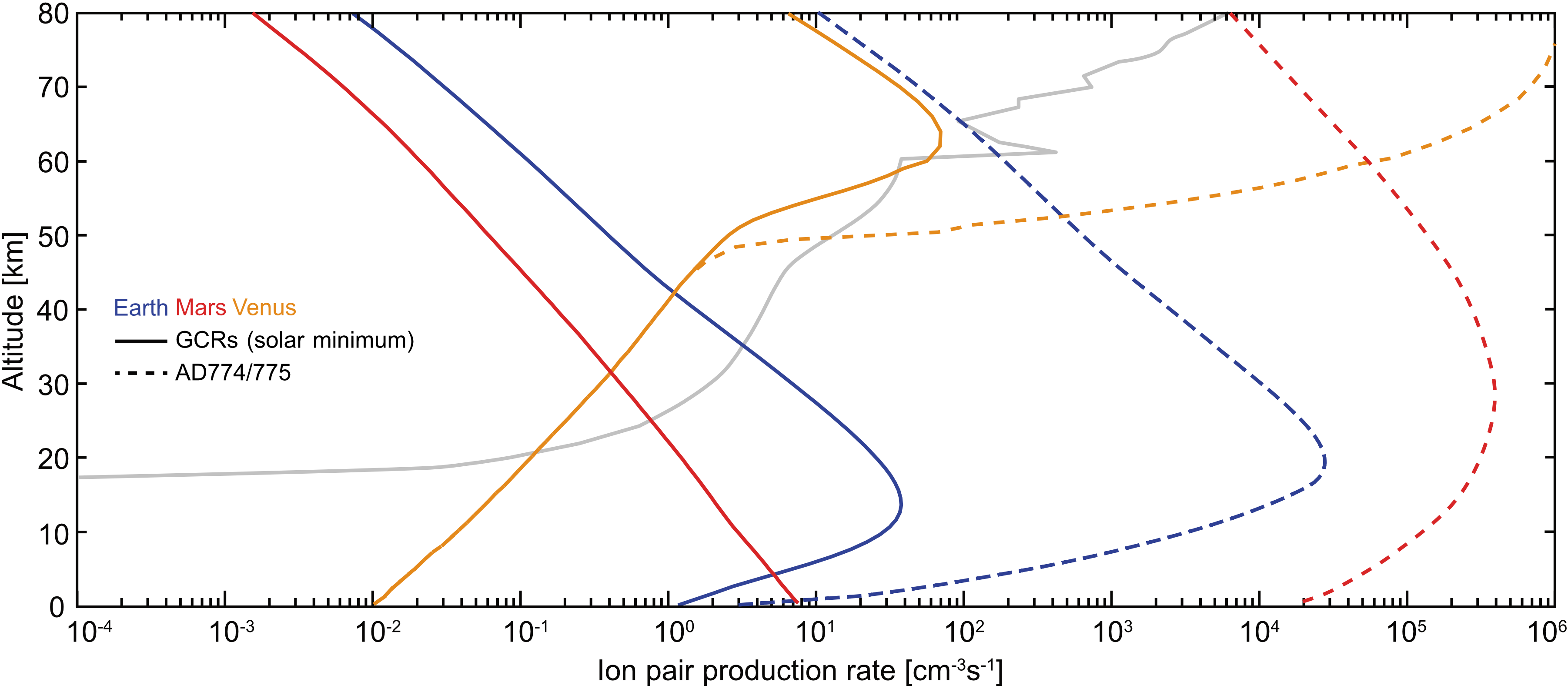}
    \caption{Cosmic ray induced atmospheric ionization profile of Earth \citep[blue; from][]{Herbst-etal-2019c}, Venus \citep[orange; from][]{Herbst-etal-2019b}, and Mars (red) caused by GCRs during solar minimum conditions (solid lines) and SEPs during the AD774/775 event (dashed lines). For direct comparison the radiation belt contribution by \citet{Frahm-1997} is shown (gray line).}
    \label{fig:5.4.6}
\end{figure}

Figure~\ref{fig:5.4.6} shows the altitude-dependent atmospheric ionization profiles of Earth (in blue), Venus (in orange), and Mars (in red). Already during solar minimum conditions (solid lines) strong differences become apparent. While being located at altitudes around 16 - 25 km in the case of the Earth, the  Regener-Pfotzer maximum \citep{Regner1935} is located at around 65 - 70 km within the thick CO$_2$-dominated Venusian atmosphere. The Martian response, however, does not show such a maximum. The highest ionization values can be found at the surface as most GCRs and SEPs can reach these low altitudes. Further shown are the ionization profiles during one historic superflare events detected in the cosmogenic radionuclides (i.e., AD774/775, dashed lines). Such strong events increase the ionization values by several orders of magnitude. While the ionization rates of Earth and Venus are not or only marginally increased, the Martian surface radiation is increased by more than three orders of magnitude. The latter has severe consequences for possible crewed mission to Mars \citep[e.g.,][]{2019JE006246, 2022arXiv220800892A, 2025SW004724}.

Studies on the atmospheric ionization impact of cosmic rays on Earth-like exoplanets were put forth by \citet{Herbst-etal-2019c, Herbst-etal-2024}, \citet{Engelbrecht24}, and \citet{2024arXiv240907274R}. \citet{Engelbrecht24} studied the time-dependent atmospheric ionization of Prox Cen b, showing that, contrary to previous conclusions, the GCR contribution is much stronger than in the case of the Earth and that stellar rotation periods provide a novel constrain on exoplanetary habitability also been reported by \citep[see also][]{LightEA25}. The latter was confirmed by \citet{RaesideEA25} who investigated the impact of cosmic rays on the early (post-impact H$_2$ dominated) Earth. They found that, similar to current conditions on Earth, atmospheric ionization caused by SEPs likely had a much stronger effect than GCRs. In addition, they reported that if the Sun had evolved as a fast rotator (i.e., $\Omega \geq 15~\Omega_\odot$), it is highly probable that SEPs significantly influenced the planetary surface during the period when life was evolving on Earth. Most recently, \citet{EH26} studied the impact of stellar magnetic activity cycles on the atmospheric ionization of Prox Cen b. They reported a non-negligable impact on the atmospheric ionization of up to 28\%.

\subsubsection{Cosmic ray induced (ion)chemistry changes} \label{particles-chemistry}
To quantify the influence of cosmic rays on the atmospheric ionization as well as the atmospheric radiation field it is further essential to know the flux of the secondary particles produced during the evolution of the electromagnetic and hadronic branches. Of special interest are the evolving secondary alpha, neutron, proton, muon, electron/positron, and gamma fluxes as they trigger chemistry and climate changes as well as radiation exposure increases within planetary atmospheres. 

The impact of SEPs on atmospheric chemistry and climate has been studied since the late 1960s. For example, \citet{Jones1973} reported that cosmic rays primarily interact with He or H within the thermosphere (90 - 500 km) while primarily interacting with N$_2$, O$_2$, and O in the mesosphere (80-90 km). Below the mesosphere, mainly N$_2$ and O$_2$ are ionized \citep{Porter1976,Rusch1981}. The interaction reactions, thereby, can be both dissociative and ionizing. Particularly in the case of atomic nitrogen, these reactions can further generate an excited electronic state (e.g., N($^2$D,$^3$P)) or ground state (e.g., N($^4$S)) \citep{Jones1973,Porter1976}, in which free electrons are being released during the ionization processes. If their energy is sufficiently high, these secondary electrons can contribute to the ionization. Thus, the average kinetic energy of the background then determines the end of the processes \citep{Sinnhuber2012}. 

The formation rate of a primary ion, $P(c^+)$, is directly linked to the ion pair production $Q$ given in Eq.~(\ref{eq:Q}) and can be described as
\begin{equation}
    P(c^+) = \frac{\sigma_c^+\cdot [C]\cdot Q}{Q_{tot}},
    \label{eq:5_4_2}
\end{equation}
where [C] represents the abundance of a target specie, $\sigma_c^+$ is the cross-section of the formation of an ion c$^+$\footnote{Such cross-sections can, for example, be found in \citep{RevModPhys.38.1}.}, and Q$_{tot}$ is the total produced charge \citep{Sinnhuber2012}. The latter is given by
\begin{equation}
    Q_{tot} = \sum_i \sigma_{tot,i} \cdot [C_i].  
    \label{eq:5_4_3}
\end{equation}
In a N$_2$-O$_2$-dominated atmosphere, this leads to the production of excited atomic states such as N($^2$D) and O($^1$D), as well as ions such as N$_2^+$, O$_2^+$, and O$^+$. In CO$_2$-dominated atmospheres, however, the most likely product will be ions and excited-state dissociation products of CO$_2$.

According to \citet{Herbst-etal-2019c}, atmospheric ionization leads to rapid chains of ion-chemistry reactions, which can subsequently trigger the formation of neutral reactive radicals, such as NO, H, and OH. These radicals play a role in catalytic reaction chains that deplete atmospheric ozone, as illustrated by, for example, the following reactions:
\begin{eqnarray}
\mathrm{NO} + \mathrm{O}_3 &\longrightarrow& \mathrm{NO}_2 + \mathrm{O}_2 \label{ionchem1}\\
\mathrm{NO}_2 + \mathrm{O}(^3\mathrm{P}) &\longrightarrow& \mathrm{NO} + \mathrm{O}_2.\label{ionchem2}
\end{eqnarray}
Incorporating molecules such as HNO$_3$, N$_2$O$_2$, or HCl can result in the formation of large ion clusters. These clusters change the distribution of nitrogen- and chlorine-containing species, which in turn affects atmospheric ozone levels. Thereby, ozone is a critical component for radiative heating in the stratosphere. Thus, a link between ozone heating and  temperatures is apparent within the stratosphere. Additionally, the Brewer-Dobson meridional circulation significantly impacts O$_3$ loss \citep{Sinnhuber-etal-2003, Jackman-etal-2009}. Since ozone is instrumental in stratospheric radiative heating, this process directly affects atmospheric temperatures and circulation patterns. In methane-rich atmospheres, like those found on Jupiter and Titan, ion chemistry can produce hydrocarbon haze from methane \citep[see, e.g.,][]{Vuitton-etal-2007, Lavvas-etal-2013, Hoerst-2017, Garcia-Munoz-2018}. Nevertheless, methane-rich and $N_2$-dominated atmospheres act as natural laboratories for examining chemical evolution \citep[see, e.g.,][]{Kobayashi-etal-2017} and its relevance to prebiotic chemistry \citep[see, e.g.,][]{Rimmer-Rugheimer-2019}. 

It has been observed that GLEs significantly affect the chemical composition of the Earth's atmosphere at altitudes above 20 km. This is especially true for nitrogen species and ozone, both of which have been continuously monitored since the 1970s \citep[e.g.,][]{Crutzen-etal-1975, Solomon-etal-1983, Jackman-etal-2005a, Jackman-etal-2005b, Rohen-etal-2005, Funke-etal-2011}. Model studies employing chemistry-transport models (CTMs) or chemistry-climate models (CCMs) have been compared against observations from instruments such as \textit{MIPAS} on \textit{ENVISAT} in several targeted assessments \citep[e.g.][]{Funke-etal-2011, Funke-etal-2017}, where a generally good agreement has been observed regarding NO$_x$ production and ozone loss during strong particle events. However, more recent studies have highlighted a strong influence of complex ion-chemistry reactions during larger atmospheric ionization events \citep{Sinnhuber2012,Verronen-Lehmann-2013}.

\begin{figure}[!t]
    \centering
    \includegraphics[width=\textwidth]{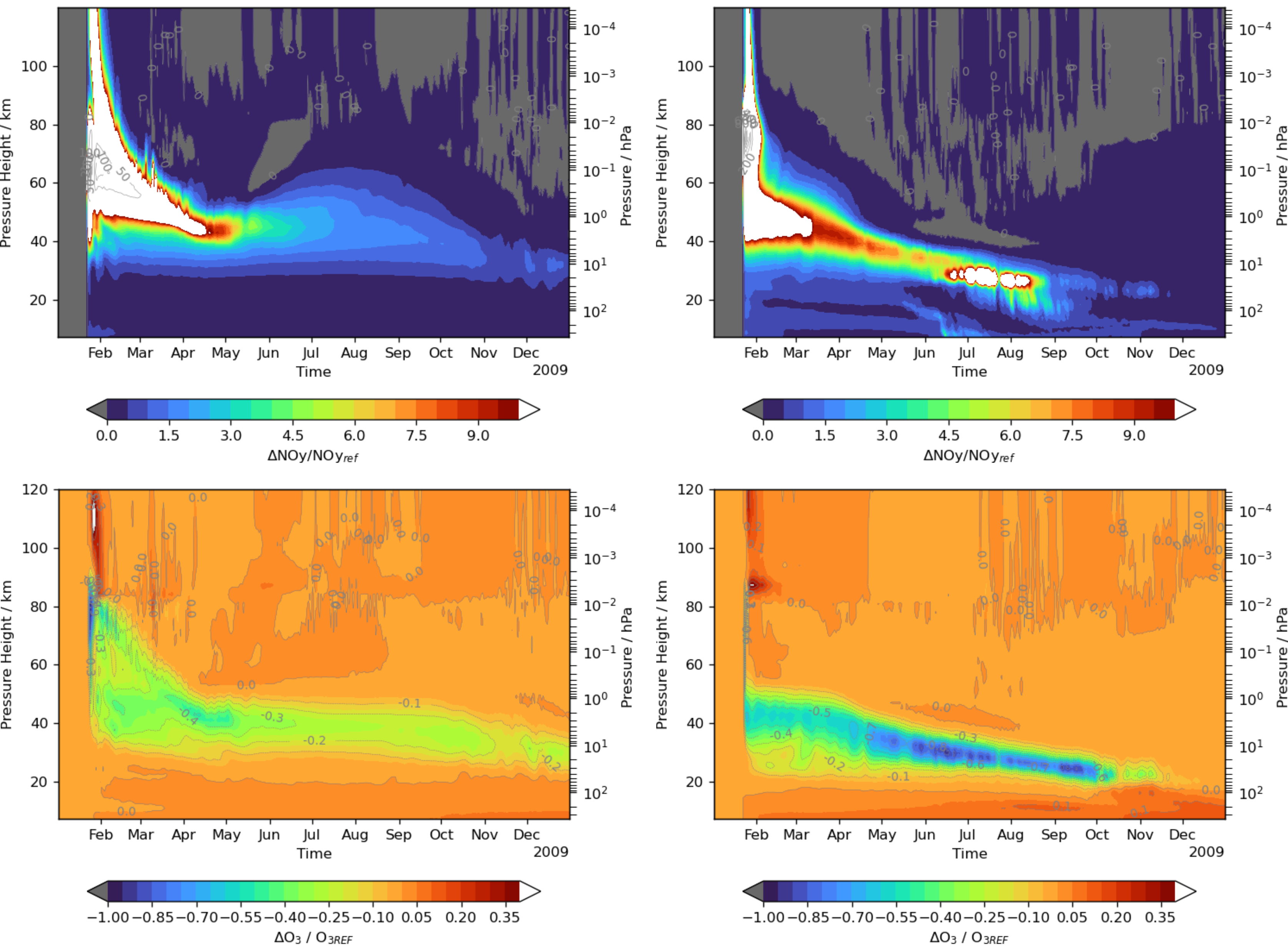}
    \caption{Upper panels: Relative NO$_y$ during the AD774/775 event changes relative to quiescence conditions. Lower panels: Corresponding Relative ozone change related to the NOy intrusion at the northern (left) and southern (right) polar caps at latitudes $>$70$^\circ$. Figure taken from \citet{acp-23-6989-2023}, used under CC BY 4.0.}
    \label{fig:5.4.7}
\end{figure}

Recently, also the impact of GLEs on Earth's middle-atmospheric chemistry has been investigated. For example, \citet{acp-23-6989-2023} investigated the impact of the AD774/775 event compared to solar quiescence conditions. Figure~\ref{fig:5.4.7} shows the relative NO$_y$ (upper panels) and O$_3$ (lower panel enhancements in the northern (left) and southern (right) polar cap regions (i.e., for latitudes $< \pm 70^\circ$). They found that an event like the AD774/775 event likely has a significant impact on the N$_2$-O$_2$ dominated terrestrial atmosphere and reported on the additional production of NO$_y$ and the induced ozone loss in both regions. Both are directly correlated with the deep penetration of the high-energy solar energetic particles. Note that both phenomena persist for at least ten months and scale with the strength of the event.

The role of cosmic rays in shaping the exoplanetary chemistry and climate has gained considerable attention in recent years. By now, several 1D and 3D studies can be found in the literature. For example, the atmospheric chemistry changes due to GCRs in Earth-like exoplanets have been first studied by \citet{2007AsBio...7..208G}.  \citet{2010AsBio..10..751S} and \citet{Tilley2019-multiflare} used X-ray–based proxies to derive the StEP fluxes in a global-mean coupled climate–chemistry column model of Earth-like N$_2$–O$_2$ dominated atmospheres around M-dwarfs, reporting that strong StEP-driven NOx production can catalytically destroy O$_3$ during intense flaring. \citet{2012AsBio..12.1109G} further showed that extreme SEPs around active M-dwarfs can remove atmospheric O$_3$. Extending this work, \citet{Tabataba-Vakili-2016} added an energetic-particle–induced HOx source, parameterizing O$_2$ breakup by StEPs and subsequent ion-cluster chemistry involving H$_2$O, and demonstrated that HOx provides an additional sink for biosignature gases such as O$_3$. \citet{Scheucher_2020} studied the impact of cosmic rays on chemistry and climate of an Earth-like Prox Cen b and found that large amounts of NO$_2$ can be produced by StEP interactions and indicated that HNO$_3$ not only is a signature for the impact of cosmic rays but also and indirect measure for the presence of water in N$_2$-O$_2$ dominated atmospheres. In a TRAPPIST-1 e study investigating the StEP impact on dry-dead, wet-dead, and wet-alive atmospheres of 0.1 and 1 bar of CO$_2$, \citet{Herbst_2024} found high event-induced amounts of NO$_2$, a reduction of the atmospheric transit depth in all water bands, a decrease in methane in the 3.0 - 3.3 $\mu$m band, and a depletion of ozone at 9.0 - 10.0 $\mu$m were found in all scenarios. They further reported that the HNO$_3$ feature was not present in both wet-dead scenarios. 

Utilizing a 3D general circulation model (GCM), \citet{2021NatAs...5..298C} determined that strong energetic particle events can result in increased concentrations of greenhouse gases, including water vapor and nitrogen oxides. Recently, \citet{Chen25} studied the effects of stellar particle events on the chemistry and climates of tidally locked exo-Earths and suggested that the O$_3$ variability has a strong impact on atmospheric temperatures. Most recently, \citet{Kobayashi_etal_2026} showed that StEP events are a potent abiotic driver of both greenhouse warming and prebiotic chemistry in primitive planetary atmospheres, and that on young planets with N$_2$–CO$_2$ dominated atmospheres, frequent events may be essential for the emergence and maintenance of surface habitability.

\subsubsection{Cosmic ray induced atmospheric radiation exposure} 
Studying the impact of cosmic rays on the terrestrial radiation environment in space, low-Earth orbit, flight altitudes, and on the ground started in the early 1970s, after visual phenomena insight the Apollo 11 spacecraft were reported by Buzz Aldrin \citep[Sec. 6.40 of][]{Debriefing1969}. 

The ionization of atoms and the energy they release can damage biological tissue, leading to breaks in DNA and DNA mutations, as well as cell malfunctions or cell death. The primary responsibilities of the \textit{International Commission on Radiological Protection} (ICRP) include quantifying radiation effects, assessing the severity of the resulting damage, and proposing acceptable exposure limits. The \textit{International Commission on Radiation Units and Measurements} (ICRU) was established in 1925 to develop a system of units and guidelines for measurement techniques and procedures essential for implementing the recommendations of the ICRP. 

Various quantities were established to promote a uniform framework for radiation protection, ensuring that measurements are standardized and comparable across diverse disciplines. Thefollowing definitions are based on \citetalias{protection1991icrp}, \citetalias{valentin2007icrp}, and \citetalias{dietze2013icrp}. Often used in dosimetry is the absorbed dose $D$, the mean energy d$\bar{\epsilon}$ deposited in a volume of mass d$m$ given as
\begin{equation}
    D = \frac{\mathrm{d}\bar{\epsilon}}{\mathrm{d}m}
\end{equation}
given in units of G (i.e., J/kg). 
As shown, $D$ is independent of the radiation type, its energy and the linear energy transfer (LET) and, thus, solely depends on the deposited energy. As such, the biological effect, for example, on cells can not be investigated. Therefore, a second quantity, the equivalent dose $H_T$ has been introduced that is given as
\begin{equation}
    H_T = \sum_R w_R \cdot D(w, R),
\end{equation}
the product of absorbed doses $D$ in tissues/organs $T$ and their corresponding radiation weighting factors $w_R$ summed up over all radiation types $R$ (in units of Sv). Here, $w_R$ describes the radiation-type dependent relative biological effectiveness. The latest release of weighting factors published in \citetalias{valentin2007icrp} are listed in Tab.~\ref{tab:4_5_1}. As can be seen, $w_R \geq 1$ for all particles. The radiation
weighting factor for neutrons is given by:
\begin{table}[h]
\caption{Radiation
weighting factors $w_{R,j}$ after \citetalias{valentin2007icrp}.}
\label{tab:4_5_1}
\begin{tabular}{l l}
\hline
\hline
Radiation type $j$ & $w_{R,j}$ \\
\hline
photons, electrons, muons & 1 \\
protons, charged pions & 2\\
$\alpha$, fission fragments, heavy ions & 20\\
neutrons & see Eq.~(\ref{eq:rwf})\\
\hline
\end{tabular}
\end{table}
\begin{equation}
    w_{R,n}=\begin{cases}
         2.5 + 18.2\cdot e^{\frac{-\left[ \ln\left(E\right)\right]^2}{6} },&\text{for } E < 1\,\text{MeV}  \\ 
         5.0 + 17.0\cdot e^{\frac{-\left[ \ln\left(2\cdot E\right)\right]^2}{6} },&\text{for } E\in\left[1\,\text{MeV}-50\,\text{MeV}\right]  \\ 
         2.5 + 3.25\cdot e^{\frac{-\left[ \ln\left(0.04\cdot E\right)\right]^2}{6} },&\text{for } E> 50\,\text{MeV} 
    \end{cases}
    \label{eq:rwf}
\end{equation}
with $E$ giving the kinetic energy of the neutron. Particles not listed in the table (i.e., photons, electrons, and muons) have an effective weighting factor of 1. 

\begin{figure}[t]
    \centering
\includegraphics[width=\textwidth]{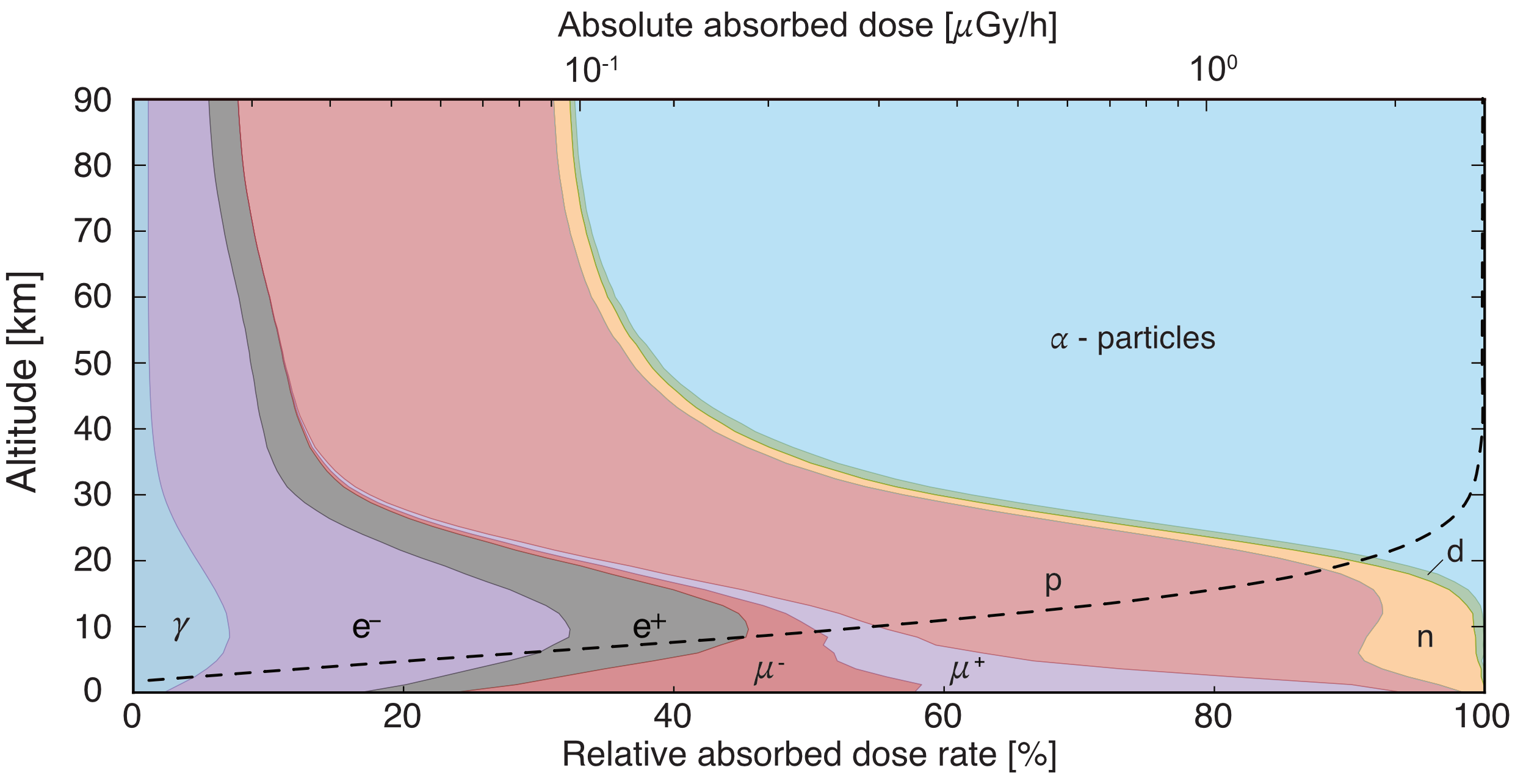}
    \caption{Altitude-dependent relative contribution of secondary particles to the absorbed dose (different colors). In addition, the the corresponding altitude-dependent  absolute absorbed dose rate profile is given by the dashed line. Figure after \citet{2019JA026622}.}
    \label{fig:5.4.9}
\end{figure}

The first numerical studies on the cosmic ray induced radiation exposure have been published by \citet{Sandmeier1981}, \citet{NYMMIK19981689}, and \citet{1208571}. Today several models to derive the cosmic ray induced radiation dose exist in the literature \citep[PHITS, e.g., ][]{Sato02012024}, \citep[CRAC:DOMO,][]{2015JSWSC...5A..10M}, \citep[AtRIS,][]{Banjac-etal-2019a}. Utilizing AtRIS, \citet{2019JA026622} investigated the relative contribution of different secondary particles to the net absorbed dose altitude profile shown in Fig.~\ref{fig:5.4.9}. As can be seen,  primary alpha particles are the dominant cause of absorbed dose at altitudes above 35~km while below (i.e., down to 10~km) protons are the strongest contributors. At 10 km a strong electron and positron contribution can be seen, while the neutron contribution is highest around 7~km. At the surface, the strongest contributors to the relative absorbed dose rate are muons. The corresponding atmospheric absolute absorbed dose profile is given by the dashed black line. Besides the terrestrial observation and model efforts, several lunar, Martian, and Venusian efforts have been put forth over the past decades discussing the solar system body radiation exposure and its differences. For example, \citet{2019JE006246} showed that, in agreement with the ionization, the radiation maximum is in the Martian regolith. In addition, the presence of hydrogen in the Martian regolith significantly influences the attenuation of the equivalent dose by modulating neutrons below an energy of 10 MeV. This efficient protection by subterranean water is also evident at the surface, offering indirect safeguarding for prospective human explorations in this area. Note that a total dose of 4 to 5 Sv has been defined as the lethal limit for 50 percent of an exposed population within 30 days\footnote{\href{https://www.nrc.gov/reading-rm/basic-ref/glossary/lethal-dose-ld}{https://www.nrc.gov/reading-rm/basic-ref/glossary/lethal-dose-ld}}.

The surface radiation dose has also been discussed in the exoplanetary context. \citet{ast.2013.1052} investigated the impact of galactic cosmic rays (GCRs) on the surface radiation dose of Earth-like planets, considering fluctuations in planetary magnetic field intensities and atmospheric densities. Their findings revealed that atmospheric thickness is the most important parameter determining surface radiation dose. Furthermore, \citet{atri2017} showed that stellar superflares are improbable to sterilize the surfaces of close-in exoplanets, however, related stellar particle events might result in recurrent radiation exposure at extinction levels, potentially influencing the evolution of prospective ecosystems (assuming life as we know it from Earth). \citet{2019ApJ...881..114Y} came to the same conclusion, further highlighting that a thick atmosphere is crucial for protecting exoplanets against lethal radiation induced by stellar superflares. Thereby, periodic strong radiation doses could make the surface uninhabitable—even on worlds in the classic habitable zone. \citet{2020MNRAS.492L..28A} conclude that for close-in exoplanets, especially those orbiting active M-dwarfs, with thin or eroded atmospheres, surface radiation levels from StEPs can be biologically catastrophic, and that only subsurface or highly radiation-resistant life could persist under such conditions. Recently, \citet{Herbst_2024} studied the atmospheric response of close-in CO$_2$-rich (1bar CO$_2$) and -poor (0.1 bar CO$_2$) atmospheres to a Carrington-like event. In the case of CO$_2$-rich atmospheres, absorbed surface dose rates remain low while in the CO$_2$-poor scenarios, particularly in the wet-alive CO$_2$-poor atmosphere could cause severe chromosomal mutations in terrestrial life forms.

\subsection{Biosignatures}\label{Biosig}
A biosignature is ``any substance, group of substances, or phenomenon that provides evidence of life'' \citep{catling2018exoplanet}. Some use biomarkers as a synonym of biosignatures \citep[e.g.][]{grenfell2011sensitivity}, but here \textit{biomarkers} are signatures of life that can be tested \textit{in situ} while biosignatures are remotely detected. Biosignatures must be identified considering the planetary context, ideally, they should be reliable, be able to survive in a given environment, and be detectable with a specific instrument \citep{meadows2018-O2-biosignature}. Here we focus on gaseous biosignatures that could be potentially detected in planetary atmospheres with telescopes. A comprehensive review of biosignatures currently discussed within the exoplanetary community is available in \citet{Schwietermann-etal-2018}. The study of biosignatures informs the requirements of future instruments to detect life on other planets and provides tools for identifying biosignatures in planetary spectra. Currently, several missions are planned for  the spectral characterization of potentially habitable planets \citep{Fujii-etal-2018,grenfell2025detectability,parenteau2026habitableworldsobservatoryliving}.

Among the various molecules identified as potential atmospheric biosignatures, three are frequently highlighted in the literature: nitrous oxide (N$_2$O), methane (CH$_4$), oxygen (O$_2$) and its photochemical byproduct, ozone (O$_3$).

\textit{Nitrous oxide}: is almost exclusively formed by nitrifying and denitrifying bacteria in Earth's atmosphere that are part of Earth's nitrogen cycle \citep{Syakila-Kroeze-2011}. Abiotic sources of N$_2$O are reactions initiated by high-energy particles \citep{Aea16ng} and the reduction of NO by ferrous iron (chemodenitrification) \citep{stanton2018-n2o}, but there are discriminants that allow to recognize false positives \citep{schwieterman2022-n2o}.

\textit{Oxygen and ozone}: Oxygen results of oxygenic photosynthesis where CO$_2$ is reduced by H$_2$O. Earth’s atmosphere oxygenation started around 2.5 Gyr during the Proterozoic. Ozone is an indirect biosignature of life, which is produced mainly through the Chapman mechanism \citep{Chapman-1930} via O$_2$ photolysis (in the stratosphere) and the smog mechanism \citep[in the troposphere, see, e.g.,][]{Haagen-Smit-1952}. The latter, among others, nitrogen oxides and UV radiation. Although being an indirect biosignature of life and having a strong spectral absorption, O$_3$ sinks can be rather complex and include, for example, catalytic cycles \citep{Bates-Nicolet-1950, Crutzen-1970}, which can involve various chemical reactions that deplete ozone in the atmosphere. This couple are considered one of the most robust biosignatures \citep{meadows2018-O2-biosignature,meadows2017reflections}.

\textit{Methane}: has significant biological sources, particularly methanogenic bacteria, as well as abiotic sources from outgassing.  However, the abiotic sources are approximately one order of magnitude less significant than the biological sources \citep{thompson2022-CH4}. Within the terrestrial atmosphere, methane is a crucial greenhouse gas that is primarily removed from the atmosphere through oxidative degradation by the reactive hydroxyl (OH) radical. \citet{Wahlen-1993} provides an overview of the global atmospheric budget inventory of methane.

\subsubsection{UV impact on atmospheric biosignatures}

\begin{figure}[ht]
\centering
\includegraphics[width=0.6\textwidth]{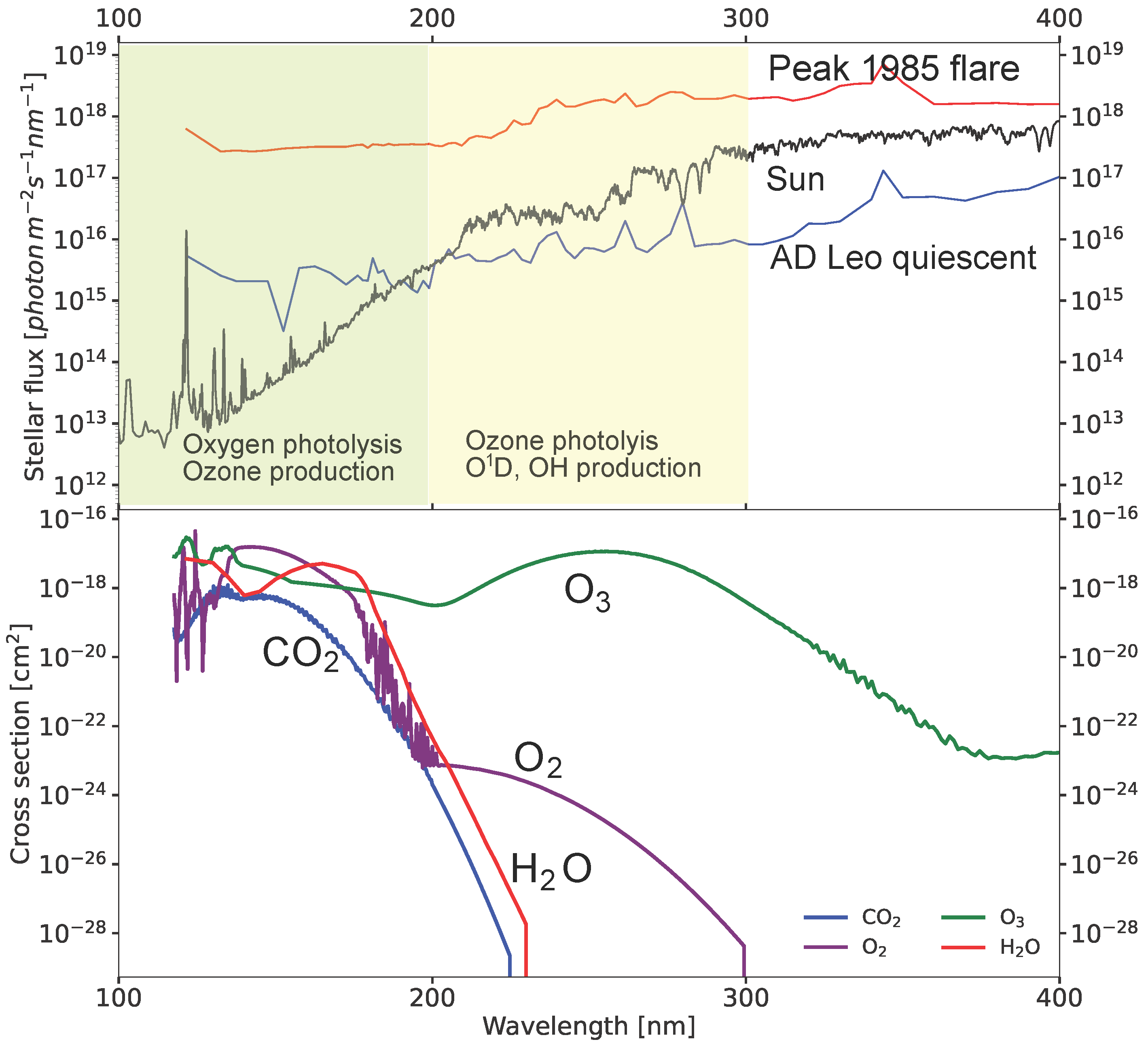}\hfill
\includegraphics[width=0.4\textwidth]{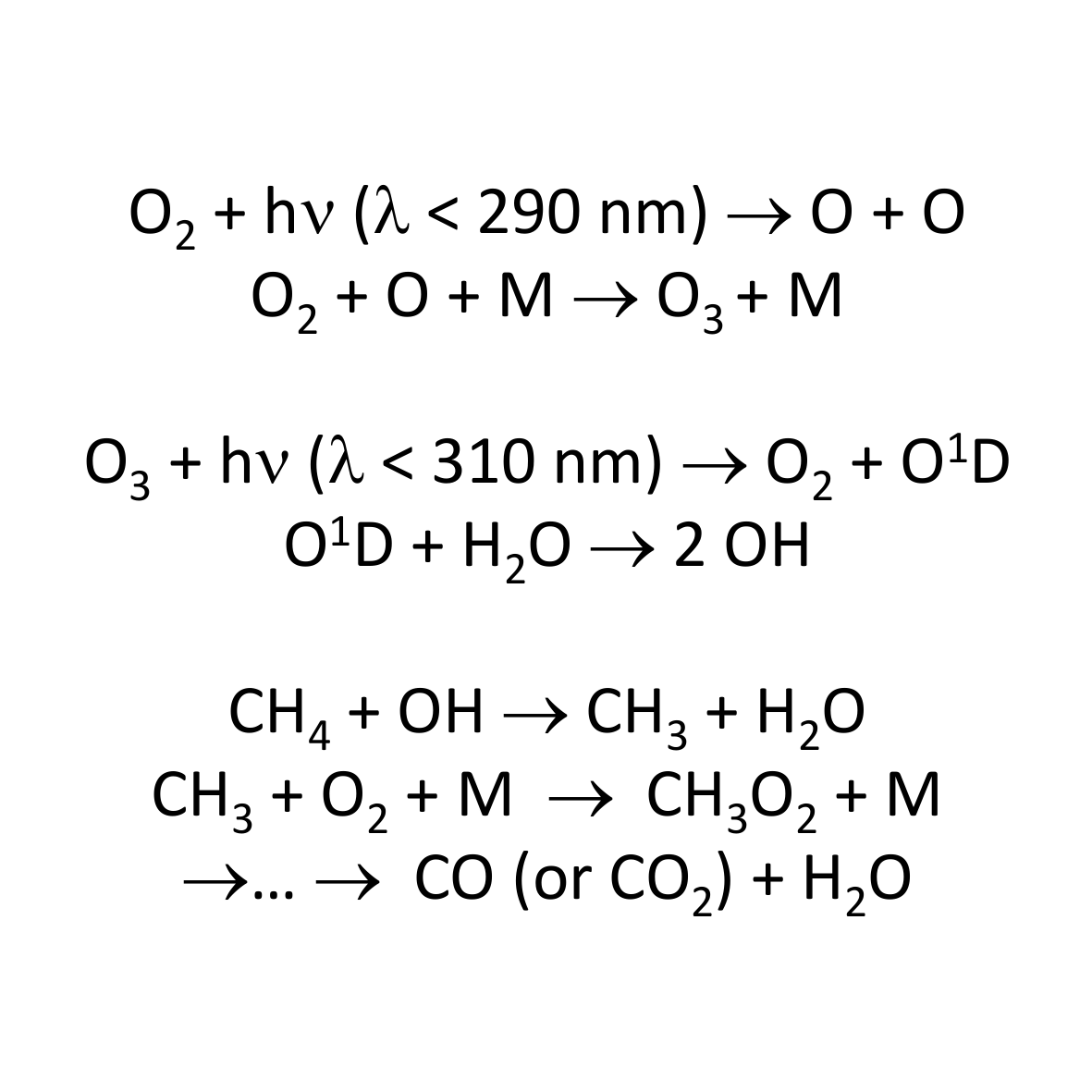} 
\caption{\textit{Upper left: }Flux from the Sun (black line) and AD Leo (red and blue lines) that arrives to Earth and an hypothetical planet in the HZ of ADLeo, respectively. Blue line is the quiescent spectrum of AD Leo and the red line the spectrum at the peak of the 1985 AD Leo Flare. Green shade is the wavelength region most relevant for O$_2$ photolysis which promotes the production of O$_3$. The wavelength range relevant for O$_3$ photolysis is marked with a yellow shade. \textit{Lower left: } cross sections of relevant atmospheric molecules (After \citealt{miranda2025-CO2flare}). 
\textit{Right: } Reactions relevant for methane destruction, M stands for any molecule in the atmosphere.}  
\label{fig:ch4-chemistry}
\end{figure}

Once a gaseous chemical compound is produced by life, its lifetime in the atmosphere depends on photochemistry and physical processes such as atmospheric escape or rainout. 
\citet{segura2003ozone,segura2005biosignatures} studied the effect of stellar UV on biosignatures from planets with atmospheres bulk compositions similar to that of present Earth witha mixing ratio of 0.21 O$_2$. Potential biosignatures where included using the estimated fluxes required on Earth to maintain the observed atmospheric abundances, for example, N$_2$O was included with a flux of $1.32 \times 10^{13}$ g of N$_2$O/year instead of a mixing ratio of $3 \times 10^{-7}$. They found that planets around hotter stars than the Sun, would produce more ozone because of their higher UV fluxes \citep{segura2003ozone}. For planets around highly active M dwarfs, CH$_4$ would have longer lifetimes in the atmosphere because of the lower production of its main sink, OH \citep{segura2005biosignatures}. The main sources of atmospheric OH are water photolysis and the reaction of O($^1$D) and water. As illustrated in Fig. \ref{fig:ch4-chemistry}, ozone photolysis produces O($^1$D), but for planets in the HZ of M dwarfs O$_3$ photolysis is inhibited which results in less O($^1$D) and consequently, less OH and a longer lifetime of CH$_4$ for these planets \citep{segura2005biosignatures}. N$_2$O is removed mainly via photolysis and reactions with electronically-excited oxygen atoms (O$^{*}$) in the Earth's atmosphere. But for planets in the HZ of K-type and early non-active M dwarfs, the low stellar UV emission results in longer lifetimes for N$_2$O, higher abundances and promising spectral features in the mid-IR (7.8 and 8.5  $\mu$m) that may be detectable by future instruments \citep{segura2005biosignatures,schwieterman2022-n2o}. 

Stellar flares emit in all wavelengths from radio to X-rays and can accelerate particles \citep{kowalski2024-flares}. For atmospheric chemistry the most relevant flare emissions are UV and particles. The first approach to study the effect of flares on compounds relevant for biosignatures was performed by \citet{segura2010effect} using the 1985 AD Leonis (AD Leo) UV flare reported by \citet{hawley1991-ADLeoFlare}. Twenty spectra for the time evolution of the flare were built from the International Ultraviolet Explorer (IUE) satellite observations in two wavelength ranges, 115-200 nm and 190-310 nm. There was no simultaneous  observations on both windows, thus, the spectrum for each flare time required extrapolation either at the short or the long wavelength range \citep{segura2010effect}. The 1D simulation considered an hypothetical planet around AD Leo with an atmosphere with O$_2$ and N$_2$ abundances equal to those of present Earth. The UV flare produced a minimal (1\%) depletion in the O$_3$ column density, when a series of flares are simulated, the maximum depletion found was 86\% of the ozone column density \citep{Tilley2019-multiflare}. \citet{ridgway2023-3dflare} simulated planets under the effect of flares using a 3D Global Circulation Model (GCM) finding that UV flares increase the ozone abundance. This as a result of ozone transport from the night-side to the day-side of the planet.

Archean-like atmospheres (CO$_2$ dominated) produce O$_2$ and O$_3$ from the photolysis of CO$_2$, which is a potential biosignature false positive. A 1D model predicts the destruction of both, oxygen and ozone, under the effect of the 1885 AD Leo flare, as long as the atmosphere contains sources of HOx there is no accumulation of O$_2$ or O$_3$ \citep{miranda2025-CO2flare}. For both, there is a small increase after the flare has ended that may be attenuated or potentiated during a series of flares.

\subsubsection{Cosmic ray impact on spectral atmospheric biosignature features}

\begin{figure}[t]
    \centering
    \includegraphics[width=\textwidth]{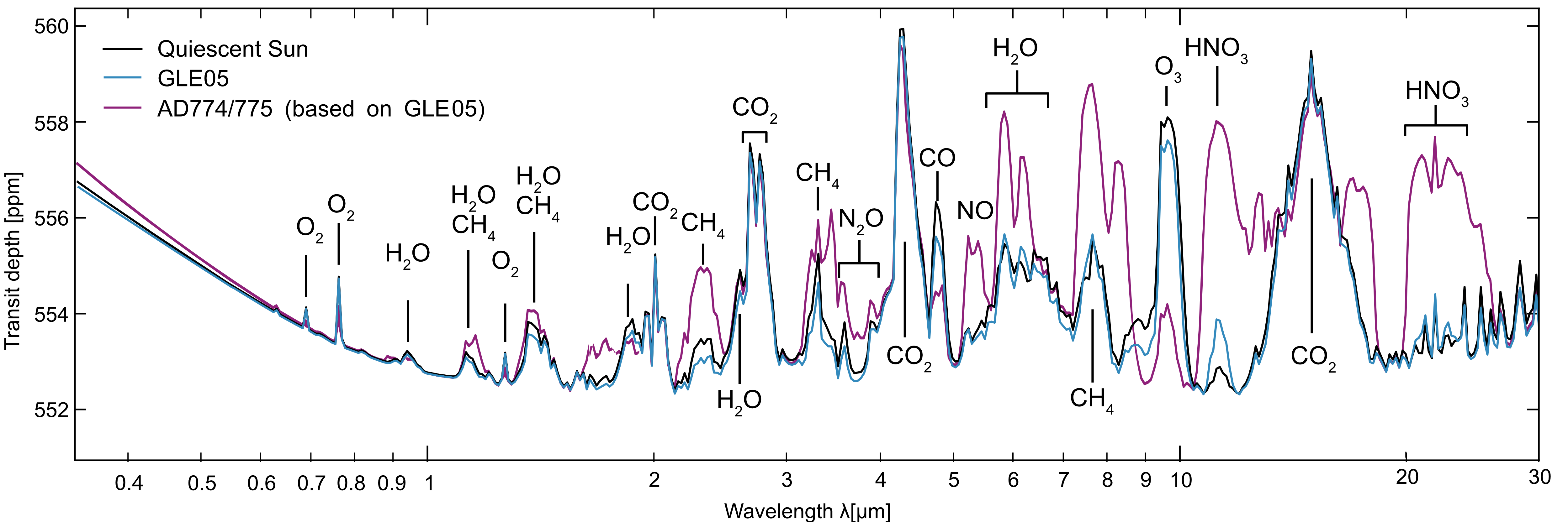}
    \caption{Modeled transit depths in parts per million over wavelength (R = 100) for Earth as an exoplanet around our G2V star. The quiescent Sun transit spectrum is shown in black, the impact of GLE05 in blue, and the impact of the AD774/775 event is shown in purple. Figure after \citet{Herbst-etal-2019c}.}
    \label{fig:5.4.8}
\end{figure}

In an Earth-as-an-exoplanet study, \citet{Herbst-etal-2019c} investigated the impact of strong solar events on the terrestrial atmosphere and the corresponding changes in the terrestrial transmission spectrum. Originally only focusing on the strongest modern GLE (i.e., GLE05), Fig.~\ref{fig:5.4.8} shows the impact of GLE05 (in blue) and AD774/775 (in purple) on the Earth's transmission spectrum. In agreement with current observations, the modern-Earth spectrum exhibits features like, for example, the Rayleigh extinction in the UV–visible, the rotational–vibrational H$_{2}$O and CH$_{4}$ bands in the NIR, the strong O$_{3}$ and CO$_{2}$ features at 9.6 $\mu$m and 15 $\mu$m, as well as weaker biosignature-related features from O$_{2}$ in the visible and N$_{2}$O (around 7.8 $\mu$m). Comparison of the quiet-Sun (in black) and the GLE05 scenario (in blue) shows moderate suppression of several features during the GLE, notably O$_{3}$ and CH$_{4}$ in the NIR, attributed to a photochemical loss triggered by the high-energy cosmic rays. \citet{Scheucher18} proposed HNO$_{3}$ as a potential cosmic-ray signature in Earth-like planets around active M dwarfs. \citet{Herbst-etal-2019c} found this absorption feature near 12 $\mu$m in Earth’s atmosphere during GLE05. \citet{Scheucher18} showed that the prominence of such CR-induced features is highly sensitive to the  atmospheric composition. The AD774/775 induced transit spectrum (purple) further reveals this feature also strongly depends on the strength of the particle event. Here, a second, even stronger, HNO$_{3}$ feature occurs at wavelengths above 20 $\mu$m.

\section{Generation, Evolution, and Habitability Implications of Planetary Magnetic Fields}\label{Magnetichhabitability}

Is planetary magnetic field essential to (exo)planetary habitability? This is the central question posed and discussed in this section. The traditional definition and classical notion of a habitable zone is primarily based on a radiative criterion -- positing that the host star’s radiation flux and the distance of the planet in question influences the latter’s equilibrium temperature and by extension, its ability to host liquid water – deemed essential for life. The magnetic habitability problem, on the other hand, is concerned with a complementary but increasingly relevant question. The question may be stated in a tractable manner open to investigation: Given the host star’s magnetic forcing and the intrinsic magnetic field of the planet (or the lack of it), does the planetary atmosphere survive and the near- or sub-surface planetary environment retain properties essential for life to originate and sustain over geologic timescales.

The problem framed in the aforementioned manner, assumes that all other conditions being conducive to life, how do two interacting magnetic systems impact habitability.  The first system, namely, the stellar magnetic engine -- rooted in the star's internal dynamo – is responsible for magnetic forcing of the (exo)planetary environment through the creation of space weather and space climate \citep{Gudel2007,Nandy2007,Airapetian2020,Nandy2021}.  The second system is the planet's intrinsic magnetic field or magnetosphere -- generated by its own internal dynamo. It is the interaction of these two systems which eventually leads us to the topic of ``magnetic habitability’’.

\subsection{Magnetohydrodynamic Foundations of Planetary Dynamos}
The generation and maintenance of planetary magnetic fields, such as Earth's geodynamo, are fundamentally governed by the self-sustaining nonlinear interactions of fluid turbulence, rapid planetary rotation, and the resultant induction of magnetic fields within an electrically conducting fluid core \citep{Tobias2021}. Because interior planetary temperatures vastly exceed the Curie temperature of ferromagnetic materials (temperature at which a ferromagnetic material loses its permanent magnetism), permanent remanent magnetism cannot account for active, planetary-scale fields \citep{jones2011}. Instead, the process relies on the conversion of kinetic energy from convection within planetary molten cores into magnetic energy and can be described by the formalisms of magnetohydrodynamics (MHD) \citep{jones2011,Tobias2021}.

The temporal evolution of the macroscopic magnetic field ($\mathbf{B}$) is governed by the magnetic induction equation, derived from Maxwell's equations and Ohm's law for a moving conductor \citep{Tobias2021}:\begin{equation}\frac{\partial \mathbf{B}}{\partial t} = \nabla \times (\mathbf{u} \times \mathbf{B}) + \eta \nabla^2 \mathbf{B}\end{equation}In this governing equation, the left-hand side ($\frac{\partial \mathbf{B}}{\partial t}$) represents the local rate of change of the magnetic field over time. This temporal evolution is dictated by the continuous physical competition between the two terms on the right-hand side. The first term, $\nabla \times (\mathbf{u} \times \mathbf{B})$, is the advective generation (or induction) term; it describes the creation, stretching, and twisting of magnetic field lines driven by the macroscopic fluid velocity field ($\mathbf{u}$). The second term, $\eta \nabla^2 \mathbf{B}$, represents Ohmic dissipation (or magnetic diffusion); it accounts for the continuous decay and smoothing of the magnetic field due to the internal electrical resistance of the medium. Here, $\eta = (\mu_0 \sigma)^{-1}$ is the magnetic diffusivity, dependent on the vacuum permeability $\mu_0$ and the electrical conductivity $\sigma$ of the collisional plasma (or fluid core).The successful operation of a dynamo requires that the advective generation term [$\nabla \times (\mathbf{u} \times \mathbf{B})$] successfully overcomes the Ohmic dissipation term [$\eta \nabla^2 \mathbf{B}$] to prevent the field from decaying away. The relative magnitude of these two competing processes is parameterized by the dimensionless magnetic Reynolds number ($Rm$), defined as the ratio of magnetic advection to magnetic diffusion. Planetary dynamos operate in the highly turbulent, high-$Rm$ regime, necessitating highly complex, three-dimensional fluid flows to avoid the mathematical constraints of Cowling's anti-dynamo theorem \citep{Tobias2021}. This fundamental theorem mathematically constrains that a purely axisymmetric magnetic field, i.e., one that is perfectly symmetric around the rotational axis, cannot be self-sustained by any axisymmetric fluid flow alone \citep{Tobias2021}. In a strictly two-dimensional or axisymmetric system, there inherently exists a circular ``neutral line'' where the poloidal magnetic field vanishes. At this specific geometric boundary, the advective electromotive contribution ($\mathbf{u} \times \mathbf{B}$) evaluates to exactly zero, leaving the system fundamentally unable to drive the electrical currents necessary to replenish the field against continuous Ohmic dissipation \citep{Tobias2021}. Consequently, for example, while Earth's observable magnetic field manifests largely as a simple, axisymmetric dipole at the surface to a crude approximation, the underlying geodynamo cannot be symmetrically simple. Instead, it strictly relies on chaotic, highly turbulent, and asymmetric three-dimensional fluid motions within the outer core; these microscopic fluid eddies break the spatial symmetry, successfully bypassing the restrictions posed by Cowling's theorem to construct the inductive requirements to sustain the macroscopic planetary field over geological timescales \citep{jones2011,Tobias2021}.

The fluid dynamics driving this magnetic induction are typically modeled using the Navier-Stokes momentum equation, subject to rotational and magnetic forces. Because the fluid velocities in the core are much slower than the characteristic sound speed, the mathematics is often formulated under the Boussinesq approximation for incompressible flows \citep{jones2011}:
\begin{equation}
\label{eq:boussinesq}
\rho \left( \frac{\partial \mathbf{u}}{\partial t} + (\mathbf{u} \cdot \nabla)\mathbf{u} \right) = -\nabla p + \rho \nu \nabla^2 \mathbf{u} - 2\rho \mathbf{\Omega} \times \mathbf{u} + (\mathbf{J} \times \mathbf{B}) + \mathbf{f}_b\end{equation}\begin{equation}\nabla \cdot \mathbf{u} = 0, \quad \nabla \cdot \mathbf{B} = 0
\end{equation}
In Equation \ref{eq:boussinesq}, the left-hand side encapsulates the total inertial acceleration of the fluid mass, comprising both the local rate of change of velocity ($\frac{\partial \mathbf{u}}{\partial t}$) and the nonlinear advective acceleration ($(\mathbf{u} \cdot \nabla)\mathbf{u}$). This inertial motion is balanced by the sum of the physical forces detailed on the right-hand side. The term $-\nabla p$ represents the isotropic pressure gradient, while $\rho \nu \nabla^2 \mathbf{u}$ accounts for viscous dissipation, where $\nu$ is the kinematic viscosity of the liquid ferromagnetic alloy. The term $\mathbf{f}_b$ encapsulates the crucial radial buoyancy forces driving the convection; for example, in the Earth's core, this includes both thermal buoyancy (from secular cooling and latent heat release) and compositional buoyancy (from the expulsion of light elements during the crystallization of the solid inner core) \citep{jones2011}. The remaining two terms on the right are what uniquely define the physical regime of planetary dynamos. The Coriolis force ($- 2\rho \mathbf{\Omega} \times \mathbf{u}$), governed by the planetary rotation vector $\mathbf{\Omega}$, plays a paramount role. In rapidly rotating bodies like the Earth, the Coriolis force strongly opposes variations in fluid flow parallel to the axis of rotation. Consequently, it organizes the chaotic convective turbulence into coherent, rotationally-aligned cylindrical structures known as Taylor columns; these helically convecting columns are the primary macroscopic structures responsible for generating the large-scale dipolar magnetic field \citep{jones2011}. Simultaneously, the Lorentz force ($(\mathbf{J} \times \mathbf{B})$), where $\mathbf{J}$ is the electrical current density, introduces a non-linear feedback mechanism into the system. As the convective fluid motion successfully generates the magnetic field via the induction equation, the resulting Lorentz force acts back upon the fluid to oppose the very motion creating it (analogous to Lenz's Law). The ultimate strength of the planetary magnetic field is therefore determined by a dynamical equilibrium where the Lorentz force grows strong enough to balance the driving Coriolis and buoyancy forces, saturating the dynamo \citep{jones2011}. 

\subsection{Long-Term Evolution of the Geodynamo}
Over geological time scales spanning $10^5$ to $10^9$ years, the secular variation of the geomagnetic field acts as a critical proxy for the thermodynamic conditions at the core-mantle boundary (CMB) \citep{aubert2010}. The overlying solid mantle strictly dictates the rate of secular cooling and the spatial distribution of heat flux escaping the outer core. Because the liquid outer core is well-mixed and largely adiabatic (a condition where no heat is transferred across the boundary by conduction alone), lateral variations in CMB heat flux, driven by deep mantle heterogeneities such as subducting slabs (cold, dense oceanic plates sinking into the mantle) or large low-shear-velocity provinces (LLSVPs) (massive, hot, and buoyant thermochemical piles at the base of the mantle), impose strong thermal boundary conditions on the geodynamo. These heterogeneous boundary conditions control the overall vigor of convective upwellings, dictate the symmetry of fluid flow, and fundamentally shape the time-averaged morphology of the magnetic field \citep{aubert2010}. 

\begin{figure}[ht]
    \centering
    \includegraphics[width=0.8\textwidth]{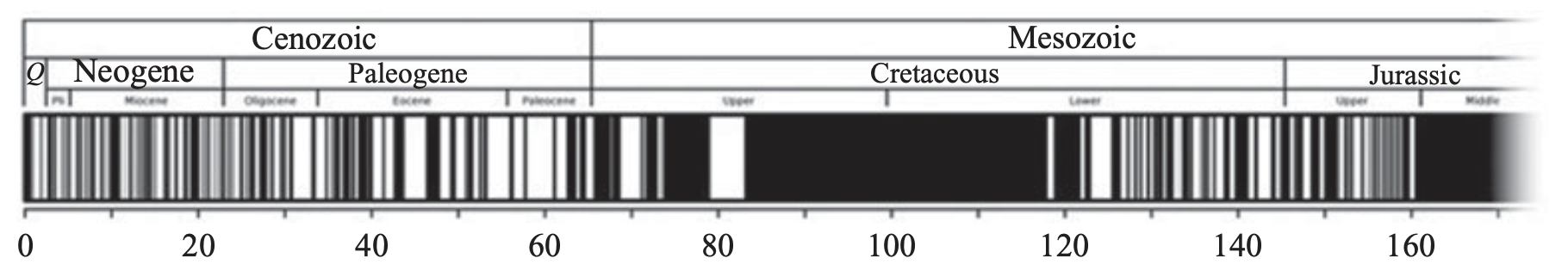}
    \caption{A composite record of Earth's magnetic polarity reversals through the Cenozoic and Mesozoic eras. Black bars indicate intervals of normal polarity (consistent with the current orientation), while white bars indicate reversed polarity. Note the distinct presence of 'superchrons' (prolonged intervals of unchanging polarity) such as the Cretaceous Normal Superchron, which represent significant departures from the high-frequency reversal behavior observed in other geological epochs (adapted from \citet{Tobias2021}).}
    \label{fig:reversal_record}
\end{figure}

Earth's current field is predominantly dipolar, exhibiting a mean surface strength of approximately $40~\mu\mathrm{T}$ with its primary axis offset by roughly $10^\circ$ from the planetary rotation axis \citep{Tobias2021}. However, the temporal expression of the geodynamo is remarkably broad, spanning sub-annual geomagnetic jerks to continuous secular evolution over billions of years \citep{aubert2010}. Analyzing the power spectrum of the dipole moment over millions of years reveals extreme variability in reversal frequencies. Magnetic polarity reversals are driven by nonlinear convective fluctuations that generate reversed magnetic flux patches at the CMB; when these patches migrate poleward and proliferate, they temporarily overwhelm the primary axial dipole \citep{Tobias2021}. Conversely, the geological record features ``superchrons'' (prolonged intervals lasting tens of millions of years), such as the Cretaceous Normal Superchron, during which the magnetic field completely ceases to reverse its polarity \citep{aubert2010, Tobias2021}. The geological record of these magnetic field reversals is summarized in the geomagnetic polarity time scale (Figure \ref{fig:reversal_record}), which illustrates the alternating intervals of normal and reversed polarity and highlights extended periods of stability. These exceptionally stable periods are theoretically linked to phases of mantle convection that either depress the total global CMB heat flux or establish highly symmetric thermal boundary conditions, suppressing the turbulent fluctuations that trigger reversals \citep{aubert2010}.

Furthermore, determining the strength and variability of the magnetic field during the Archaean and Proterozoic epochs provides vital constraints on the deep thermal history of the Earth. It is hypothesized that the primitive dynamo was driven entirely by secular cooling (and potentially trace radiogenic heating) prior to the nucleation of the solid inner core. The eventual crystallization of the inner core marked a regime shift in core dynamics. As iron alloys solidified at the planet's center, the continuous rejection of lighter elements (e.g., oxygen, silicon, and sulfur) at the inner core boundary introduced compositional buoyancy as a potent, primary convective driver \citep{aubert2010}. This critical phase transition from purely thermal to robust thermochemical convection likely rejuvenated the geodynamo, ensuring its energetic persistence against the gradual thermal equilibration of the planet.

\subsection{Paleomagnetic Records and Tectonic Reconstructions}
Retrieving the deep-time history of the geomagnetic field relies extensively on the paleomagnetic record—the fossilized orientation of the magnetic field at the time of rock formation—preserved in igneous and sedimentary archives \citep{dallanave2020}. In igneous rocks, this record is fixed during cooling as magnetic minerals pass through their Curie temperature, locking in a Thermoremanent Magnetization (TRM) parallel to the ambient field. Conversely, sedimentary rocks acquire a Detrital Remanent Magnetization (DRM) as magnetic grains align with the Earth’s field during deposition and subsequent consolidation. Paleomagnetism remains the sole quantitative methodology capable of testing for the lateral motions of tectonic plates across ancient Earth's surface \citep{evans2008}. By determining the paleolatitudes of continents via the geocentric axial dipole (GAD) hypothesis, which posits that, when averaged over $10^4$ to $10^5$ years, the time-averaged geomagnetic field is equivalent to a dipole aligned with the Earth's rotation axis, apparent polar wander paths (APWPs) can be mapped to track the relative trajectories of specific landmasses with respect to an absolute geomagnetic reference frame \citep{evans2008,dallanave2020}. This transforms paleomagnetism into a predictive tool for global tectonic reconstruction, effectively serving as the primary coordinate system for deep-time geophysics. 

\begin{figure}
\centering
\includegraphics[width=0.8\textwidth]{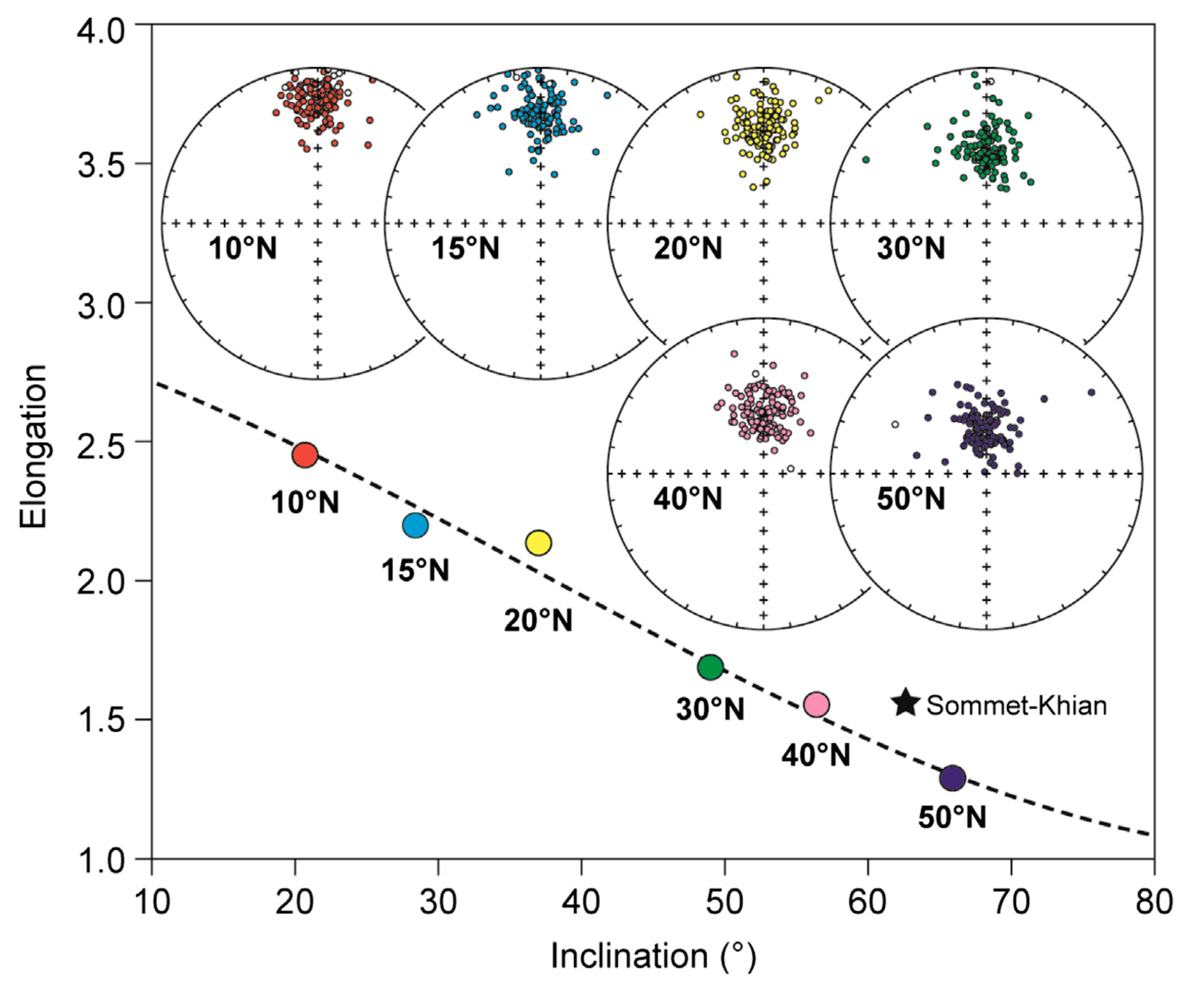}
\caption{Predictive E/I Distributions from the TK03.GAD Model. This figure displays simulated sets of paleomagnetic directions as they would appear in stereographic projections at different latitudes. The dashed line defines the expected elongation/inclination (E/I) relationship for an undistorted magnetic record. Comparing actual sedimentary datasets against this curve allows to identify data that have been skewed by tectonic deformation \citep[adapted from][]{dallanave2020}.}
\label{fig:elonginc}
\end{figure}

Applying rigorous quality constraints to global databases has helped document extensive surface motions of cratons (the stable, ancient interiors of continents) dating back beyond 2775 million years ago, although analytically distinguishing the differential motion of independent plates from true polar wander (TPW), a geophysical phenomenon where the entire solid mantle and crust rotate in unison relative to the fluid outer core and spin axis, remains a significant challenge in the oldest Archean datasets \citep{evans2008}. The reliability of these paleogeographic reconstructions is inextricably linked to the physical fidelity of the magnetic archives, which are prone to secondary overprinting and mechanical deformation \citep{dallanave2020}. Continuous temporal datasets derived from sedimentary rocks are particularly valued for their high resolution, yet they are notoriously susceptible to inclination flattening and tectonic strain \citep{dallanave2020}. Inclination flattening occurs during the post-depositional compaction of sediments, where the long axes of magnetic grains are preferentially rotated toward the horizontal plane, causing the measured inclination to be shallower than the true paleofield inclination. Paleosecular variation (PSV) models, such as the widely utilized TK03.GAD model, describe the statistical wandering of the magnetic poles over millennial timescales. These models dictate that the distribution of paleomagnetic directions should exhibit a predictable, latitude-dependent elongation, which is best visualized through inclination-versus-elongation (E/I) plots. Figure \ref{fig:elonginc} illustrates the theoretical expectation for this scattering. At high latitudes, the magnetic field is steep, resulting in a distinctively elongated distribution of directions as the pole wanders; conversely, at low latitudes, the field is shallower and the distribution is more circular. The dashed line in the figure represents the mathematical E/I relationship predicted by the TK03.GAD model. If a set of sedimentary paleomagnetic data is pristine, meaning it has not been affected by post-depositional tectonic strain, its measured E/I values will fall directly upon or very near this theoretical curve.

When sedimentary basins undergo post-depositional tectonic deformation, the magnetic grains within the rock fabric are mechanically rotated, typically resulting in "inclination flattening." This causes the measured paleomagnetic directions to shift horizontally or vertically away from the TK03.GAD curve, violating the expected latitudinal elongation. By plotting the E/I values of a rock collection against this model, one can quantitatively diagnose whether the data have been compromised. Data points that deviate significantly from the predicted E/I trajectory can be discarded, ensuring that subsequent tectonic reconstructions are founded exclusively upon unaltered, primary magnetic remanence that accurately reflects the ancient geomagnetic field.

\subsection{Geomagnetic Shielding and Planetary Habitability}
The longevity, intensity, and topological stability of planetary dynamos have profound implications for sustained planetary habitability. The internal geodynamo generates a macroscopic magnetospheric barrier that continuously deflects the incident solar wind plasma and significantly attenuates the surface flux of high-energy particles \citep{jones2011}. The absence of an actively sustained magnetic field renders a planet's atmosphere highly vulnerable to severe erosion via non-thermal escape mechanisms, including ion pick-up and atmospheric sputtering. These physical processes are thought to have contributed to the atmospheric collapse and desiccation of Mars following the cessation of its ancient internal dynamo \citep{jones2011}. For Earth, the uninterrupted persistence of the geomagnetic field over more than three billion years has potentially minimized the catastrophic stripping of its water inventory and volatile elements to space \citep{Tobias2021}. Extreme long-term fluctuations in the geodynamo, such as intervals of high reversal frequency or ultra-weak field intensity, temporarily reduce the shielding efficacy of the magnetosphere \citep{aubert2010, Tobias2021}. Such variations in magnetic shielding influence atmospheric chemistry and act as an evolutionary stressor for exoplanetary atmospheres.

\subsection{Stellar magnetism as the primary external influence}

Stellar magnetic fields are generated by a MHD dynamo mechanism operating in stellar interiors \citep{Charbonneau2020, Hazra2023}. The strength of this magnetic activity depends on various factors including the rotation and differential rotate rate of the star, the depth and thermal flux of the stellar convection zone and large-scale internal plasma flows. These physical properties of a star evolves as a star ages, mediated via varying energy generation rate, changes in composition, rotational braking mediated via angular momentum losses via stellar plasma winds \citep{Nandy2004,Brun2015}. On the one hand, diverse stars (corresponding to different spectral classification) exhibit very different magnetic output. On the other hand, this activity output changes with stellar evolution. Chapter~1 and Chapter~2 in this series and earlier sections of this Chapter present a detailed discussion of stellar magnetic activity and its evolution, highlighting the extreme magnetic environment of young solar-type stars, the decline of this activity as the star ages and the emerging understanding of a magnetically quiescent phase of older solar-type stars \citep[see e.g.,][]{Gudel2007,Reinhold2020,Tripathi2021}. 

This understanding makes it abundantly clear that the magnetic output of stars vary over Gyrs and varies from one type of star to another.  This in turn results in a variable flux of stellar radiation, energetic particles, magnetized plasma winds and the severity and frequency of flares and CMEs (see, e.g, Chapter~2 of this collection). All of these have profound implications for the forcing of exoplanetary environments.

These considerations are especially  important for planets orbiting M dwarfs, because their habitable zones are close to the star and such (typically close-in) planets confront stronger stellar-wind pressure, stronger interplanetary magnetic fields, larger flare and energetic-particle fluxes, and in many cases tidal locking `-- all of which come together to exacerbate close-in star magnetic interactions affecting habitability (see Chapter~3 for a comprehensive discussion on close-in star planet interactions). It is worth noting that the abundance  and longer lifetimes of M-dwarfs make (hot Jupiter type) exoplanets around them compelling observational targets for exploration; however, their exotic magnetic environment and tidal locking also imply their habitability is compromised \citep{Shields2016}. There is no question, therefore, rocky, Earth-like exoplanets with atmosphere -- although difficult to detect -- remain the most viable candidates for hosting life.

\subsection{Planetary magnetism as an X-factor}

Like stars, planets also host a diverse range of magnetic fields; some have strong intrinsic magnetic fields, some weak and some none. Before we go on to discuss their importance, it is important to appreciate how they arise, how they vary and how their origins are tied to the planetary system being considered. A long-lived dynamo requires a conducting fluid region in the planetary interior, adequate convective flux through the existence of a thermal or compositional gradient that sustains core cooling or inner-core growth and rotation. Diverse mechanisms are thought to generate magnetic fields in distinct types of planets, e.g., liquid iron cores for terrestrial bodies, metallic hydrogen or ionic water layers for gas and ice giants \citep{Stevenson2003}. The field's magnitude and geometry are determined by the convective vigour, core conductivity, rotation rate, and thermal boundary conditions. 

The tilt of the planetary dipole (important from the context of star-planet interactions) depends on the relative orientation of the planet’s internal rotation axis and the scaling of the dynamo generated magnetic field strength depends on the interplay of convective flux and rotation \citep{Christensen2006,Olson2006}. While this suggest that massive, rapidly-rotating terrestrial planets (super-Earths) might sustain relatively strong magnetospheres, the actual relationship is confounded by the material properties of the mantle and the presence or absence of plate tectonics, both of which regulate core heat extraction and thus dynamo efficiency. 

Tidal heating further complicates the picture for close-in exoplanets. \citet{Driscoll2015} show that tidal dissipation around low-mass stars can either maintain or disrupt thermal and magnetic evolution, depending on orbital eccentricity and circularization history. A strong field may therefore be easier for some planets and harder for others, even at the same incident stellar flux. Intriguingly, numerical simulations demonstrate that thermally-driven convection -- fuelled by radiogenic heating, primordial heat, or tidal dissipation -- can sustain long-term dynamos even in slowly-rotating worlds, producing dipole fields comparable the Earth's surface field \citep{Zuluaga2012}. Nonetheless, the viability of such dynamos depends critically on core size, composition, and age, introducing substantial uncertainties or variations.

One does not need to look beyond the solar system to appreciate the diversity of planetary types, planetary magnetic field strength, dipole field orientation and the lack of an intrinsic magnetosphere in a planet that likely hosted one in the past (e.g., Mars). The discussion above highlights two facts. First, there is indeed a wide range of magnetic behavior in planets and second, there is considerable uncertainty in our predictive capability of the nature and strength of planetary magnetism. The situation is confounded by a third fact; exoplanetary magnetic fields are super hard to detect!

Direct detection of exoplanetary magnetic fields remains one of the outstanding challenges of observational astrophysics. However, three approaches appear promising. The first is the Cyclotron maser instability (CMI) emission from magnetospheric electron precipitation is the primary emission mechanism at Earth, Jupiter, and Saturn. The LOFAR and forthcoming SKA telescopes are designed to detect analogous emission from exoplanets, with predicted flux densities of about 0.01–10 mJy corresponding to  hot Jupiters and close-in super-Earths. The second possibility is transit observations. Magnetospheric stand-off distances can --in principle -- be detected as early ingress or extended transit asymmetries in UV or soft X-ray bands, where the magnetosphere's bow shock absorbs or scatters stellar emission. The third and most theoretically well studied mechanism for planetary magnetism detection is star-planet magnetic interactions. Chromospheric hot spots and enhanced flaring rates phase-locked to planetary orbital periods provide indirect evidence for the presence of exoplanetary magnetic fields (see Chapter~3 in this series for further details).

Why does knowledge of planetary magnetic fields matter in the context of our discussion? This brings us to the heart of the matter -- the importance of planetary magnetism for habitability.

The only planet with evidence of life -- Earth -- offers the primary motivation. Its dynamo generated magnetosphere holds the solar wind and storms at bay beyond the magnetopause (whose location is dynamic and varies between 6-10 Earth radii) and within which the solar plasma and energetic particles cannot normally penetrate \citep{Das2019}. This protects our thin, approximately 60 km atmosphere from the onslaught of the magnetized solar wind. Paleomagnetic evidence indicates that our geodynamo existed for at least 3.4-3.45 Gyr, and possibly earlier, when the young Sun’s stronger magnetized plasma wind, high-energy radiation and more frequent magnetic storms was battering an young Earth \citep{Tarduno2010,Tarduno2014}. This provokes the idea that planetary magnetism is the hidden X-factor for habitability, helping Earth retain its atmosphere, water and habitability \citep{Tarduno2025}. The fact that Mars -- with its extinct dynamo and consequently no intrinsic magnetosphere -- has no significant atmosphere at present proves the converse and lends further credence to this idea \citep{Basak2021}. 

Yet the story is not so simple! Venus has no intrinsic magnetosphere either, yet, it retains a massive atmosphere -- perhaps due to its larger size and enhanced gravity. Thus, this field is nuanced – influenced by myriad factors, conclusions have caveats and views are hotly debated. In the following sections we shall delve in to the complexities of habitability vis-à-vis planetary magnetism.

\subsection{How planetary magnetic fields can support habitability?}

The fundamental debate we are dealing with is whether an intrinsic planetary magnetic field is a prerequisite for habitability, or whether a thick atmospheres, volcanic outgassing, and other factors (such as strong gravity) can retain an atmosphere and support habitability over geologic timescales.

The case for ``magnetic protection’’ has developed over the last several decades and now stands on firm physical foundations based on solar system studies. A global-scale magnetosphere, which is significantly larger than the atmosphere within, acts as a protective cage reducing or entirely stopping stellar wind access to most of the inner atmosphere. This restricts charge particle access and atmospheric erosion at biologically relevant altitudes. This is well proven for our home planet Earth. While the location of the magnetopause varies with the solar wind speed and the nature of the incoming solar wind magnetic field, numerical simulations \citep{Das2019} and observations \citep{Nguyen2022,Grimmich2023} show that the magnetopause does not typically shrink below five Earth radii. The Earth’s magnetospheric stand-off distance -- where the Earth’s compressed magnetic field stops the solar wind – is on average about ten times the planet’s radius, i.e., about thousand times thicker than the Earth’s atmosphere (60 km) shielding the Earth’s atmosphere from erosion mediated via solar wind and plasma storms and protecting life forms.

For close-in terrestrial exoplanets, particularly around M dwarf stars, it has been suggested that the planetary environment could be bombarded by intense cosmic ray flux and stellar energetic particles if their intrinsic magnetic field are weak \citep{Griessmeier2005, Griessmeier2009}; atmospheric pressure and magnetic moment jointly set the surface radiation environment. MHD simulations by \citet{Vidotto2013} and  \citet{Cohen2015} have explored the ion escape rates from planets with various magnetic field strengths embedded in realistic M-dwarf wind environments, finding that even modest planetary fields can reduce total ion escape rates relative to the unmagnetized case. However, it is to be noted that polar outflow channels persist regardless of field strength. A detailed exposition of star planet interactions in close-in and far-out systems and consequent atmospheric forcing is presented in Chapter~3 of this series; readers are referred therein for the dynamics at play in such systems. 

While magnetic shielding, gravity and upper-atmosphere structure and the nature of the atmosphere together control planetary vulnerability to solar and stellar forcing \citep{Lundin2007}, once all other ingredients are in place, current understanding posits that magnetic field helps preserve the atmosphere and upper atmospheric composition (e.g., Ozone shielding), global atmospheric density, pressure and plausibly surface water -- especially during early evolution of the star-planet system when stellar activity is very high (see Chapter 1 of this series).

Is the converse true, i.e., is there evidence that the lack of an intrinsic magnetic field results in (catastrophic) atmospheric losses? Mars is an excellent neighboring laboratory for testing this scenario. It is thought that Mars had an internal dynamo powering a magnetosphere early in its history, which was subsequently lost about 3--3.5 Gyr ago \citep{Mittelholz2020,Mittelholz2022}. Independent studies indicate that the early Martian atmosphere was also lost around a similar time \citep{Pepin1994}; indeed observations from modern spacecrafts like MAVEN indicate an ongoing mass loss that is significant \citep{Jakosky2018}. Numerical simulations specifically geared towards exploring solar wind forcing of the Martian environment, see, e.g., \citet{Basak2021} and references therein, demonstrate a link between the lack of a global magnetosphere and atmospheric erosion. The latter is expected to have led to a complete loss of the -- (magnetically) unprotected -- large-scale atmosphere of Mars. So at least in the case of Mars, which is open to relatively more detailed investigations than extrasolar planets, the narrative remains consistent. 

There are nonetheless caveats and nuances in the dynamics of star planet magnetic interactions that makes this narrative nuanced, and we turn our attention to these complexities in the next section.

\subsection{Can strong planetary magnetic fields be a hindrance to habitability?}

The paradox of planetary magnetism as a protective shield is confounded by solar system analogues. Like Mars, Venus lacks an intrinsic dynamo, has only an induced magnetosphere and yet it sustains a massive C0$_2$ atmosphere. This is possibly a result of volcanic outgassing and gravity compensating for solar activity induced atmospheric erosion. From the specific context of magnetism, it is possible that an induced magnetosphere can sometimes provide protection against ion-loss by virtue of its geometry, shutting down specific polar outflow channels and disfavoring magnetic reconnection that typically occurs in the presence of a significant intrinsic planetary field. 

Magnetospheres do exclude stellar wind access to some regions close to the planet, however, they also foster magnetic reconnection and open-up planetary magnetic field lines, host current sheets, power outflows from auroral caps and drive plasma blobs out of the upper ionosphere and magnetotail regions. \citet{Gunell2018} argue that, therefore, intrinsic magnetization is not a sufficient condition for protection of planetary atmospheres. \citet{Blackman2018} further notes that magnetospheres enable mass, energy and momentum coupling between the stellar wind and the planetary atmosphere, which can either mitigate or enhance atmospheric erosion. In the terrestrial context, recent work shows that in typical (nominal) solar-wind conditions, ion escape may be partly balanced by solar wind ion deposition, thereby not allowing the Earth’s atmosphere to be depleted over geologic timescales \citep{Hinton2025}. These considerations add layers of mystique to star planet interaction mediated atmospheric evolution.  

Close-in planets in the habitable zone around M dwarfs offer a distinct, more exotic laboratory for testing the role of magnetism in atmospheric evolution. Their orbital proximity to the host star often places them inside dense, magnetized stellar winds and expose them to more frequent flares (and possibly CMEs).  \citet{Airapetian2017} estimates that enhanced ionising radiation around {\it{Proxima Cen}} like stars can increase polar outflows channelled along magnetic fields by orders of magnitude, favouring atmosphere erosion. \citet{Meadows2018} contend that in a planet like {\it{Proxima Cen b}}, the nature of the initial atmospheres, volatile delivery, stellar history, magnetic shielding and escape processes all factor in to eventually determine evolution of habitability.  In the TRAPPIST-1 planets, studies show that ion escape forced by the extreme space weather around M-dwarfs could be significant over long timescales \citep{Dong2018}. Such conclusions are further supported by simulations of stellar CME induced magnetospheric forcing of hot Jupiter-like systems \citep{Hazra2025}. 

A comprehensive exploration of of this star-planet magnetic interplay in the context of habitability is based on theory and global MHD simulations of atmospheric erosion in which both stellar and planetary magnetic fields are varied -- allowing a more complete appreciation of the diversity of possibilities. In the context of a far-out, Sun-Earth like star-planet system, \citet{Gupta2023} finds that increasing the stellar magnetic-field strength or weakening the planetary field compresses the dayside magnetopause, opens the magnetotail, creates Alfv\'en wing-like structures, and enhances mass loss through reconnection-mediated pathways. Conversely, given a stellar wind magnetic field, stronger intrinsic planetary field favors atmospheric retention in an Earth-like planet. However, and this is where things get really interested, \citet{Gupta2023} perform a detailed parameter space study to demonstrate that when the stellar and planetary magnetic field are both made stronger, atmospheric losses become larger due to larger reconnection surfaces coupling the atmosphere to the stellar wind; they go on to establish a semi-analytic relationship that qualitatively predicts atmospheric erosion as a function of stellar and planetary magnetic field strength -- illuminating a diversity of possibilities governed by magnetism.   

These studies challenge the simple assumption that a stronger intrinsic planetary field always means better atmospheric retention and indicates the question posed at the beginning of this section is more nuanced and layered than we can perhaps fully comprehend at present. Such results do not make magnetic fields irrelevant but rather suggest that the detailed nature of the stellar and planetary magnetic field – combined with other factors -- makes the answer non-trivial.

\subsection{Outstanding issues and future outlook}

It would be obvious to any reader now that several issues remain unsettled. Is an intrinsic planetary magnetosphere (and thus a internal dynamo) necessary for long-term surface habitability, or can sufficient atmospheric mass, gravity and stellar wind replenishment compensate? How do induced magnetospheres compare with intrinsic magnetospheres when escape is integrated over billions of years of coupled star-planet interaction? What planetary field strength, what dipole tilt is most suitable for atmospheric retention? Is a more complex, multi-polar planetary magnetic field a better shield or worse compared to a dipole field configuration? How do magnetic reversals, sustained phases of weakened magnetic field impact atmospheric erosion? For planets around M dwarfs the largest uncertainty may be the stellar forcing itself, with poor understanding of surface magnetic field distribution and variability, flare and CME intensity and frequency, uncertainty about stellar wind speeds and their evolution over timescales relevant for habitability.  

What is certain is this. On the one hand, magnetic fields directly mediate stellar wind and CME interaction with planetary magnetosphere-atmosphere systems and consequently atmospheric evolution. On the other hand, planetary magnetism indirectly influences radiation and thermally driven outflows from planets and influence the geometry of ion pick-up and losses. 

The biological threshold also depends on other factors that work in conjunction in star planet magnetic interactions. These are related to the planetary properties. While stellar radiation, energetic particles and magnetized wind access impacts atmospheric evolution and chemistry, a relatively thick-atmosphere biosphere system, sub-surface and oceanic habitats are less prone to stellar forcing, at least on short timescales. Another related debate concerns the timescales of the problem. Instantaneous atmospheric escape rates can be misleading if they are not considered relative to volatile supply, volcanic outgassing, impact delivery, photochemical feedback and climate dynamics occurring over geologic timescale. A planet may lose lighter elements like hydrogen but retain habitability if water is abundant, or it may retain a massive atmosphere that is uninhabitable because of runaway greenhouse conditions. Likewise, while intense flare related high energy radiation may be harmful over a sustained period of planetary evolution, in the early phases when life is taking shape -- ultraviolet radiation may catalyse prebiotic chemistry.

Magnetic habitability must therefore be cast as a coupled star-planet evolution problem along with other habitability-determining factors, rather than a standalone atmospheric shielding metric. While this outlook complicates the central theme of habitability, it also makes this problem richer and points to future interdisciplinary pathways for exploring planetary habitability. 

\section{Concluding remarks}\label{conclusion}

The apparent simplicity of the requirement for liquid water on a terrestrial planet to be considered habitable is deceptive. In reality, this requirement is contingent on the evolution of the interior and atmosphere of the planet, commencing with the emergence of a secondary atmosphere. It follows that the most comprehensive scenarios for habitability must encompass stellar evolution and the response of a planetary atmosphere throughout its evolution, including the potential migration of the HZ due to changes in stellar bolometric luminosity. 

The efforts to build those scenarios are mostly supported by numerical simulations given our present limitations to characterize the interiors and atmospheres of terrestrial exoplanets. On the other hand, stars are much better known than exoplanets, which helps to inform our numerical models to a certain point. Stellar spectra and radiation emitted by chromospheric and coronal activity can be measured, but the stellar wind, high-energy particles, and the configuration of their magnetic fields are much more difficult to constrain. Going from large to smaller scales, an ideal model for planetary habitability should be able to reproduce the effects of space weather on a given planet, including transient stellar events and plausible magnetic field configurations for the planet. At the planetary level, we need to calculate the interaction of high-energy radiation heating the upper atmosphere, promoting atmospheric loss, including the cooling mechanisms depending on the atmospheric composition, and particles that ionize lower atmospheric layers and result in atmospheric chemistry and climate changes. Current atmospheric chemistry models that include particles and radiation provide a comprehensive scenario that is relevant to understand the potential for prebiotic chemistry or for the detection of biosignatures. 1D and 3D models provide complementary information considering the uncertainties and assumptions adopted by each model. Intercomparisons between models help to identify the effect of those assumptions and uncertainties. 

The other crucial factor is time. The HZ moves with the luminosity evolution of the star, which implies that planets located in the HZ could have lost their atmospheres or water in the past. As young stars are more active, their XUV and UV luminosity is more intense, and flares as well as CMEs are more frequent and energetic, further indicating a much harsher stellar energetic particle environment. The combination of these factors may lead to the loss of the atmosphere, and its recovery depends on both the energy stored in the planetary mantle and the planet's original volatile budget. 

Most studies of planetary habitability have been focused on terrestrial planets with CO$_2$-N$_2$-H$_2$O atmospheres, but large terrestrial exoplanets would be able to hold H$-2$ atmospheres. Another intriguing prediction is the massive O$_2$ atmospheres left after a greenhouse runaway. The evolution of those atmospheres and their potential for habitability has yet to be fully explored.

\backmatter
\section*{Declarations}

\bmhead{Acknowledgments}
None for now

\bmhead{Funding}
AS acknowledges support from the UNAM–PAPIIT grant IN114026. KH acknowledges support from the Research Council of Norway through the Centres of Excellence funding scheme, project number 332523 (PHAB). RE's research was carried out at the Jet Propulsion Laboratory, California Institute of Technology, under a contract with the National Aeronautics and Space Administration (80NM0018D0004).

\bmhead{Conflict of interest/Competing interests}
Authors declare no competing interests

\bmhead{Ethics approval and consent to participate}
Not applicable

\bmhead{Consent for publication}
Not applicable

\bmhead{Data availability}
Not applicable

\bmhead{Materials availability}
Not applicable

\bmhead{Code availability}
Not applicable

\bmhead{Author contribution}
All authors contributed to the original draft, the order reflects the extent of the contribution to the draft and editing. A. Segura supervised all editions, organized the contributions and provided the structure of the draft. All authors reviewed and approved the final version of the manuscript.

\bibliography{Chapter4-bibliography}

\end{document}